%% file: main_arxiv.tex
\documentclass[useAMS,usenatbib]{mn2e}
\usepackage{graphicx}
\usepackage{subfig}
\pdfoutput=1
\usepackage{tikz}
\usepackage{times}
\bibpunct{(}{)}{;}{a}{}{,}
\newcommand{\RadioAppName}[1]{{\sc#1}}

\newcommand{\algorithmName}[1]{{\sc#1}}
\newcommand{\OpticalAppName}[1]{{\sc#1}}
\newcommand{\OtherAppName}[1]{\emph{#1}}
\newcommand{\subfigletter}[1]{\emph{#1}}

\title[Giant Radio Galaxies: I. Intergalactic Barometers]{Giant Radio Galaxies: I. Intergalactic Barometers}

\author[J. M. Malarecki et al.]
  {J. M.~Malarecki,$^1$\thanks{E-mail: jurek.malarecki@icrar.org}
  L.~Staveley-Smith,$^{1,2}$ L.~Saripalli,$^3$\newauthor R.~Subrahmanyan,$^{2,3,4}$ D.~H.~Jones,$^5$ A.~R.~Duffy$^{1,6}$ and M.~Rioja$^{1,7}$ \\
  $^1$International Centre for Radio Astronomy Research, M468, The University of Western Australia, Crawley, WA 6009, Australia\\
  $^2$ARC Centre of Excellence for All-Sky Astrophysics (CAASTRO)\\
  $^3$Raman Research Institute, C V Raman Avenue, Sadashivanagar, Bangalore 560080, India\\
  $^4$National Radio Astronomy Observatory, Socorro, NM, USA\\
  $^5$School of Physics, Faculty of Science, Monash University, Clayton, Victoria 3800, Australia\\
  $^6$School of Physics, University of Melbourne, Parkville, Victoria 3010, Australia\\
  $^7$Observatorio Astron\'{o}mico Nacional (OAN), Apartado 112, E-28803, Alcala de Henares, Espa\~{n}a
  }
\date{Accepted 2013 March 9.  Received 2013 February 12; in original form 2012 October 22}

\pagerange{\pageref{firstpage}--\pageref{lastpage}} \pubyear{2012}

\begin{document}

\include{body}

\end{document}

%% file: body.tex
\label{firstpage}

\maketitle

\begin{abstract}
We present new wideband radio observations with the Australia Telescope Compact Array of a sample of 12 giant radio galaxies. The radio observations are part of a larger radio-optical study aimed at relating the radio structures with the ambient medium on large scales. With projected linear sizes larger than 0.7 Mpc, these objects are ideal candidates for the study of the Warm-Hot Intergalactic Medium (WHIM). The sample includes sources with sizes spanning 0.8 to 3.2 Mpc and total powers of $1.2\times10^{24}$ to $4.0\times10^{26}$ W Hz$^{-1}$ at 2.1 GHz. Redshifts were limited to $z\leq0.15$ to permit spectroscopic observations of the hosts and neighbouring galaxies, which were obtained using the AAOmega spectrograph on the Anglo-Australian Telescope. We derive lobe energy densities from the radio observations via equipartition arguments. The inferred pressures in the lobes of the giant radio sources, which range from $1.1\times10^{-15}$ to $2.0\times10^{-14}$ Pa ($80$ to $1500$ cm$^{-3}$ K), are lower than previously inferred from X-ray observations of dense filaments. Comparison with the OverWhelmingly Large Simulations (OWLS) suggests that the WHIM in pressure balance with the radio lobes has a temperature in excess of $\sim$$10^{6.5}$ K or a particle overdensity in the range 50 to 500. This study highlights the capability of next generation surveys, such as the Evolutionary Map of the Universe (EMU) survey with the Australian Square Kilometre Array Pathfinder (ASKAP), to study populations of giant radio sources at lower surface brightness and thereby discriminate between models for the cosmological evolution of the intergalactic medium and examine the validity of cosmological hydrodynamical simulations.
\end{abstract}


\section{Introduction}
A census of observable baryon density in the present day Universe~\citep[e.g.][]{Shulletal2012} reveals that approximately half of all baryons have not been observed. The missing baryons are thought to reside in galaxy filaments, part of the large-scale structure of the Universe, as tenuous gas with temperatures of $10^5$--$10^7$ K. This Warm-Hot Intergalactic Medium (WHIM) is a generic prediction of hydrodynamic simulations in Lambda-Cold Dark Matter ($\Lambda$CDM) cosmology~\citep[e.g.][]{Daveetal2010}. The WHIM gas has not yet been conclusively detected due to its relatively low density~\citep{Daveetal2001,Werneretal2008,Dietrichetal2012}. There have been a number of absorption line studies (OVI, OVII, CV and HI-Ly$\alpha$) in the FUV/X-ray spectrum of bright background extragalactic sources with potential detections of the WHIM gas~\citep[e.g.][]{Nicastroetal2005Nat,Zappacostaetal2012arXiv}. However, limitations due to low signal-to-noise ratio or uncertainties in the composition of the WHIM (such as metallicity) make it difficult to attribute these detections definitely to the WHIM~\cite[e.g.][]{FraserMcKelvieetal2011,Dietrichetal2012,Zappacostaetal2012arXiv}.

With linear extents well beyond the interstellar medium and host galaxy halos, \emph{giant radio galaxy} lobes (GRG; source sizes \textgreater~0.7 Mpc) are truly unique as they evolve within the tenuous and pervasive intergalactic medium (IGM). The radio lobes expand into the ambient medium allowing them to act as tracers of the large-scale structure (LSS) of the Universe. Their location and orientation relative to the LSS in the local galaxy distribution can modify their morphology. The large extents of GRGs provide an opportunity to investigate the distribution and physical state of the gas associated with galaxy filaments.

A number of known GRGs exhibit asymmetric structures over their megaparsec extents~\citep{Schoenmakersetal2000b,Laraetal2001,Machalskietal2001,Saripallietal2005}. This may indicate significant interaction with the ambient medium during the evolution of their lobes. As with more typical radio galaxies a few hundred kiloparsecs in size, these asymmetries may result from an asymmetric distribution of dense gas~\citep[e.g.][]{McCarthyetal1991} or different pressure gradients encountered by the jets and lobes~\citep[e.g.][]{Burns1998}.

Two recent studies have explored the use of GRGs in probing the WHIM~\citep{Subrahmanyanetal2008,Safourisetal2009}. In both cases, sensitive radio continuum images of a GRG (in total intensity, polarisation and distribution of spectral index) were used to relate the extended morphology with the ambient medium on the megaparsec scale using spectroscopic observations of galaxies distributed in a 2$^{\circ}$ field surrounding the radio galaxy hosts. Both studies showed clear correspondence between the properties of the giant synchrotron lobes and the neighbouring galaxy distributions demonstrating the potential of using GRGs in studying the WHIM gas within which their lobes evolve and interact.

We have embarked on a program using a larger sample of GRGs in the study of the WHIM properties. In this paper, we present the sample of GRGs used in our study. We have made new radio images of most of the sources in the sample to obtain good representations of their extended structures. Here we describe these radio observations and examine the structures revealed as well as some of the properties relevant to our WHIM study. The plan of the paper is as follows: we outline the GRG sample in \textsection\ref{sec:SampleCriteria} and describe the recent wideband observations with the Australia Telescope Compact Array (ATCA) in \textsection\ref{sec:ATCA}. Optical spectra corresponding to the host galaxies are presented in \textsection\ref{sec:Optical}. The data analysis procedure used in reducing the radio data is detailed in \textsection\ref{sec:Imaging}. We then provide notes on each of the sources in \textsection\ref{sec:Results}. Pressure estimates in the diffuse parts of the radio lobes are determined from measurements of surface brightness and equipartition arguments in~\textsection\ref{sec:LobePressures}. Finally, in~\textsection\ref{sec:Discussion} we discuss the properties of the sample.

\section{Source Selection}
\label{sec:SampleCriteria}
We have assembled a sample of 17 GRGs in the redshift range $0.05$--$0.15$ from several complete radio source samples~\citep{JonesMcAdam1992,Subrahmanyanetal1996,Saripallietal2005}, which were themselves compiled from the 408~MHz Molonglo Reference Catalogue~\citep[MRC;][]{Largeetal1981} and the 843~MHz Sydney University Molonglo Sky Survey~\citep[SUMSS;][]{Bocketal1999,Mauchetal2003}, and including one GRG from the Australia Telescope Low Brightness Survey~\citep[ATLBS;][]{Subrahmanyanetal2010}. Our final sample (19 GRGs) for the main study contains two more giant radio galaxies taken from the case studies of~\citet{Subrahmanyanetal2008} and~\citet{Safourisetal2009}, MSH 05-22 and MRC B0319--454 respectively.

The main criterion for choosing these sources besides linear size (\textgreater 700 kpc) is redshift: only those that were nearby, with $z$ less than approximately 0.15, were included to enable spectroscopic observations of neighbouring galaxies using the AAOmega at the Anglo-Australian Telescope (AAT). Of the 19 sources, 12 required further radio observations to image their extended structure with better sensitivity. Table~\ref{tbl:ATCA} lists the sources observed with the ATCA and Table~\ref{tbl:properties} gives the basic physical parameters of each source, including classification in terms of symmetry. A source is described as asymmetric if the ratio between the distances from the core to the outermost extent of each lobe exceeds 1.3. We adopt a flat cosmology with parameters $\Omega_m = 0.27$, $\Omega_{\Lambda} = 0.73$ and a Hubble constant of $H_0 = 71\mbox{ km s}^{-1}\mbox{Mpc}^{-1}$. The sources in our sample span a range of total power from $1.2\times10^{24}$ to $4.0\times10^{26}\mbox{ W Hz}^{-1}$ (2.1 GHz).

\section{Radio Observations}
\label{sec:ATCA}
\begin{figure*}
  \centering
  \begin{tabular}{cc}
      \includegraphics[width=\columnwidth]{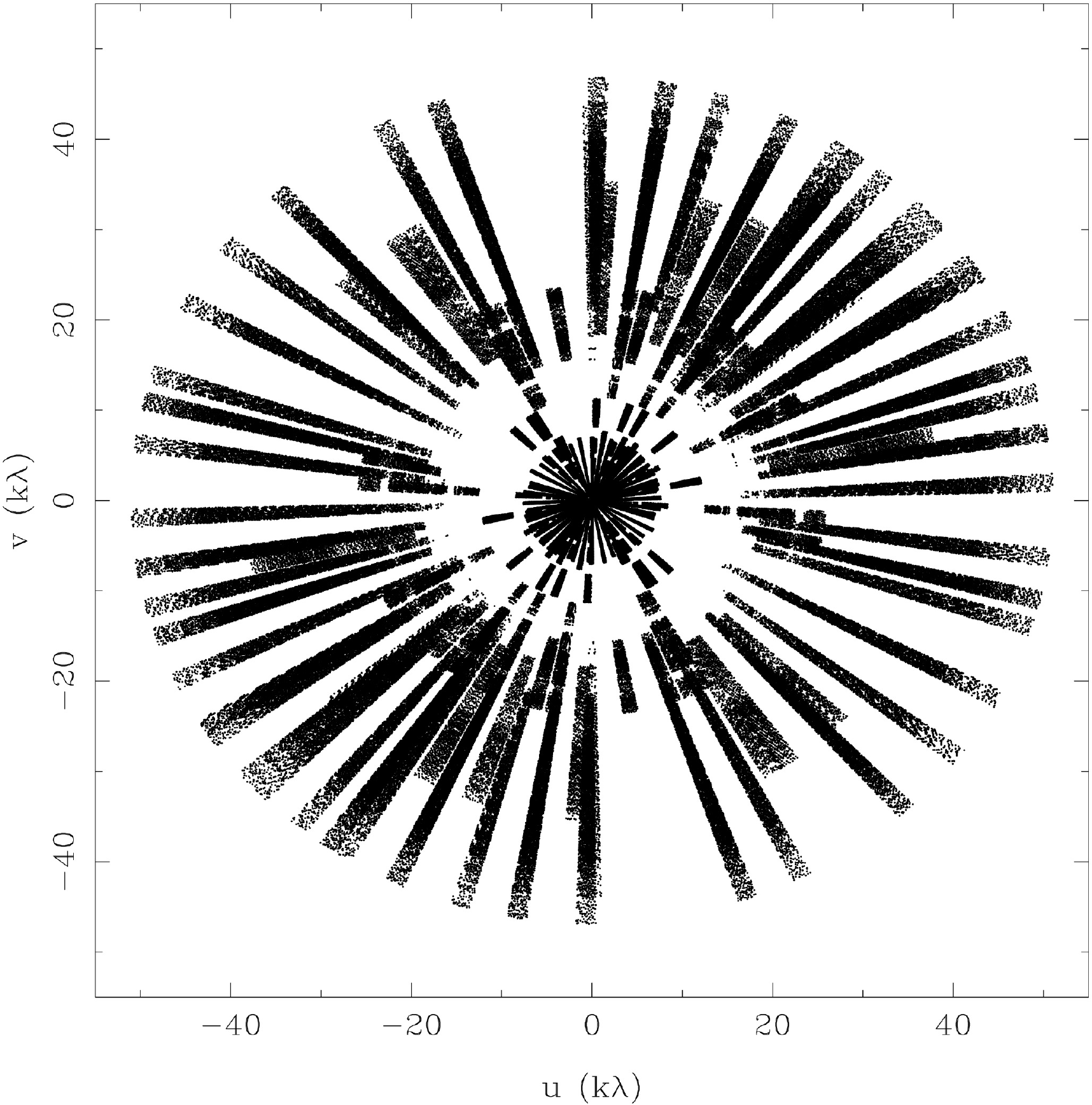}
      &
      \includegraphics[width=\columnwidth]{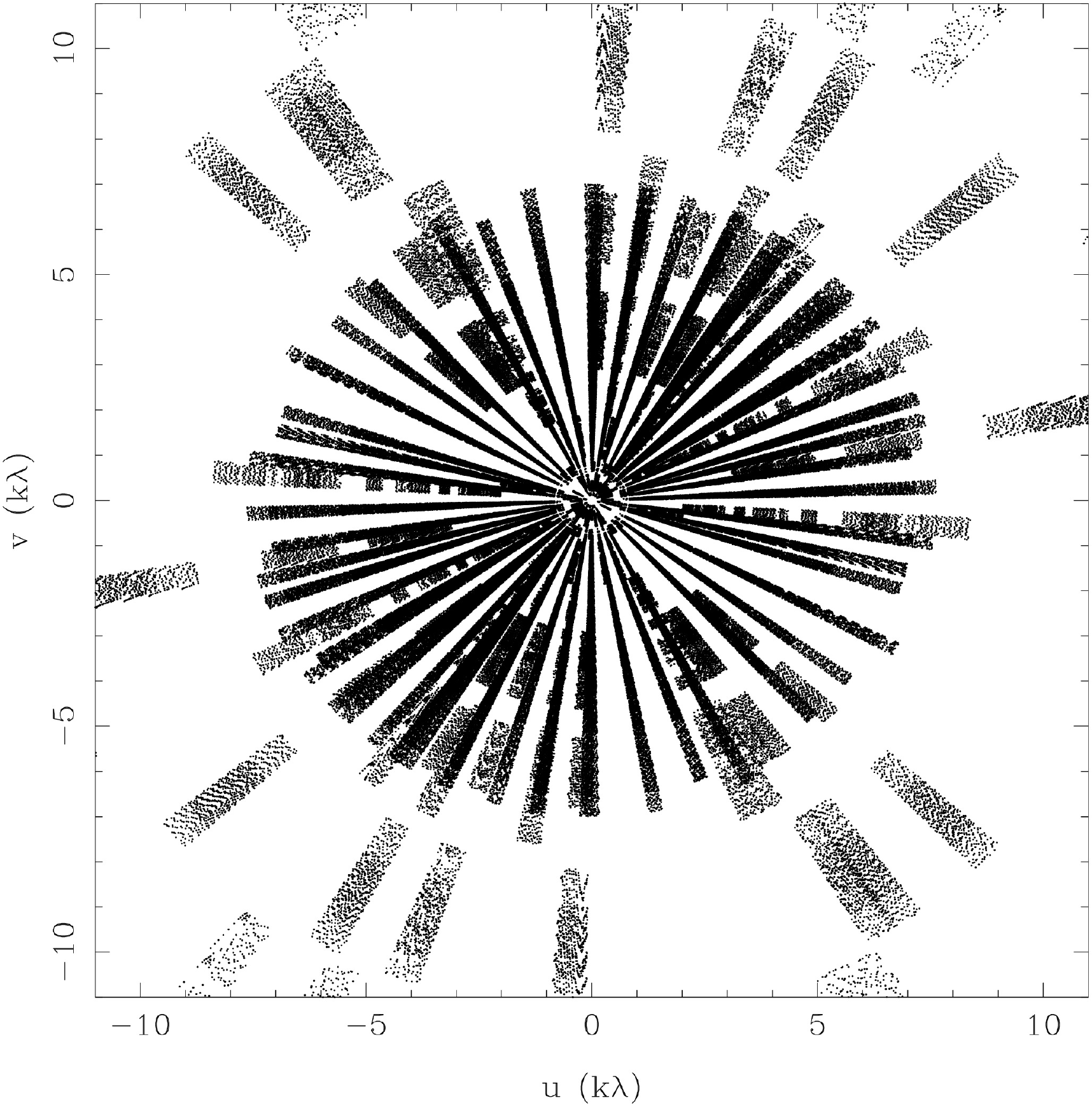}
\end{tabular}
\caption{Total u--v coverage (left panel) and central u--v coverage (right panel) for J0034--6639 using multi-frequency synthesis.}
\label{fig:J0034UVcoverage}
\end{figure*}

We observed 12 GRGs using the ATCA in order to obtain better images of their extended radio morphologies. Observations of this sample were made using the 1.5D array configuration at 1.5 and 2.3 GHz, and the EW352 configuration at 1.5 GHz in three consecutive 24 hr sessions during July 2010. Antenna 4 was not used for the period of these observations as it lacked a suitable receiver for the required frequencies. Additional observations were made in October 2011 with the 750C configuration utilising the complete 2 GHz bandwidth capability of the Compact Array Broadband Backend (CABB) correlator centred at 2.1 GHz. The frequency resolution is 1 MHz per channel. The total observing time for each source in each configuration is listed in Table~\ref{tbl:ATCA} and the typical u--v coverage achieved is shown in Fig.~\ref{fig:J0034UVcoverage}.

\begin{table}
 \caption{A list of objects observed with the ATCA. Array configuration, frequency, on-source observing time and RMS noise is also listed.}
 \label{tbl:ATCA}
 \begin{tabular}{lcccc}
  \hline
  Name & Array & Frequency & Time & RMS noise \\
              &            & (GHz)         & (hr)    & ($\mu$Jy) \\
  \hline
  J0034--6639 & 1.5D & 1.5 & 1.8 & 40 \\
             & 1.5D & 2.3 & 2.1 & \\
             & EW352 & 1.5 & 1.8 & \\
             & 750C & * & 4.7 & \\
  J0331--7710 & 1.5D & 1.5 & 2.4 & 85 \\
             & 1.5D & 2.3 & 2.6 & \\
             & EW352 & 1.5 & 2.1 & \\
             & 750C & * & 5.3 & \\
  J0400--8456 & 1.5D & 1.5 & 2.4 & 100 \\
             & 1.5D & 2.3 & 3.5 & \\
             & EW352 & 1.5 & 2.2 & \\
             & 750C & * & 6.0 & \\
  J0459--528  & 1.5D & 1.5 & 0.8 & 150 \\
             & 1.5D & 2.3 & 1.4 & \\
             & EW352 & 1.5 & 1.3 & \\
             & 750C & * & 3.2 & \\
  B0511--305  & 1.5D & 1.5 & 1.2 & 1250 \\
             & 1.5D & 2.3 & 1.2 & \\
             & EW352 & 1.5 & 1.8 & \\
             & 750C & * & 3.3 & \\
  B0703--451  & 1.5D & 1.5 & 1.3 & 250 \\
             & 1.5D & 2.3 & 1.1 & \\
             & EW352 & 1.5 & 1.6 & \\
             & 750C & * & 2.9 & \\
  J0746--5702 & 1.5D & 1.5 & 1.3 & 120 \\
             & 1.5D & 2.3 & 1.3 & \\
             & EW352 & 1.5 & 1.9 & \\
             & 750C & * & 3.1 & \\
  J0843--7007 & 1.5D & 1.5 & 1.7 & 95 \\
             & 1.5D & 2.3 & 2.0 & \\
             & EW352 & 1.5 & 2.8 & \\
             & 750C & * & 3.3 & \\
  B1302--325  & 1.5D & 1.5 & 1.3 & 800 \\
             & 1.5D & 2.3 & 1.3 & \\
             & EW352 & 1.5 & 1.3 & \\
             & 750C & * & 2.8 & \\
  B1308--441  & 1.5D & 1.5 & 1.6 & 400 \\
             & 1.5D & 2.3 & 1.7 & \\
             & EW352 & 1.5 & 1.4 & \\
             & 750C & * & 3.6 & \\
  J2159--7219 & 1.5D & 1.5 & 2.1 & 75 \\
             & 1.5D & 2.3 & 2.4 & \\
             & EW352 & 1.5 & 2.4 & \\
             & 750C & * & 5.5 & \\
  B2356--611  & 1.5D & 1.5 & 1.3 & $1.5\times10^{4}$ \\
             & 1.5D & 2.3 & 1.2 & \\
             & EW352 & 1.5 & 1.0 & \\
             & 750C & * & 4.2 & \\
  \hline
 \end{tabular}

 \medskip
 *2 GHz wideband (1 MHz/channel centred at 2.1 GHz) CABB data\\
 1.5D 1.5 GHz July 12, 2010 03:31 LST - July 13, 2010 3:25 LST\\
 1.5D 2.3 GHz July 13, 2010 03:28 LST - July 14, 2010 3:56 LST\\
 EW352 1.5 GHz July 15, 2010 3:27 LST - July 16, 2010 4:13 LST\\
 750C* October 28, 2011 6:55 LST - November 2, 2011 6:39 LST
\end{table}

\begin{table*}
 \caption{Observed properties of the GRG sample obtained using the ATCA at 2.1 GHz and the 2dF/AAOmega spectrograph on the AAT.}
 \label{tbl:properties}
 \begin{tabular}{llccccccc}
  \hline
  Name                       & Other names                                & \multicolumn{2}{c}{Host position (J2000)} & Redshift   & Largest angular size & Linear size & Asymmetric \\
                                    &                                                        & RA                 & Dec                                          &                   & (arcmin)                       & (Mpc)          &                       \\
  \hline
  J0034--6639.......... &2MASX J00340560--6639354 &$00~34~06$ &$-66~39~35$                         &$0.1103$  &$16.8$                          &$2.00$         & no                 \\
  J0331--7710.......... &2MASX J03313973--7713189 &$03~31~40$ &$-77~13~19$                         &$0.1446$  &$21.4$                          &$3.21$         & yes               \\
                                    &PKS B0333--773                         &                        &                                                  &                   &                                       &                      &                      \\
                                    &PMN J0331--7709                      &                        &                                                  &                   &                                       &                      &                      \\
                                    &PMN J0331--7719                      &                        &                                                  &                   &                                       &                      &                      \\
  J0400--8456.......... &PMN J0401--8457                      &$04~01~18$ &$-84~56~36$                         &$0.1033$ &$8.0$                             &$0.90$         & no                \\
  J0459--528..........   &2MASX J04590828--5252072 &$04~59~08$ &$-52~52~08$                         &$0.0957$  &$10.2$                          &$1.08$         & no                \\
  B0511--305..........  &PKS B0511--305                         &$05~13~32$ &$-30~28~50$                         &$0.0576$  &$17.5$                          &$1.16$         & yes               \\
  B0703--451..........  &PKS B0703--45                           &$07~05~31$ &$-45~13~12$                         &$0.1242$  &$8.4$                             &$1.11$         & yes               \\
  J0746--5702.......... &PMN J0746--5703                     &$07~46~19$ &$-57~02~59$                         &$0.1301$  &$7.4$                             &$1.02$         & no                \\
                                    &PMN J0746--5702                     &                        &                                                  &                    &                                       &                      &                    \\
                                    &PMNM 074521.2--565530       &                        &                                                  &                    &                                       &                      &                   \\
  J0843--7007.......... &PMN J0843--7006                     &$08~43~06$ &$-70~06~56$                         &$0.1390$  &$6.9$                             &$1.01$         & no             \\
  B1302--325..........   &2MASX J13045851--3249162&$13~04~59$ &$-32~49~14$                         &$0.1528$  &$8.4$                            &$1.33$         & no             \\
                                    &PKS B1302--325                       &                        &                                                  &                    &                                      &                      &                    \\
  B1308--441..........   &PKS B1308--441                       &$13~11~24$ &$-44~22~40$                          &$0.0507$  &$16.0$                         &$0.93$         & yes             \\
  J2159--7219.......... &PMN J2159--7220                    &$21~59~10$ &$-72~19~07$                          &$0.0967$   &$11.7$                         &$1.24$         & no              \\
  B2356--611..........   &PKS B2356--61                         &$23~59~04$ &$-60~55~00$                          &$0.0962$   &$8.0$                              &$0.84$         & no               \\
  \hline
 \end{tabular}
 
\medskip
Notes -- Units of right ascension are hours, minutes, and seconds, and units of declination are degrees, arcminutes, and arcseconds. The largest angular size of each source was measured from its total intensity radio contours at the $12\sigma$-level (or $6\sigma$ for J0331--7710 and B0703--451 due to morphology), where $\sigma$ is the RMS noise in the 2.1 GHz low resolution image. Projected linear sizes were calculated using angular distance scale values courtesy~\citet{Wright2006}.
\end{table*}

\section{Optical Observations}
\label{sec:Optical}
Optical spectra for 13 of the 19 giant radio galaxies have been published previously (B0319--454:~\citet{BryantHunstead2000}; J0331--7710, J0400--8456, J0746--5702, J0843--7007, J2159--7219:~\citet{Saripallietal2005}; B0503--286:~\citet{SubrahmanyaHunstead1986}; J0515--8100:~\citet{Subrahmanyanetal2006}; B1302--325:~\citet{Simpsonetal1993}; B1308--441, B1545--321, B2014--558:~\citet{Simpsonetal1996}; B2356--611:~\citet{KoekemoerBicknell1998}). As part of the larger study aimed at examining the relationship between the properties (including asymmetry) of GRGs and the ambient medium, we re-observed most of the sample (15 GRGs) and obtained optical spectroscopic data with the 2dF/AAOmega spectrograph on the 3.9 m Anglo-Australian Telescope. In this paper where we define the GRG sample and present the new radio images for 12 we also present, for completeness and convenience, the host optical spectra and redshifts for these 12 (as well as the remaining 3 GRGs) obtained in a consistent manner using our optical observations. For 5 GRGs (J0034--6639, J0459--528, B0511--305, B0703--451 and B0707--359) the optical spectra are being presented for the first time. The spectra were obtained with separate red and blue cameras and gratings, and together span the range $\sim$400--900 nm. The spectral dispersion varies from 0.1 to \mbox{0.16 nm pixel$^{-1}$} and the spectral resolving power $\left(\frac{\lambda}{\Delta\lambda}\right)$ is $R=1300$ for a given wavelength ($\lambda$). We reduced the raw data with the software package \OtherAppName{2dFDR}, and used the program \OpticalAppName{runz} (modified from the original version written by W. Sutherland) to classify the resulting spectra and assign redshifts to reliable spectra.

\section{Data Analysis}
\label{sec:Imaging}
The radio data were reduced in \RadioAppName{miriad}~\citep{Saultetal1995} following the standard procedure in the \RadioAppName{miriad} user guide; multifrequency synthesis (MFS) was used for the complete ATCA data set. The primary calibrator PKS B1934--638 was used in bandpass and flux calibration with its flux density taken as 12.58 Jy at 2.1 GHz. It was also used to derive the instrumental polarisation leakages for some of the 2011 observations where there was inadequate parallactic angle coverage of the phase calibrators due to the allocated observing sessions. This procedure permitted the Stokes parameters Q and U to be determined from the phase calibrators in all cases. Radio frequency interference (RFI) in the 2010 data set was manually flagged in \RadioAppName{miriad} while an automated Stokes V-based procedure was used to flag the wideband 2011 data.

Two sets of radio maps for the Stokes parameters I, Q and U were produced either including or excluding antenna 6, corresponding to the longest baselines. These 6 and 1.5 km maximum baselines represent resolutions of 6 and 24 arcsec, respectively. We chose the Brigg's robustness parameter to be $-1$ as it provided the best compromise between recovering extended structures and introducing artifacts due to gaps in the u--v coverage. This weighting scheme was used for all images unless otherwise noted. Initial deconvolution was performed using the~\algorithmName{clean} algorithm with Clark clean iterations and MFS enabled. The phase self-calibration method outlined in the \RadioAppName{miriad} user guide was used to improve the quality of the images. A negative value cutoff for Stokes I was chosen for deconvolution to create a reliable model for the self-calibration and a 3$\sigma$ cutoff was used for Stokes Q and U, where $\sigma$ is the RMS noise in the initial Stokes V dirty map. The best model was subtracted from the calibrated visibilities and further flagging was carried out on the residuals. The resulting flags were applied to the calibrated data and the self-calibration procedure was repeated to produce the final images in \RadioAppName{miriad}. The lower resolution images underwent a final deconvolution stage in \RadioAppName{casa} using a Multi-Scale Multi-Frequency Synthesis (MS-MFS)~\algorithmName{clean} algorithm~\citep[see][]{Richetal2008,RauCornwell2011} to preserve any diffuse extended structure.

A spectral index distribution map was generated for each source using the corresponding 843 MHz SUMSS image, with a $45 \times 45$ cosec$(\delta)$ arcsec beam, and a 2.8 GHz ATCA image consisting of a 2.7 to 2.9 GHz subband. However, in the case of B0703-451 a pointing offset precluded the use of higher frequency ATCA data and so a subband centred at 1.55 GHz was used. In all cases, each pair of images was smoothed to a common resolution. Spectral index, $\alpha$, is defined here by the relation $S_{\nu} \propto \nu^{\alpha}$, where $S$ is flux and $\nu$ is frequency. An estimate of core spectral index for J0331-7710 was calculated from flux density measurements in two subbands, 1.39 to 1.43 GHz and 2.7 to 2.9 GHz. Uniform weighting was applied to the visibilities in this case to enhance the compact structures.

We generated polarised intensity, fractional polarisation and polarisation position angle images for each source using subbands centred at 1.55 and 2.8 GHz, respectively. The polarised intensity images were masked according to the $3\sigma$ level in total intensity. The polarisation position angle images at 1.55 and 2.8 GHz were combined to produce Faraday rotation-corrected position angle images. A polarisation map was produced for each source showing electric field vectors that have been corrected for Faraday rotation. In each map, vector lengths represent polarised intensity rather than fractional polarisation as low-level flux in a few subband images resulted in spurious percentage values. These vectors were overlaid on the corresponding wideband, total intensity contours and greyscale pixel map for comparison.

\section{Results}
\label{sec:Results}
A number of properties of the set of 12 GRGs observed with the ATCA are listed in Table~\ref{tbl:properties}. The radio images are shown in Fig.~\ref{fig:J0034}--\ref{fig:B2356} in addition to the optical spectra of the GRG host galaxies. The remaining host spectra obtained to date are presented in Fig.~\ref{fig:optical}. Significant features in the spectra, which were used to estimate redshifts, are indicated by dashed lines and spurious features due to night-sky lines are marked in magenta. Although most of our sample have measured redshifts in the literature (J0034--6639:~\citet{Saripallietal2012}; J0116--473 = B0114--476:~\citet{Saripallietal2002}; B0511--305, B1545--321, B2356--611:~\citet{Subrahmanyanetal1996}; B0319--454:~\citet{Safourisetal2009}; J0331--7710, J0400--8456, J0746--5702, J0843--7007, J2159--7219:~\citet{Saripallietal2005}; B0503--286:~\citet{Subrahmanyanetal2008}; J0515--8100:~\citet{Subrahmanyanetal2006}; B1302--325:~\citet{Simpsonetal1993}; B1308--441, B2014--558:~\citet{JonesMcAdam1992}), we give for consistency, redshifts measured using our observations. Redshifts derived for the 12 sources observed with ATCA are listed in Table~\ref{tbl:properties} and redshifts for the complete GRG sample are included in Table~\ref{tbl:FluxPressure}. Spectra for companion galaxies will be presented in a future paper. Below we give notes on each of the sources based on our radio observations.

\subsection{J0034--6639 (Fig.~\ref{fig:J0034})}
This giant radio source ($z=0.1103$), observed as part of the ATLBS survey~\citep{Subrahmanyanetal2010,Saripallietal2012}, is over 2 Mpc in linear extent. It has symmetric, relic lobes that maintain an almost uniform width along their megaparsec lengths (Fig.~\ref{fig:J0034}). It is a restarted jet galaxy with a recessed hotspot in its northern lobe but there is no indication of a corresponding feature in the southern lobe. A linear, diffuse emission feature extends well beyond the hotspot in the northern lobe. The source is asymmetric in polarisation with more polarised flux in the southern lobe. There is some polarised flux around the northern hotspot of approximately 9 per cent (Fig.~\ref{fig:J0034}\subfigletter{c}). The optical spectrum of the host galaxy in Fig.~\ref{fig:J0034}\subfigletter{e} shows absorption line features typical of an early-type galaxy.

\begin{figure*}
  \centering
  \begin{tabular}{cc}
      \includegraphics[width=0.5\hsize]{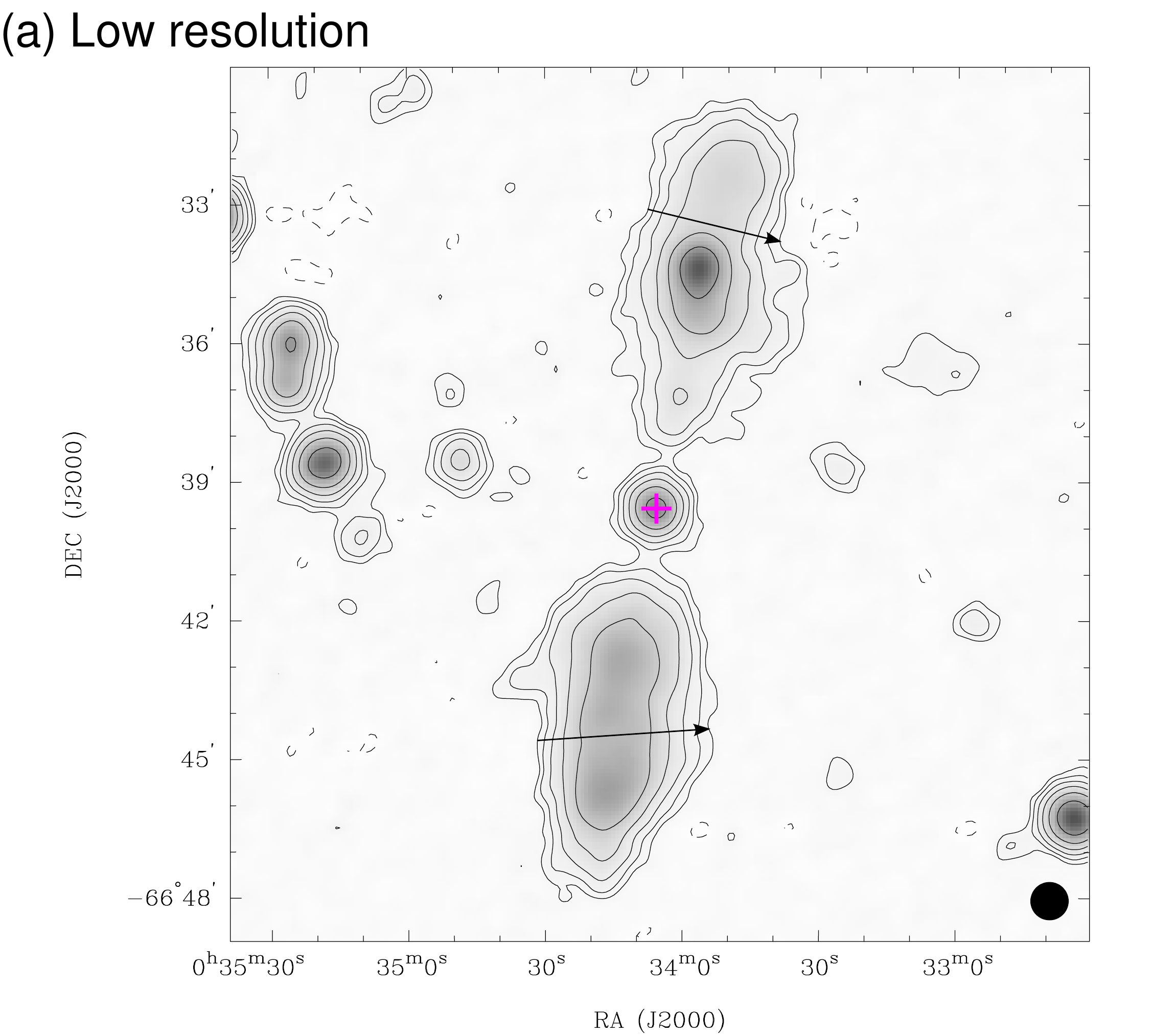} &
      \includegraphics[width=0.5\hsize]{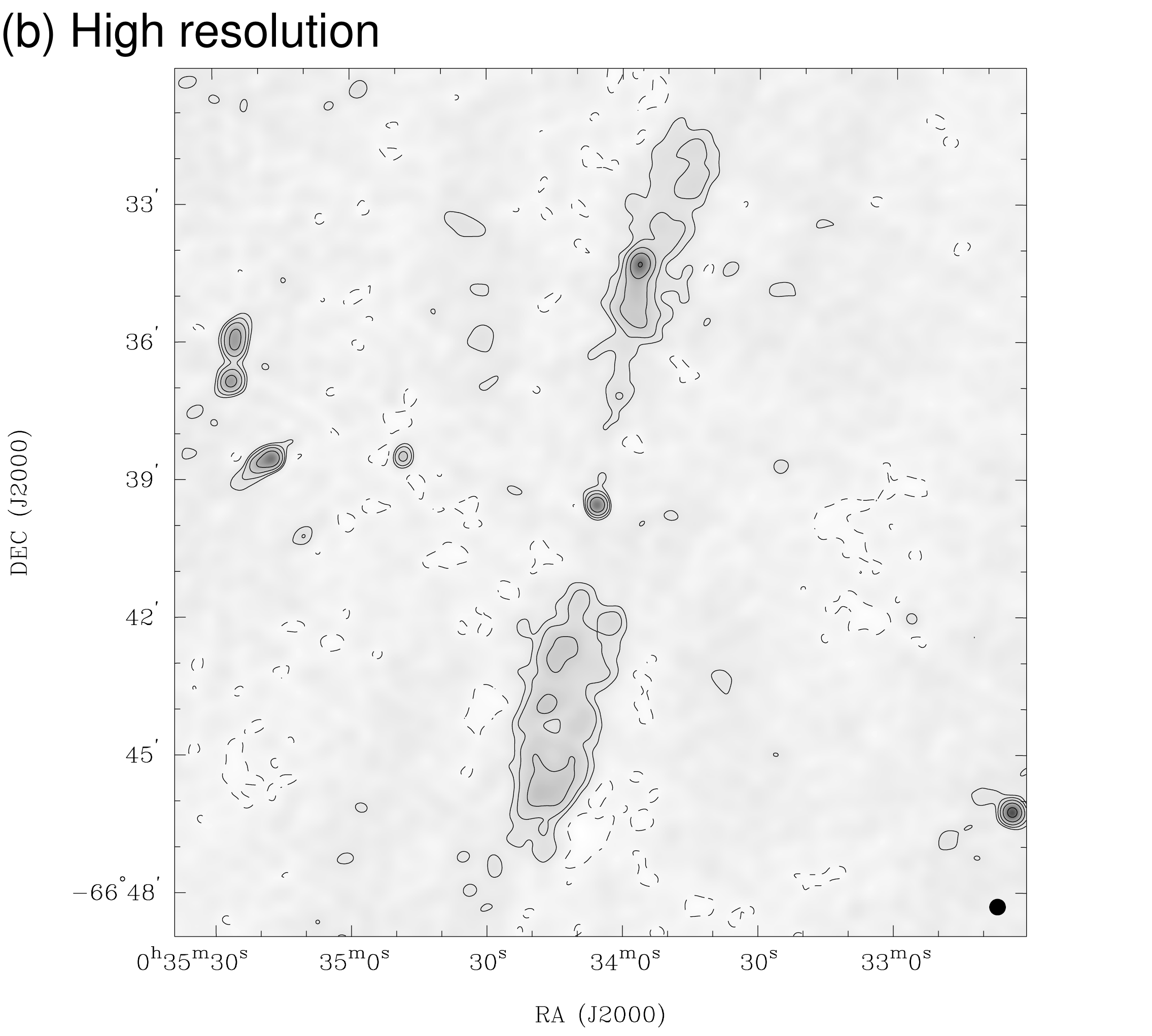} \\
      \includegraphics[width=0.5\hsize]{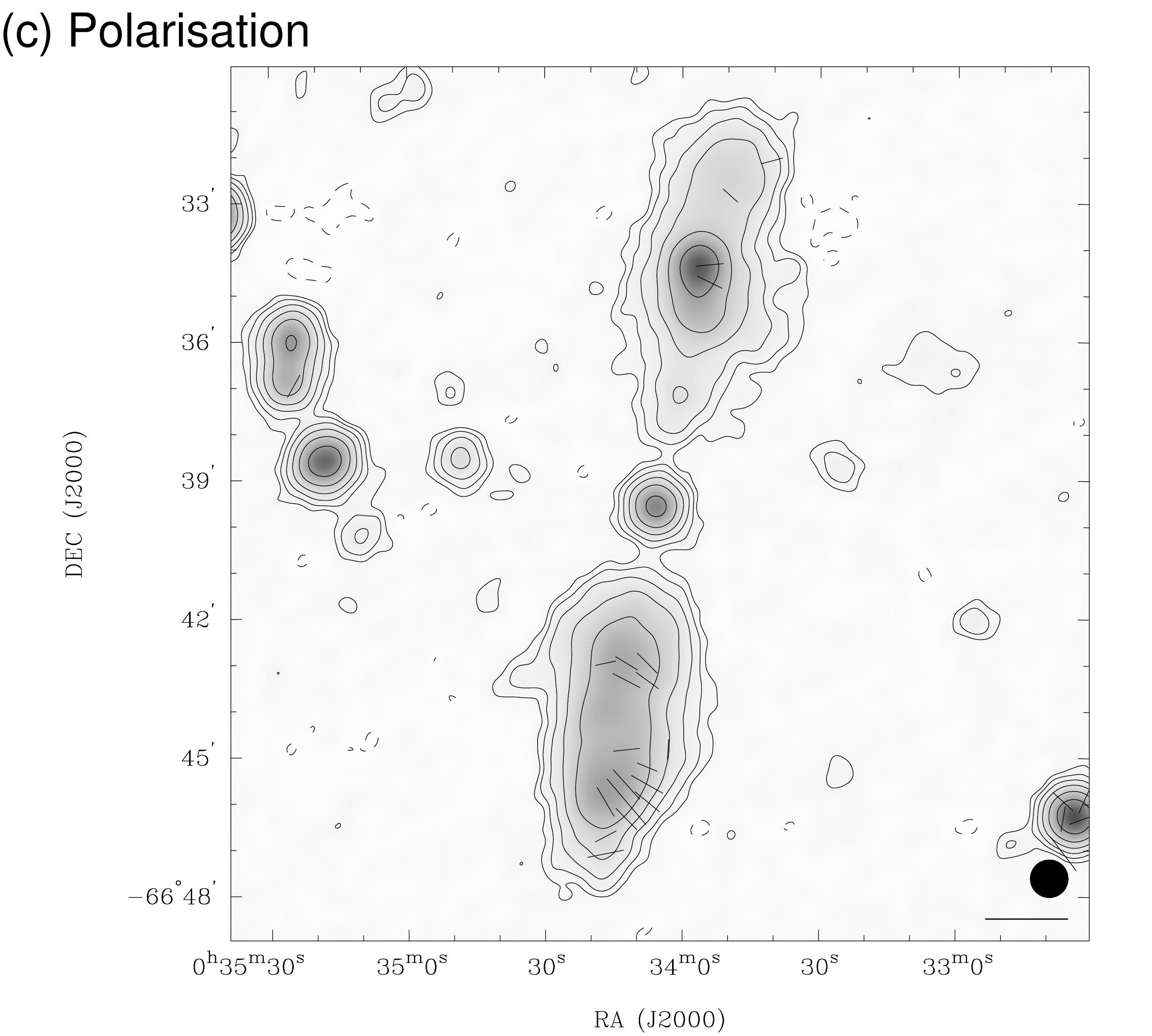} &
      \includegraphics[width=0.5\hsize]{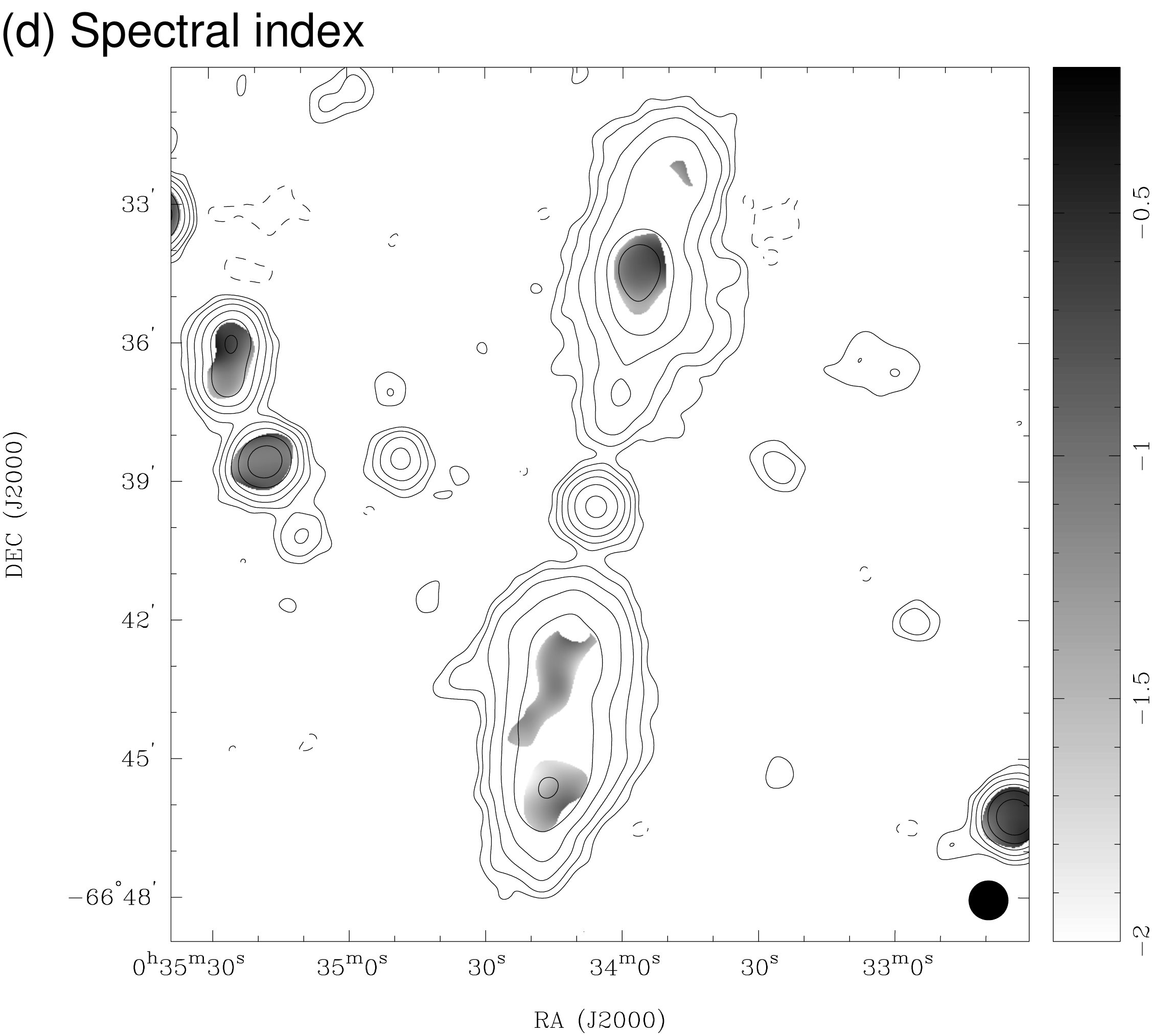}
\end{tabular}
\caption{(\subfigletter{a}) Low-resolution, wideband, total intensity image of J0034--6639 at 2.1 GHz in greyscale with contours at -3, 3, 6, 12, 24, 48, and 96 $\times$ 40 $\mu$Jy beam$^{-1}$ and a symmetric beam of FWHM 48\arcsec. The lowest contour is at a level of 3 times the RMS noise in the image. The half-power size of the synthesized beam is displayed in the lower right corner of this image, as well as all subsequent radio maps. The arrows indicate the directions of the slice profiles used to derive energy densities, see \textsection\ref{sec:LobePressures}, and the position of the GRG host is marked with a cross, which are also present in all following low resolution radio maps. In the online version the host positions are shown in magenta (\textcolor{magenta}{+}). (\subfigletter{b}) High resolution, total intensity image at 2.1 GHz with contours at -3, 3, 6, 12, 24, and 48 $\times$ 45 $\mu$Jy beam$^{-1}$ and FWHM 20\arcsec~beam. (\subfigletter{c}) Distribution of polarised intensity of J0034--6639 at 2.8 GHz shown as electric field vectors with lengths representing polarised intensity and position angles corrected for Faraday rotation. This map and all following polarised intensity images include a scale-bar in the lower right corner that represents a reference polarised flux level, in this case 1 mJy, and are overlaid on the corresponding low resolution, total intensity contours levels and greyscale pixel map for comparison. (\subfigletter{d}) Distribution of spectral index computed from narrow band images at 843 MHz (SUMSS) and 2.8 GHz (ATCA) with beam FWHM 50\arcsec. Spectral index follows the definition $S_{\nu} \propto \nu^{\alpha}$ and is shown as greyscale in the range $-2$ to $-0.2$. The overlaid total intensity contour levels are -3, 3, 6, 12, 24, 48, and 96 $\times$ 40 $\mu$Jy beam$^{-1}$.}
\label{fig:J0034}
\end{figure*}

\begin{figure}
  \centering
      \includegraphics[width=\hsize]{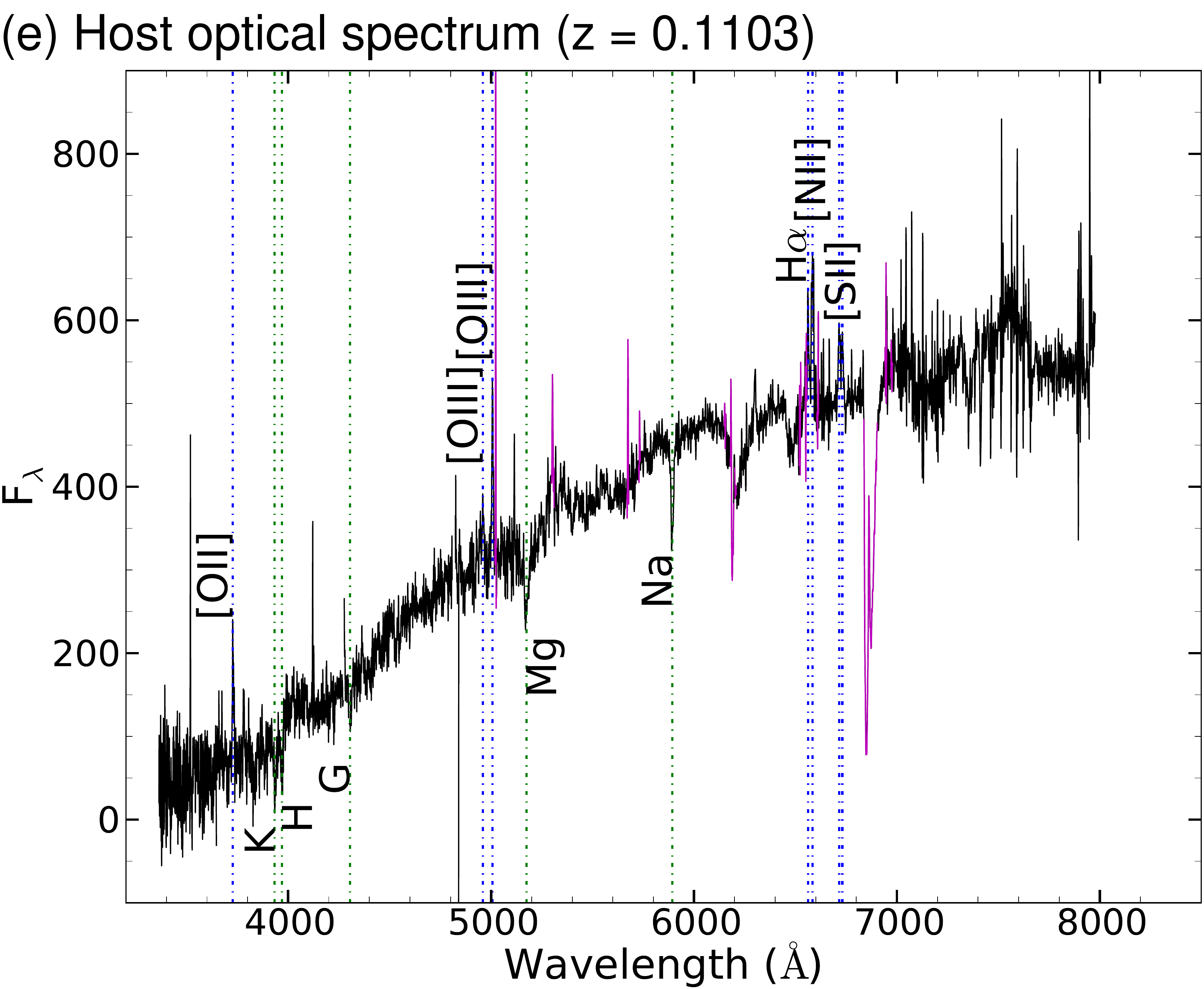}
  \contcaption{(\subfigletter{e}) An optical spectrum from AAOmega on the AAT of the host galaxy of J0034--6639. The spectrum has been shifted to rest frame using the measured redshift of 0.1103. This spectrum, as well as all following optical spectra, shows: common galaxy absorption features marked with dotted green vertical lines; emission-line features with dotted blue vertical lines; and spectral regions traced in magenta indicating non-zero sky-subtraction residuals from telluric emission and absorption features. The flux scale is arbitrary.}
\end{figure}

\subsection{J0331--7710 (Fig.~\ref{fig:J0331})}
This GRG ($z=0.1446$) is highly asymmetric in lobe morphology, brightness and separation. In particular, there is a prominent gap in emission between the core and the fainter southern lobe (Fig. \ref{fig:J0331}). In contrast, there is a much brighter lobe in the north that appears confined to a nearly constant width. The new ATCA images reveal detailed structure in these lobes for the first time. Both lobes despite differing in many respects are found to have compact hotspots at their extremities. The southern lobe however exhibits a `bottle-necked' structure not seen at the northern lobe end. Interestingly, our new images reveal a significantly fainter, 2 arcminute emission feature extending well beyond the bright edge of the northern lobe, as also seen in J0034--6639. The only previous radio image of the entire source (843 MHz SUMSS; Fig.~\ref{fig:J0331}\subfigletter{c}) does not detect this extended feature.

\citet{Saripallietal2005} detected a weak core at the inner edge of the northern lobe, which is confirmed by the high-resolution radio map using all available baselines in Fig.~\ref{fig:J0331}\subfigletter{b}. The core has a highly inverted spectral index of +0.37. The spectral indices of nearby point sources were confirmed to have reasonable values ($\alpha\approx-0.5$), including the small wide-angle tail source to the east of the core. In particular, the wide-angle tail (WAT) source has steep indices away from its peaks as would be expected from the plumes of an FRI source. The northern lobe of J0331--7710 has steeper spectral indices than the fainter southern lobe and the indices steepen away from the northern hotspot along the lobe length as expected for an edge brightened lobe (Fig.~\ref{fig:J0331}\subfigletter{e}).

J0331--7710 is also asymmetric in polarisation properties with the northern lobe having much higher polarised flux. The mean fractional polarisation in the northern lobe is 25 per cent, while in the southern lobe it is approximately 40 per cent. Both lobes show evidence of the magnetic field running parallel with the lobe edges (Fig.~\ref{fig:J0331}\subfigletter{d}). The host optical spectrum in Fig.~\ref{fig:J0331}\subfigletter{f} is typical of an early-type galaxy.

\begin{figure*}
  \centering
  \begin{tabular}{cc}
      \includegraphics[width=0.5\hsize]{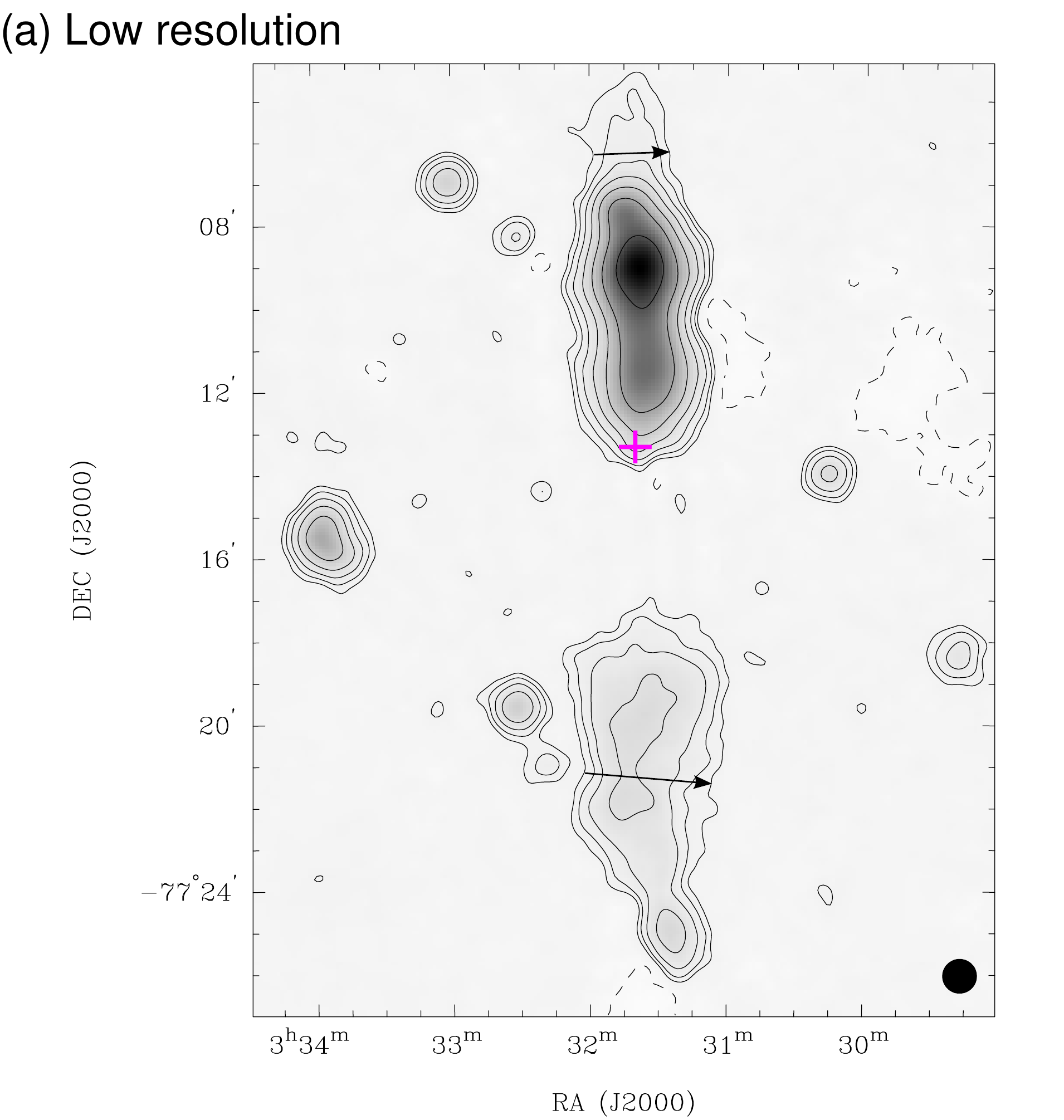} &
      \includegraphics[width=0.5\hsize]{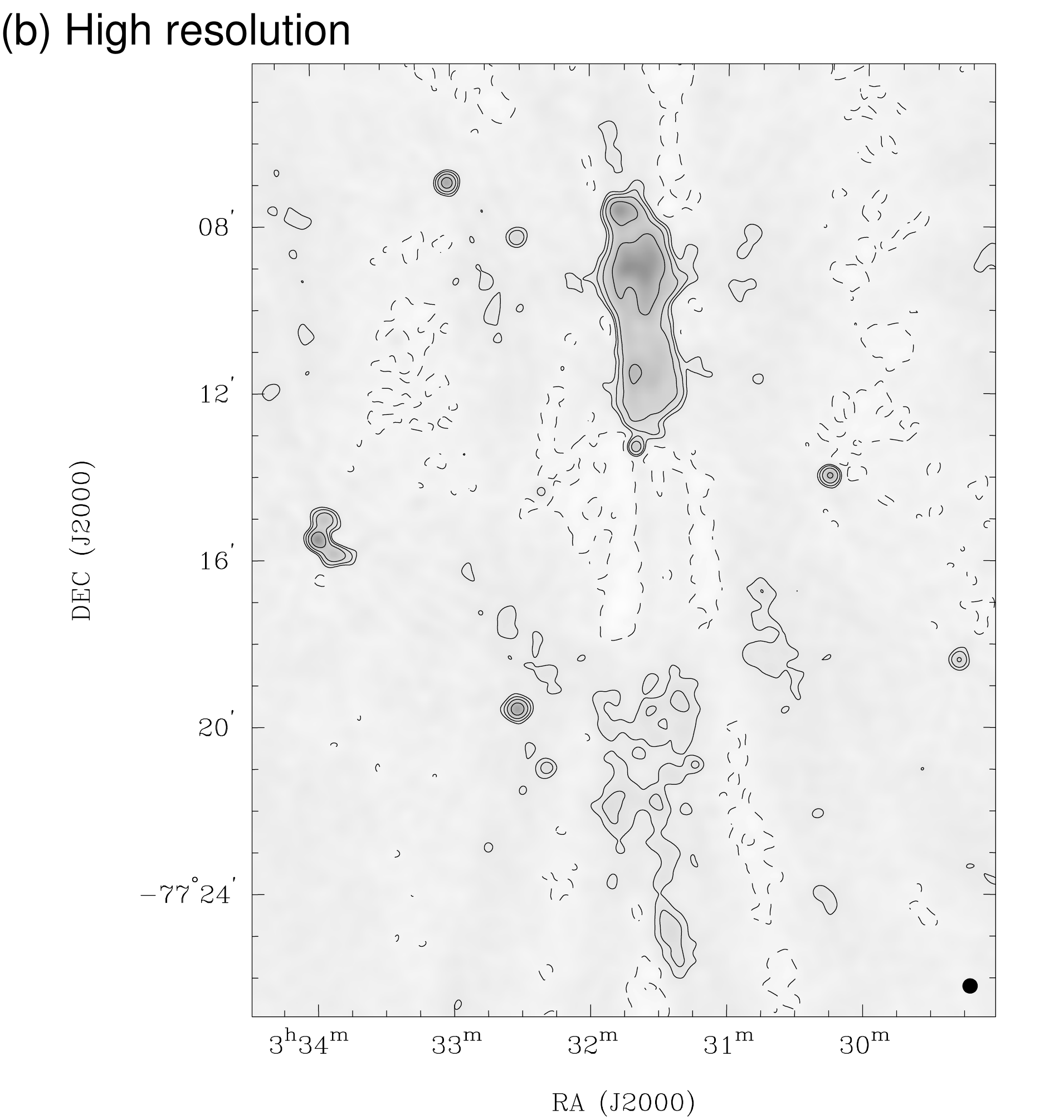} \\
      \includegraphics[width=0.5\hsize]{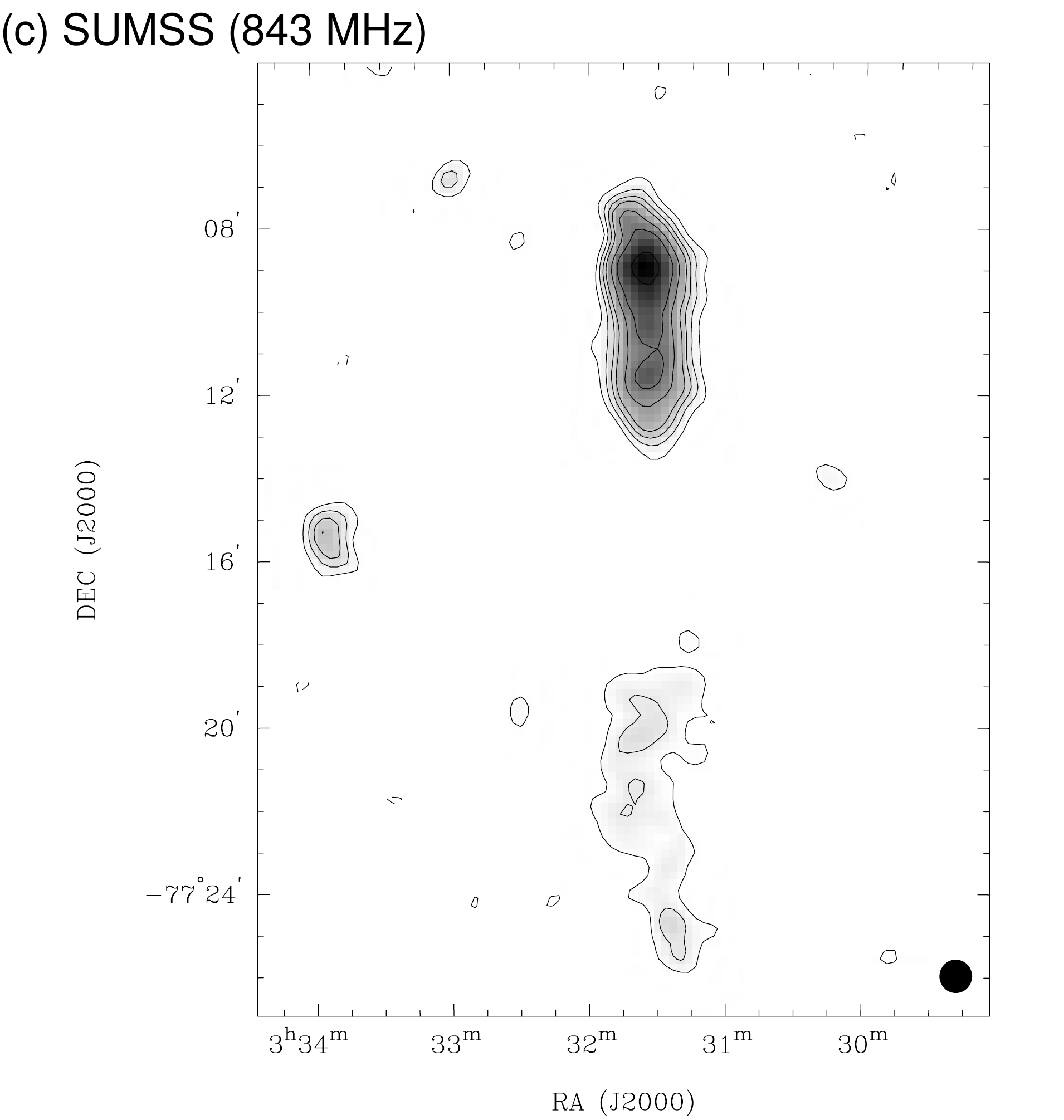} &
      \includegraphics[width=0.5\hsize]{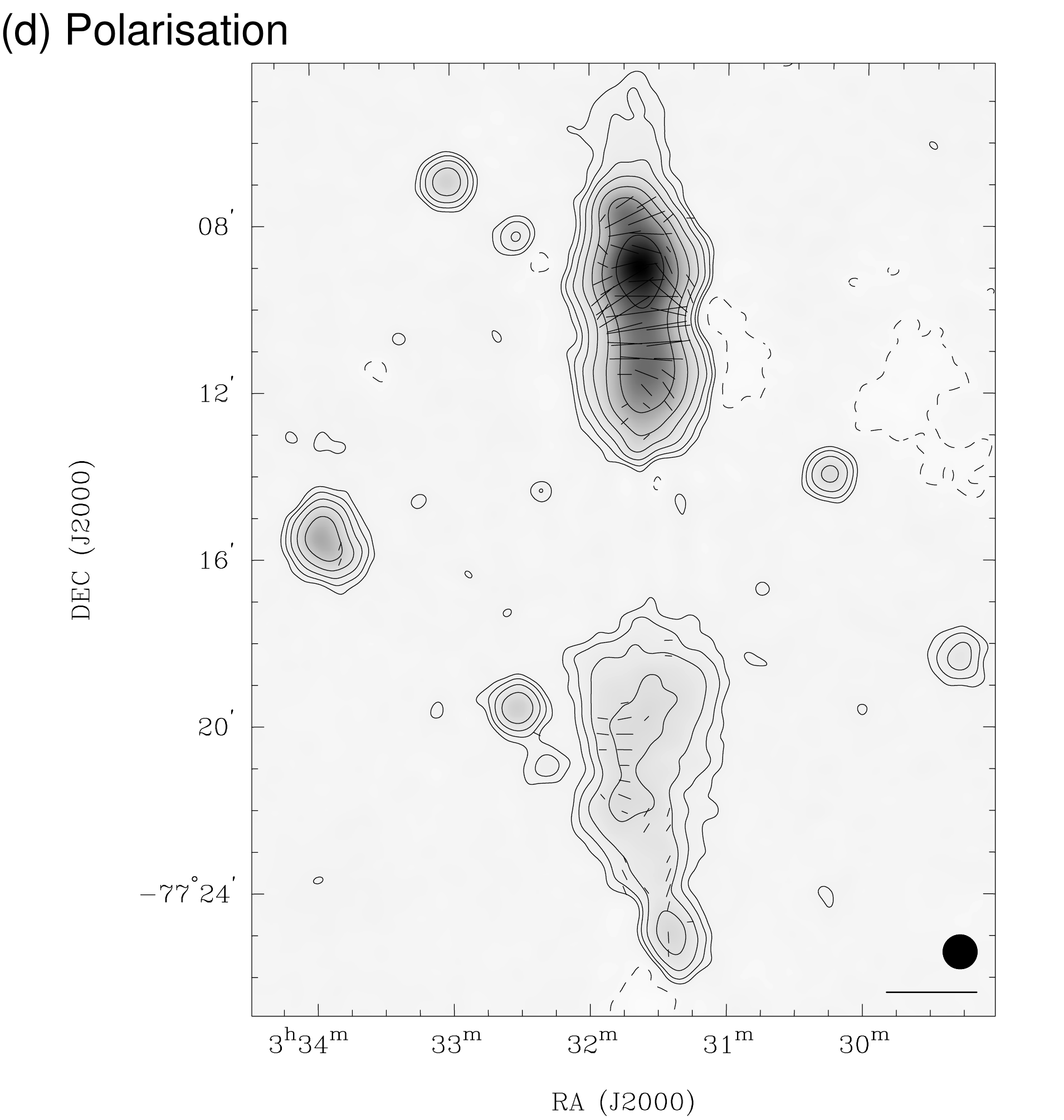}
\end{tabular}
\caption{(\subfigletter{a}) Low-resolution, wideband, total intensity image of J0331--7710 at 2.1 GHz with contours at -3, 3, 6, 12, 24, 48, 96, and 192 $\times$ 85 $\mu$Jy beam$^{-1}$ and a beam of FWHM 48\arcsec. It shows a faint diffuse emission region to the north of the brightened edge. (\subfigletter{b}) The high resolution image shows  the core below the northern lobe and has a beam of FWHM 20\arcsec. (\subfigletter{c}) 843 MHz SUMSS image included for comparison with a beam of FWHM 45\arcsec$\times$46\arcsec, total intensity contours at -1, 1, 2, 3, 4, 6, 8, and 12 $\times$ 3.9 mJy beam$^{-1}$ and greyscale in the range 3.9--53.1 mJy beam$^{-1}$. (\subfigletter{d}) Distribution of polarised intensity at 2.8 GHz represented by the lengths of the overlaid Faraday rotation-corrected, electric field vectors. The scale-bar represents 4 mJy.}
\label{fig:J0331}
\end{figure*}

\begin{figure*}
  \centering
  \begin{tabular}{cc}
      \includegraphics[width=0.5\hsize]{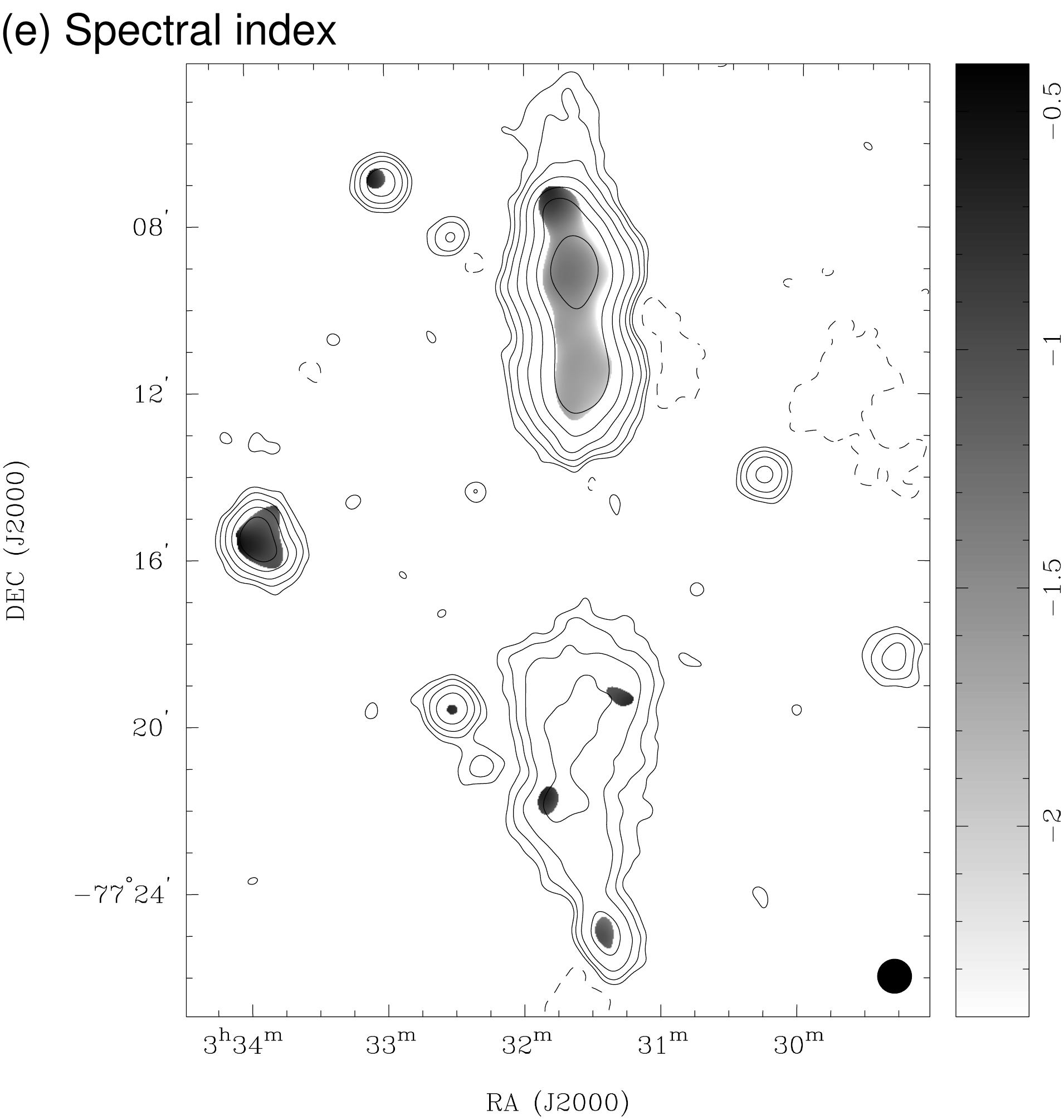} &
      \includegraphics[width=0.5\hsize]{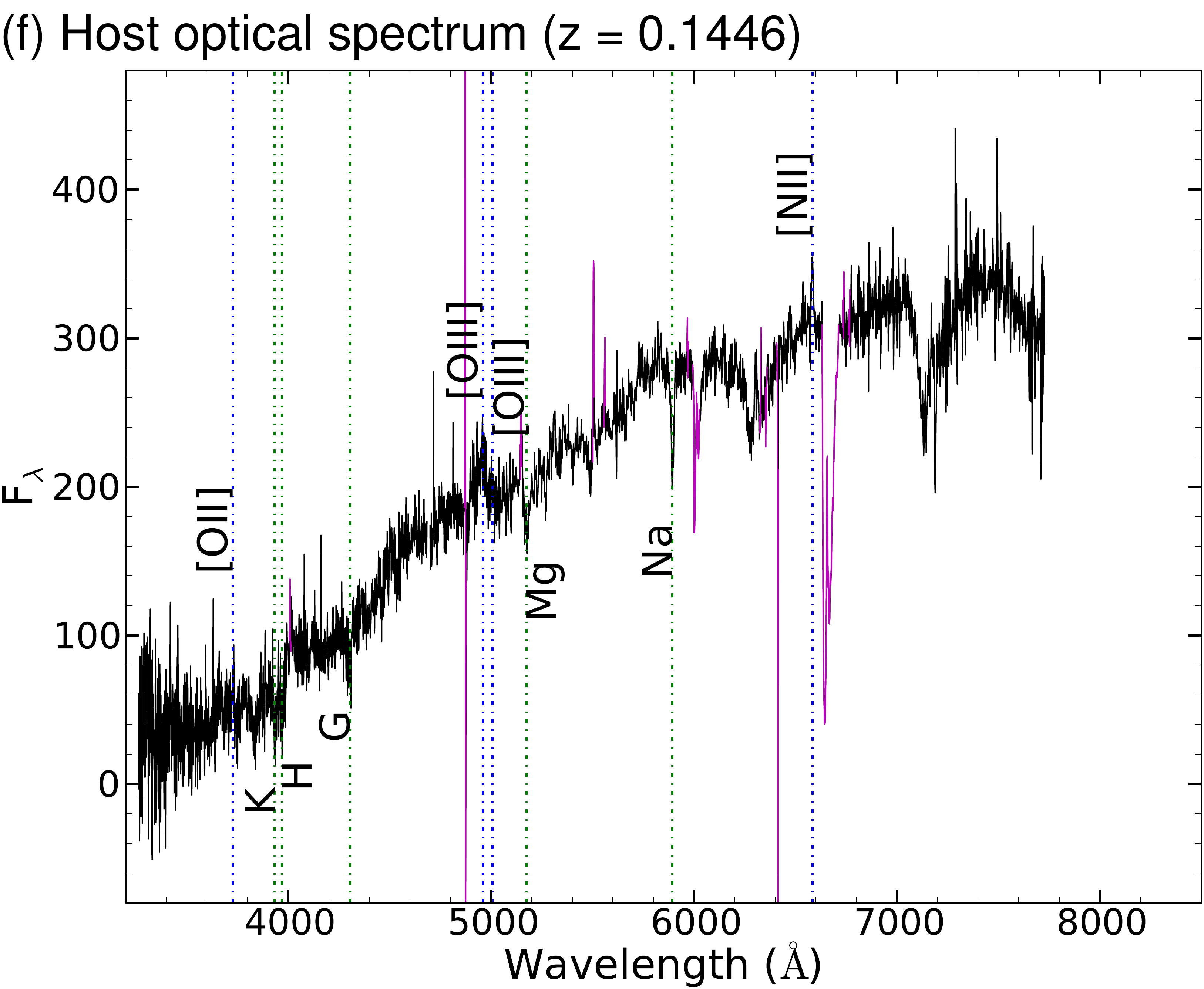}
  \end{tabular}
  \contcaption{(\subfigletter{e}) Distribution of spectral index between 843 MHz and 2.8 GHz shown as greyscale in the range $-2.4$ to $-0.4$ with a beam of FWHM 48\arcsec~and total intensity contours at -3, 3, 6, 12, 24, 48, 96, and 192 $\times$ 85 $\mu$Jy beam$^{-1}$. (\subfigletter{f}) An optical spectrum from AAOmega on the AAT of the host galaxy of J0331--7710. The spectrum has been shifted to rest frame using the measured redshift of 0.1446.}
\end{figure*}

\subsection{J0400--8456 (Fig.~\ref{fig:J0400})}
Low and high-resolution images of this source ($z=0.1033$) are shown in Fig.~\ref{fig:J0400}. These images reveal structures noted in previous ATCA images~\citep{Saripallietal2005}. 

This giant radio source is symmetric in terms of separation and brightness with respect to the hotspots along the source axis oriented in a north-south direction. However, these are both displaced towards the west. The two symmetric features in a north-south orientation on either side of a blended core, seen in the previous ATCA image, clearly suggest that the jets are already deflected after the initial stages on their way to the hotspots. 

The source combines structural characteristics seen in FRIs (bright core with twin jets), FRIIs (hotspots at the ends) and WATs (the much fainter extensions to the two hotspots). These extensions to the west of the northern hotspot and to the south-east of the southern hotspot, in this rather symmetric radio galaxy, are severely bent away from the source axis. Intriguingly, these extensions are not expanding plumes as observed in WATs but are edge-brightened and resemble lobes of powerful radio galaxies. Our images also show the magnetic field perpendicular to the jet as expected in a low power source (Fig.~\ref{fig:J0400}\subfigletter{c}). The overall characteristics of this source suggest multiple epochs of nuclear activity. The extreme non-collinearity of this source needs to be understood. The high signal-to-noise optical spectrum of the host is typical of an elliptical galaxy (Fig.~\ref{fig:J0400}\subfigletter{e}).

\begin{figure*}
  \centering
  \begin{tabular}{cc}
      \includegraphics[width=0.5\hsize]{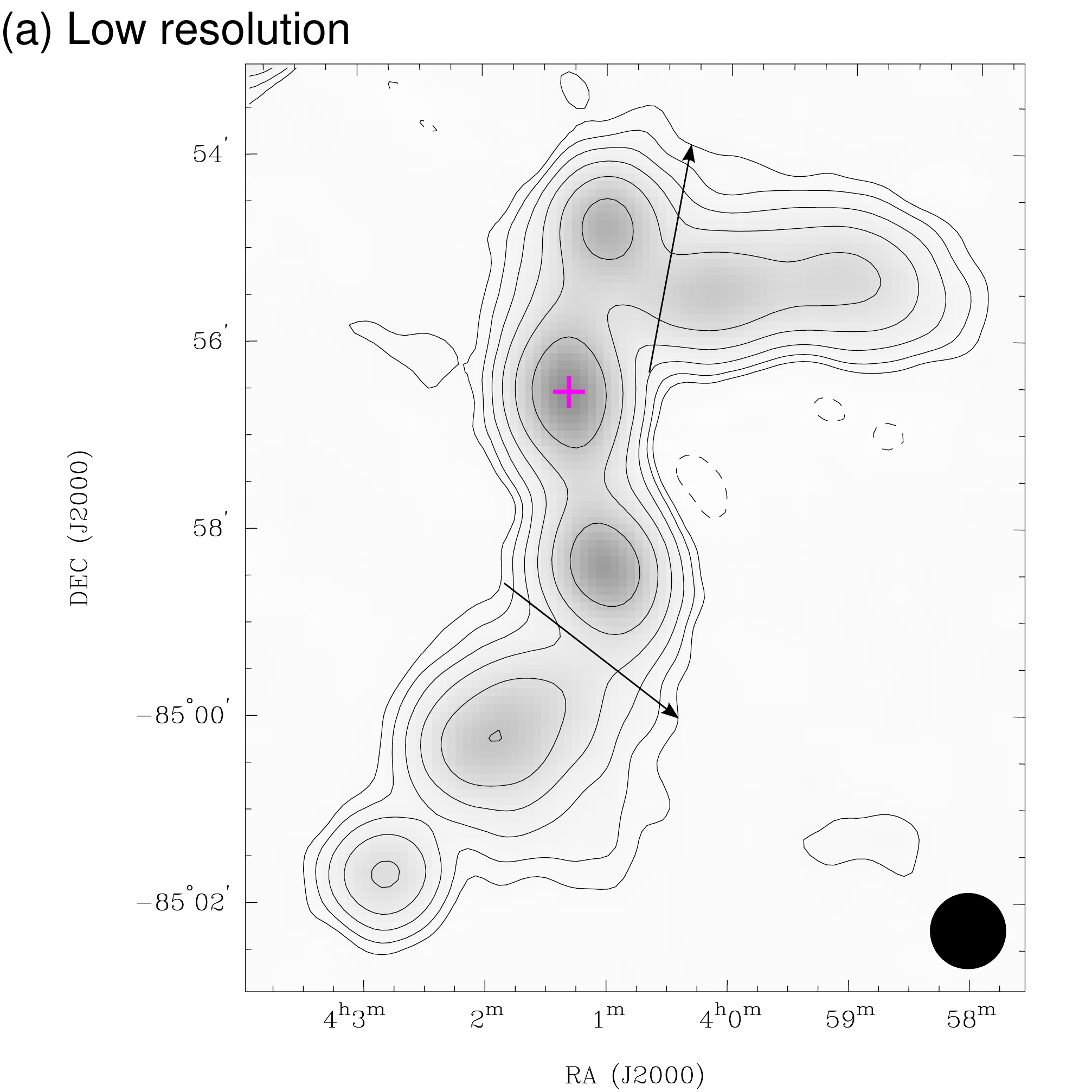} &
      \includegraphics[width=0.5\hsize]{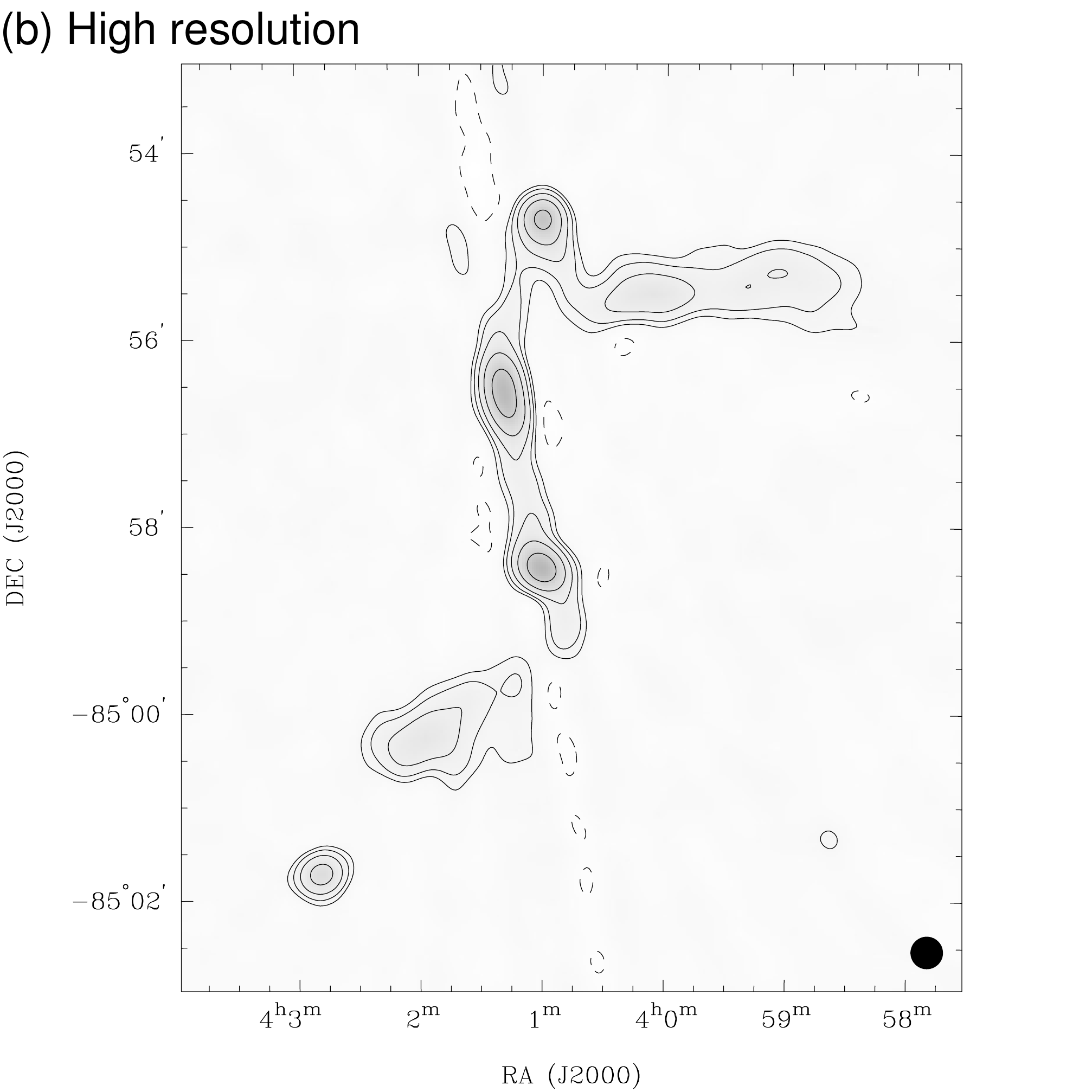} \\
      \includegraphics[width=0.5\hsize]{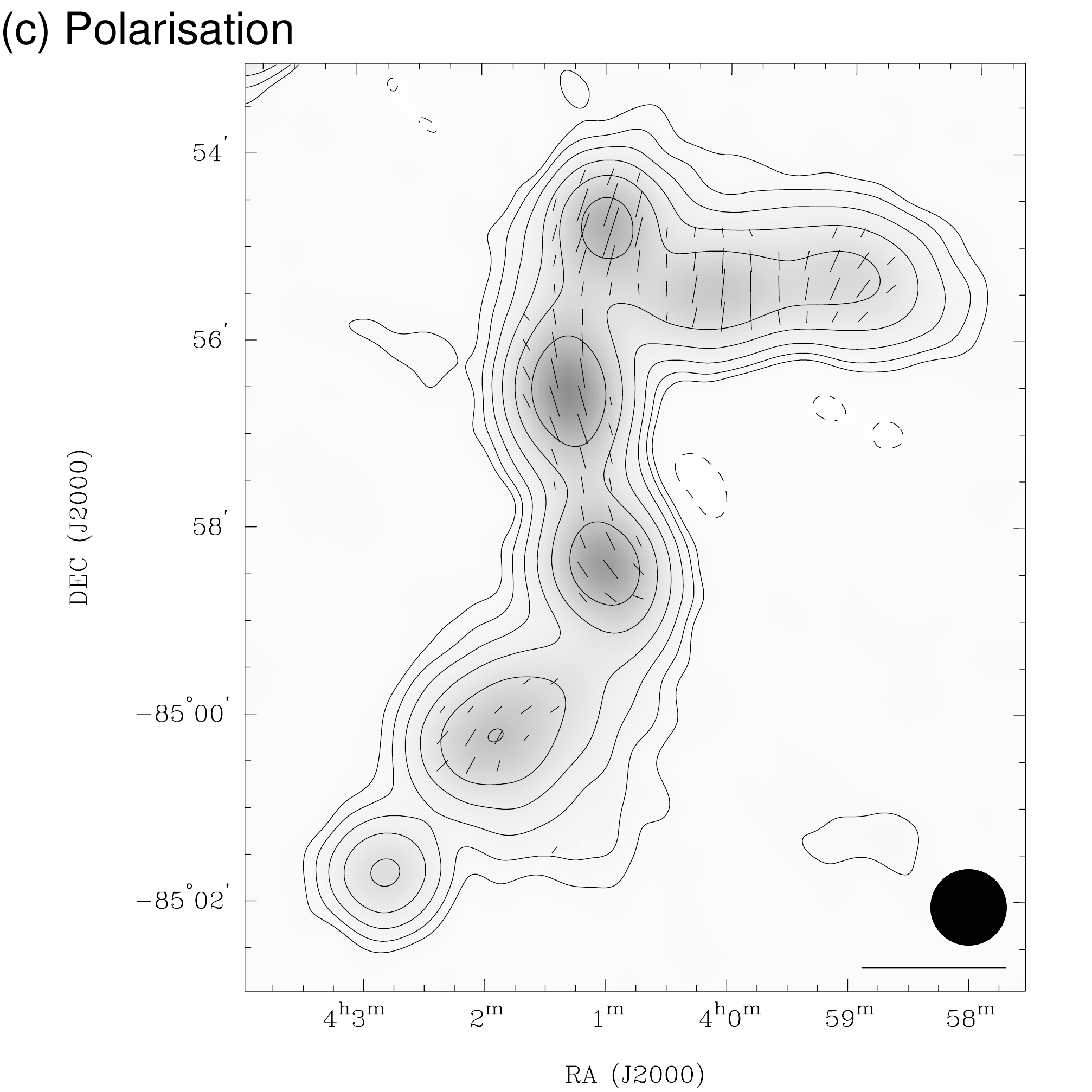} &
      \includegraphics[width=0.5\hsize]{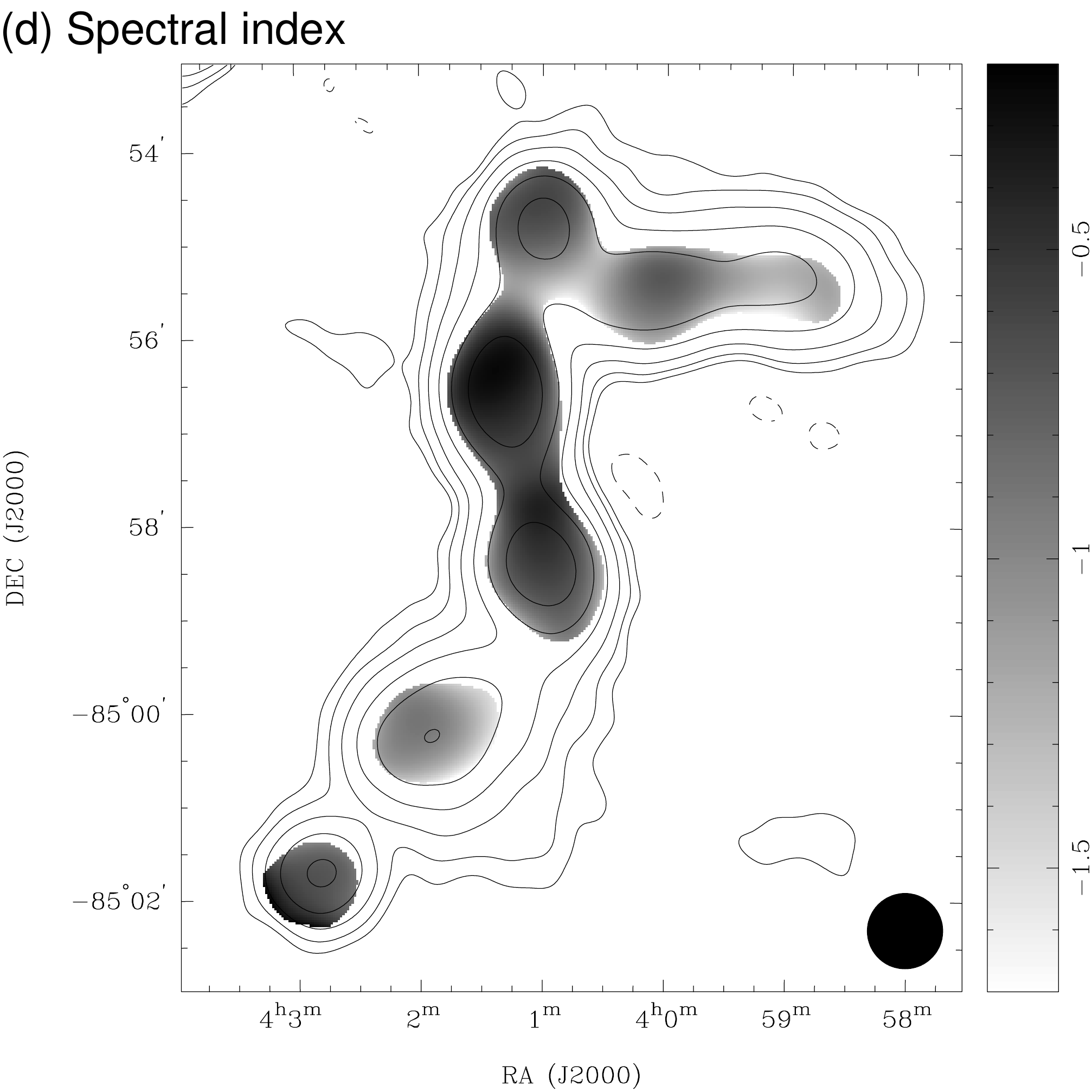}
\end{tabular}
\caption{(\subfigletter{a}) Low-resolution, wideband, total intensity image of J0400--8456 at 2.1 GHz with contours at -3, 3, 6, 12, 24, 48, and 96 $\times$ 0.1 mJy beam$^{-1}$ and a beam of FWHM 48\arcsec. (\subfigletter{b}) High resolution ATCA image of J0400--8456 at 2.1 GHz with contours at -3, 3, 6, 12, 24, 48, 96, 192, and 384 $\times$ 0.15 mJy beam$^{-1}$ and a beam of FWHM 20\arcsec. (\subfigletter{c}) Distribution of polarised intensity at 2.8 GHz represented by the lengths of the overlaid Faraday rotation-corrected, electric field vectors. The scale-bar represents 5 mJy. (\subfigletter{d}) Distribution of spectral index between 843 MHz and 2.8 GHz shown as greyscale in the range $-1.7$ to $-0.2$ with a beam of FWHM 48\arcsec. The levels of overlaid total intensity contours are -3, 3, 6, 12, 24, 48, and 96 $\times$ 0.1 mJy beam$^{-1}$.}
\label{fig:J0400}
\end{figure*}

\begin{figure}
  \centering
      \includegraphics[width=\hsize]{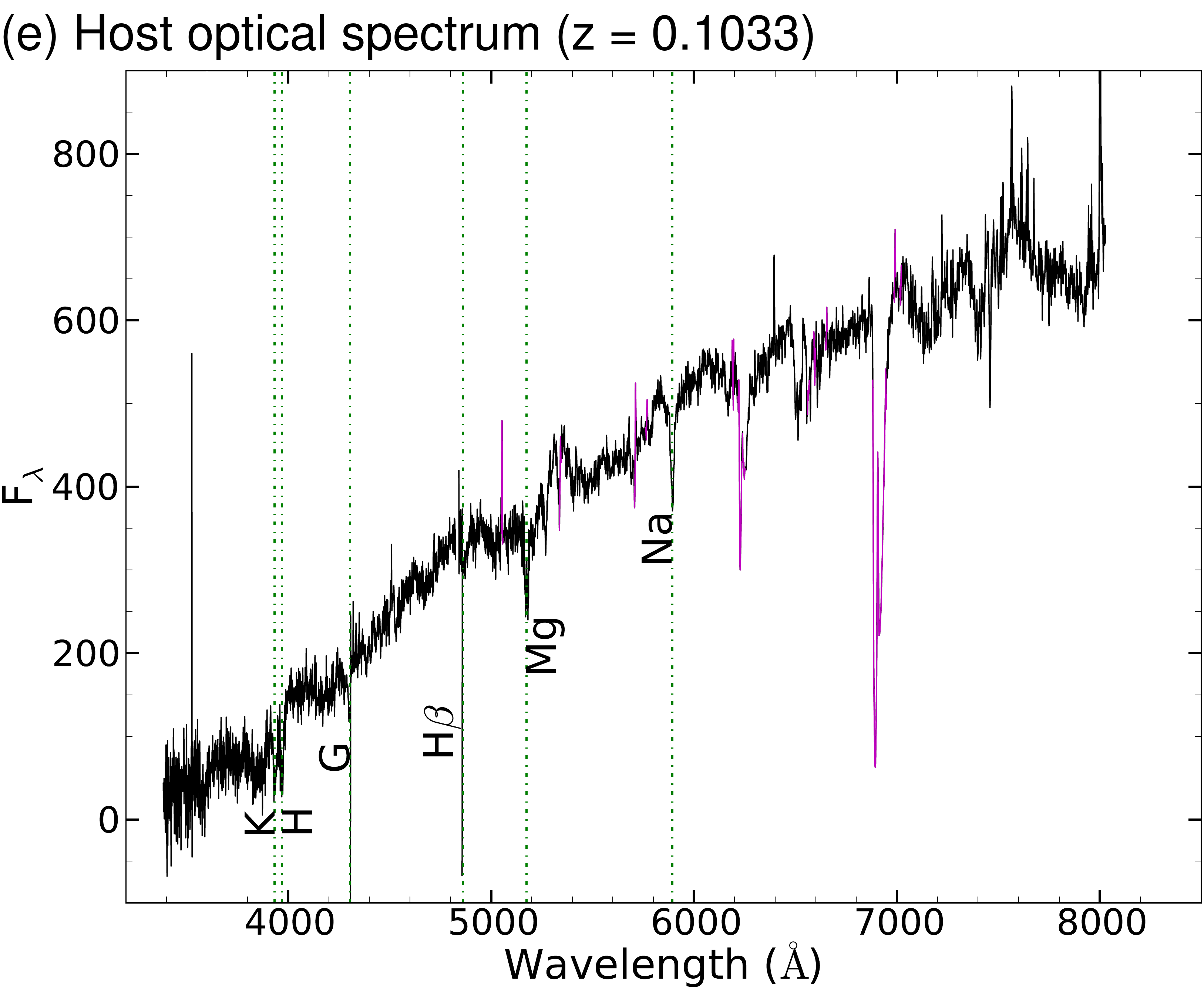}
  \contcaption{(\subfigletter{e}) An optical spectrum from AAOmega on the AAT of the host galaxy of J0400--8456. The spectrum has been shifted to rest frame using the measured redshift of 0.1033.}
\end{figure}

\subsection{J0459--528 (Fig.~\ref{fig:J0459})}
This SUMSS giant radio source ($z=0.0957$) is an FRII source with an unusually bright radio core. In the high resolution ATCA image, the core is seen to be extended east-west and is accompanied by an edge-brightened, inner lobe to the east (Fig.~\ref{fig:J0459}\subfigletter{b}). There is no corresponding inner lobe to the west although there is a faint emission seen. The axis of the extended core and the inner lobe is skewed with respect to the outer radio axis. The structures in this source indicate that this symmetric GRG may be exhibiting restarted activity. The optical spectrum of the host galaxy in Fig.~\ref{fig:J0459}\subfigletter{e} shows the absorption line features of an elliptical. The deep trough red-ward of 8000\AA~is an artifact of fringing within the fibre feeding the light of this galaxy into the spectrograph, and not astrophysical in origin.

\begin{figure*}
  \centering
  \begin{tabular}{cc}
     \includegraphics[width=0.5\hsize]{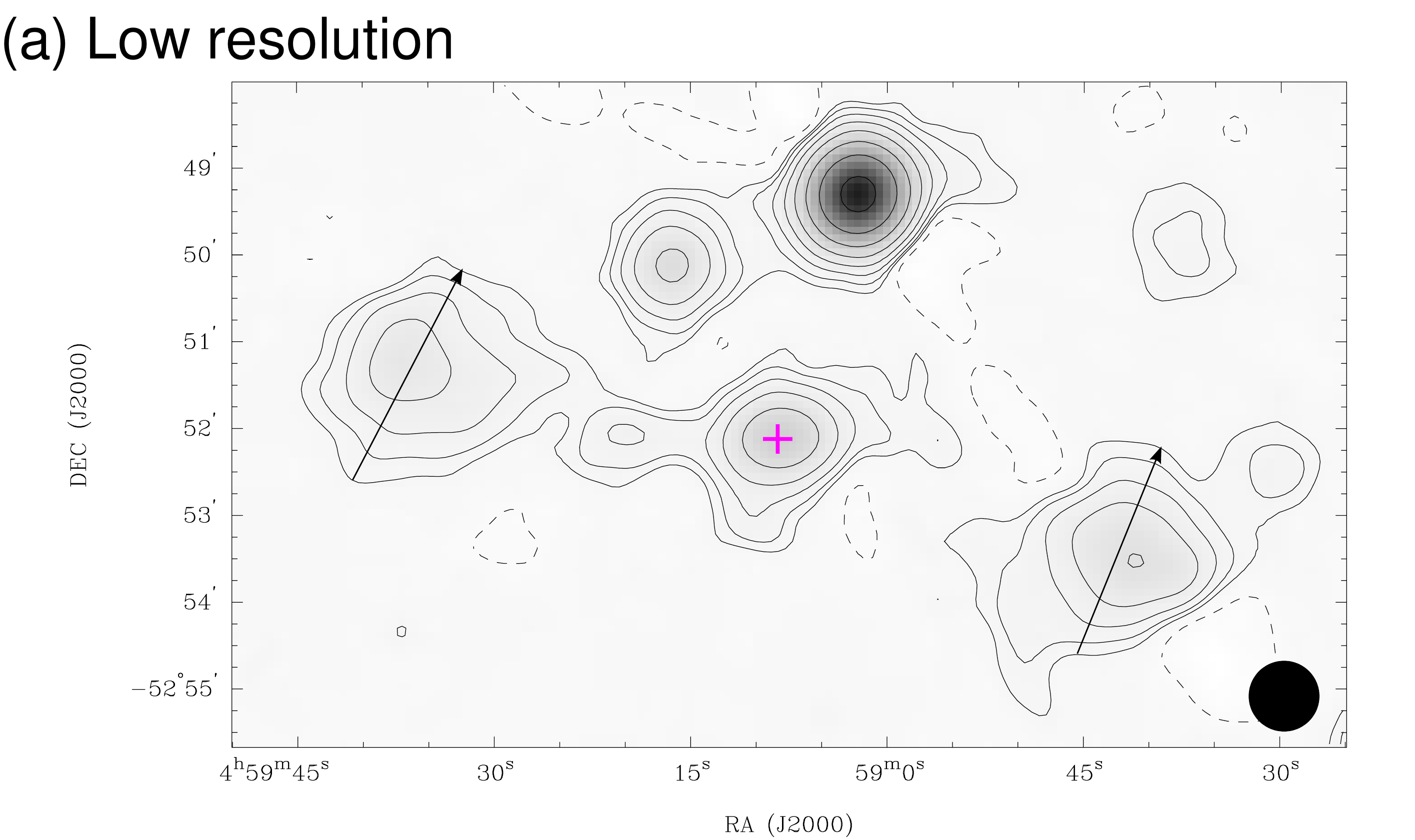} &
     \includegraphics[width=0.5\hsize]{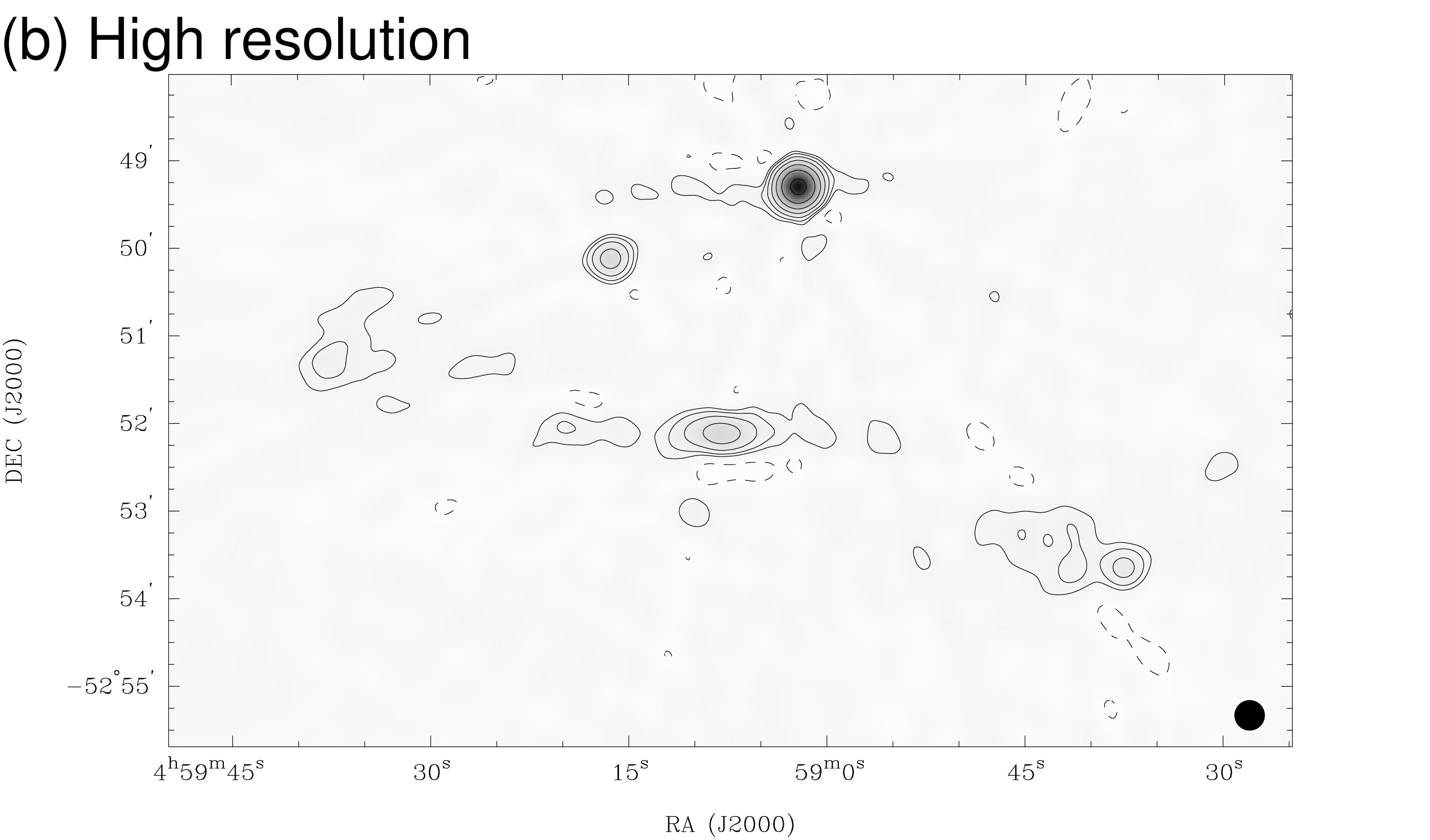} \\
     \includegraphics[width=0.5\hsize]{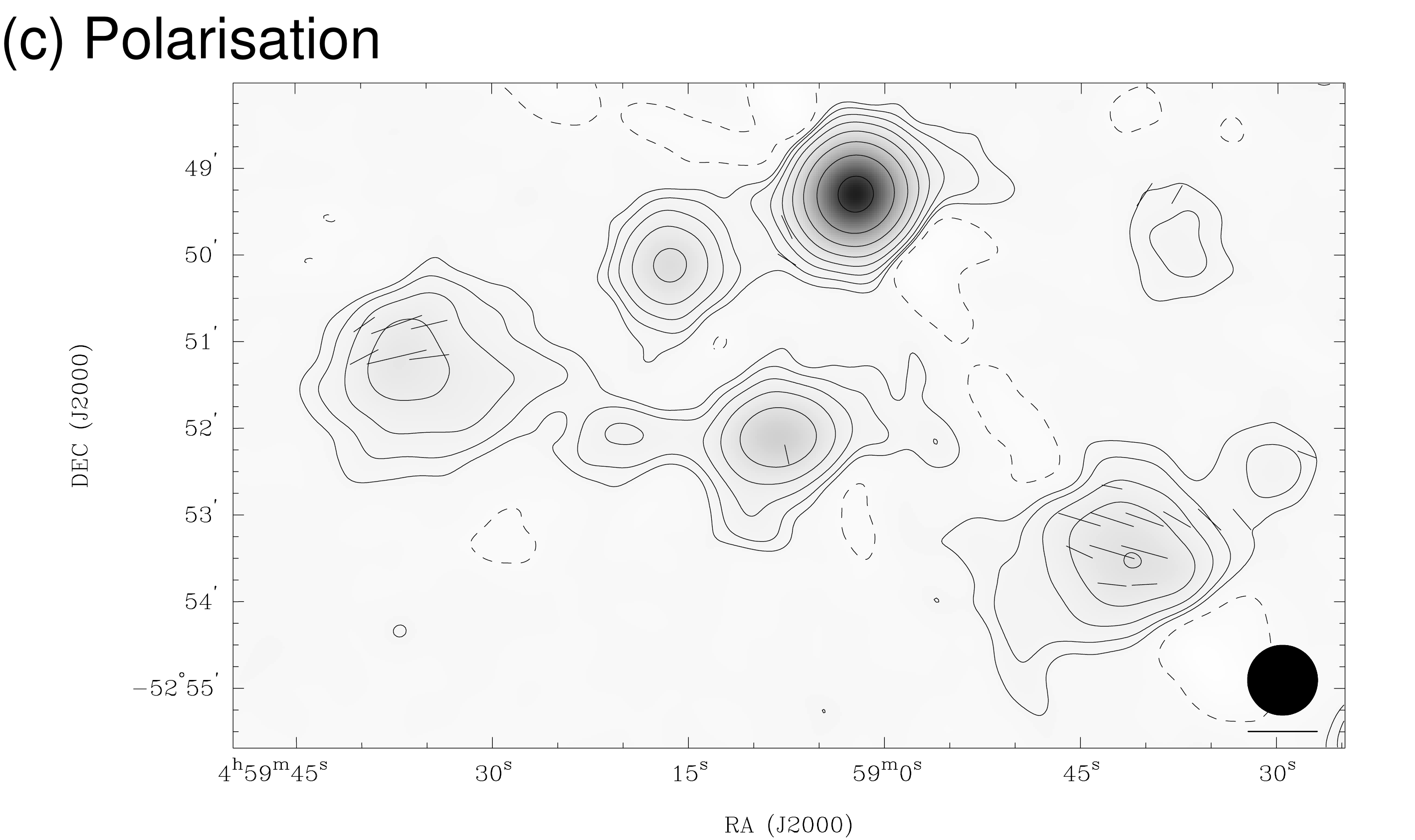} &
     \includegraphics[width=0.5\hsize]{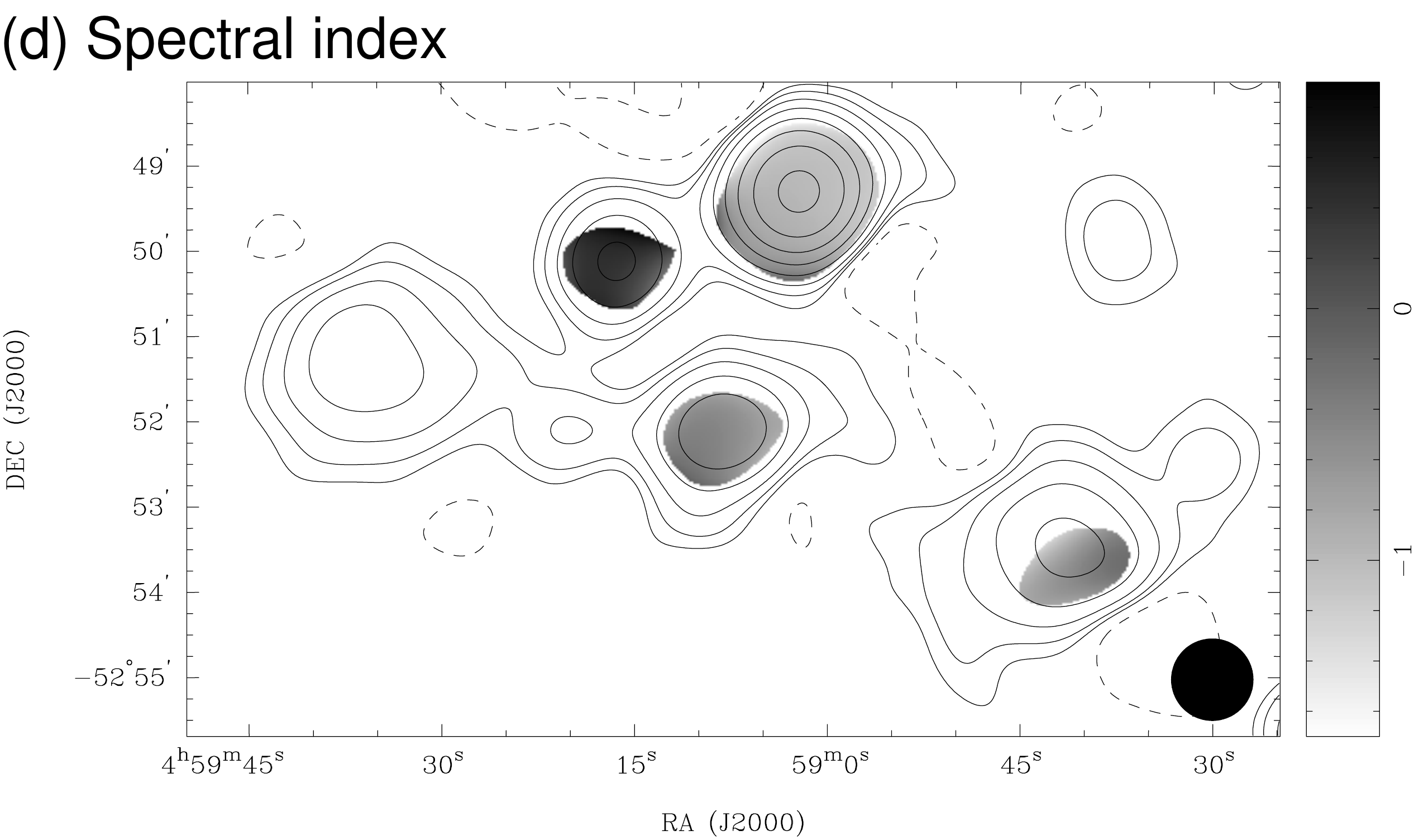}
\end{tabular}
\caption{(\subfigletter{a}) Low-resolution, wideband, total intensity image of J0459--528 at 2.1 GHz with contours at -3, 3, 6, 12, 24, 48, 96, 192, and 384 $\times$ 0.15 mJy beam$^{-1}$ and a beam of FWHM 48\arcsec. (\subfigletter{b}) High resolution, wideband ATCA image with a FWHM 20\arcsec~beam and contour levels at -3, 3, 6, 12, 24, 48, 96, 192, and 384 $\times$ 0.3 mJy beam$^{-1}$. (\subfigletter{c}) Distribution of polarised intensity at 2.8 GHz represented by the lengths of the overlaid Faraday rotation-corrected, electric field vectors. The scale-bar represents 1 mJy. (\subfigletter{d}) Distribution of spectral index computed from narrow band images at 843 MHz (SUMSS) and 2.8 GHz (ATCA) shown as greyscale in the range $-1.7$ to $0.9$ with a beam of FWHM 57\arcsec. The levels of overlaid total intensity contours are -3, 3, 6, 12, 24, 48, 96, 192, and 384 $\times$ 0.15 mJy beam$^{-1}$.}
\label{fig:J0459}
\end{figure*}

\begin{figure}
  \centering
    \includegraphics[width=\hsize]{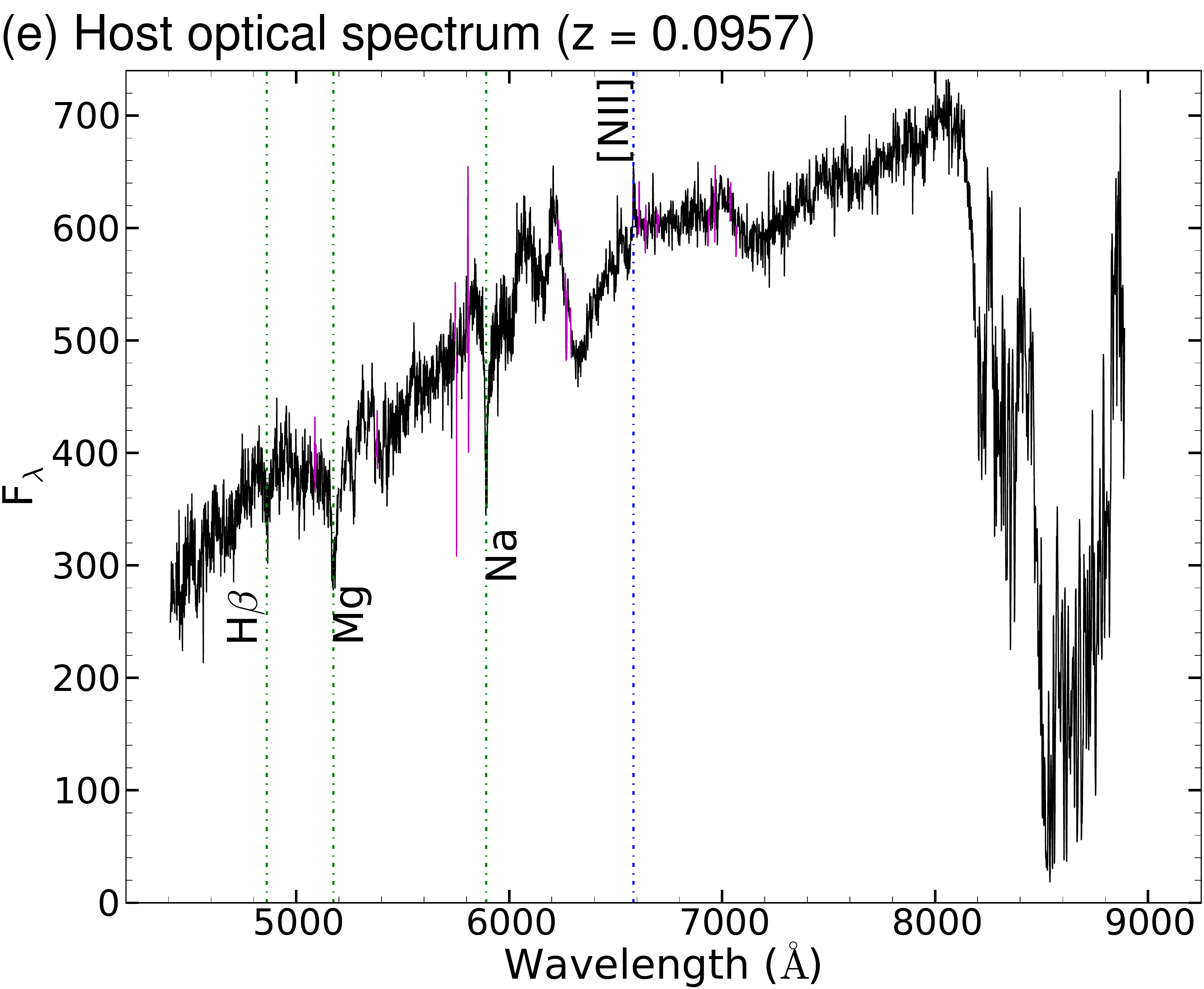}
  \contcaption{(\subfigletter{e}) An optical spectrum from AAOmega on the AAT of the host galaxy of J0459--528. The spectrum has been shifted to rest frame using the measured redshift of 0.0957.}
\end{figure}

\subsection{B0511--305 (Fig.~\ref{fig:B0511})}
Previously this giant radio galaxy ($z=0.0576$) had been imaged by~\citet{Ekersetal1989},~\citet{JonesMcAdam1992} and~\citet{Subrahmanyanetal1996}. While no core was detected, the extended structure in these images showed bright compact hotspots at the lobe extremities. In our new ATCA images (Fig.~\ref{fig:B0511}) we make several advances in our knowledge of this source: we detect a compact core at the location previously suggested by~\citet{Subrahmanyanetal1996} and in addition we detect faint radio emission extended on the scale of the full angular extent of the source. Much of this faint emission is seen to the south of the southern hotspot extending to more than half the angular size of this giant radio galaxy. A similar faint emission region is seen to the north but unlike in the south this feature is seen only to the west of the northern lobe, all along its length and it is also much fainter (Fig.~\ref{fig:B0511}\subfigletter{a}). The detection of faint and extended radio emission beyond the known extent of the radio galaxy is similar to what we image in J0034--6639 and J0331--7710. Moreover, as in the previous giants this emission is seen more prominently on the side of the shorter lobe.

We note that the hotspots in this giant are not collinear with the core. The southern hotspot is closer to the core and the southern lobe is brighter than the northern lobe. This giant is asymmetric in lobe separation but skewed.

The detection of faint extended emission in this source suggests that it has had a restarted nuclear activity. It should be noted that the host galaxy is disturbed with an extended east-west feature, as discussed in~\citet{JonesMcAdam1992},~\citet{Subrahmanyanetal1996} and~\citet{Govonietal2000b}. In the optical, the spectrum of the host is early-type, although a number of emission line features (including H$\alpha$, [SII], [OII], and [OIII]) are also prominent (Fig.~\ref{fig:B0511}\subfigletter{f}).

\begin{figure*}
  \centering
  \begin{tabular}{cc}
     \includegraphics[width=0.5\hsize]{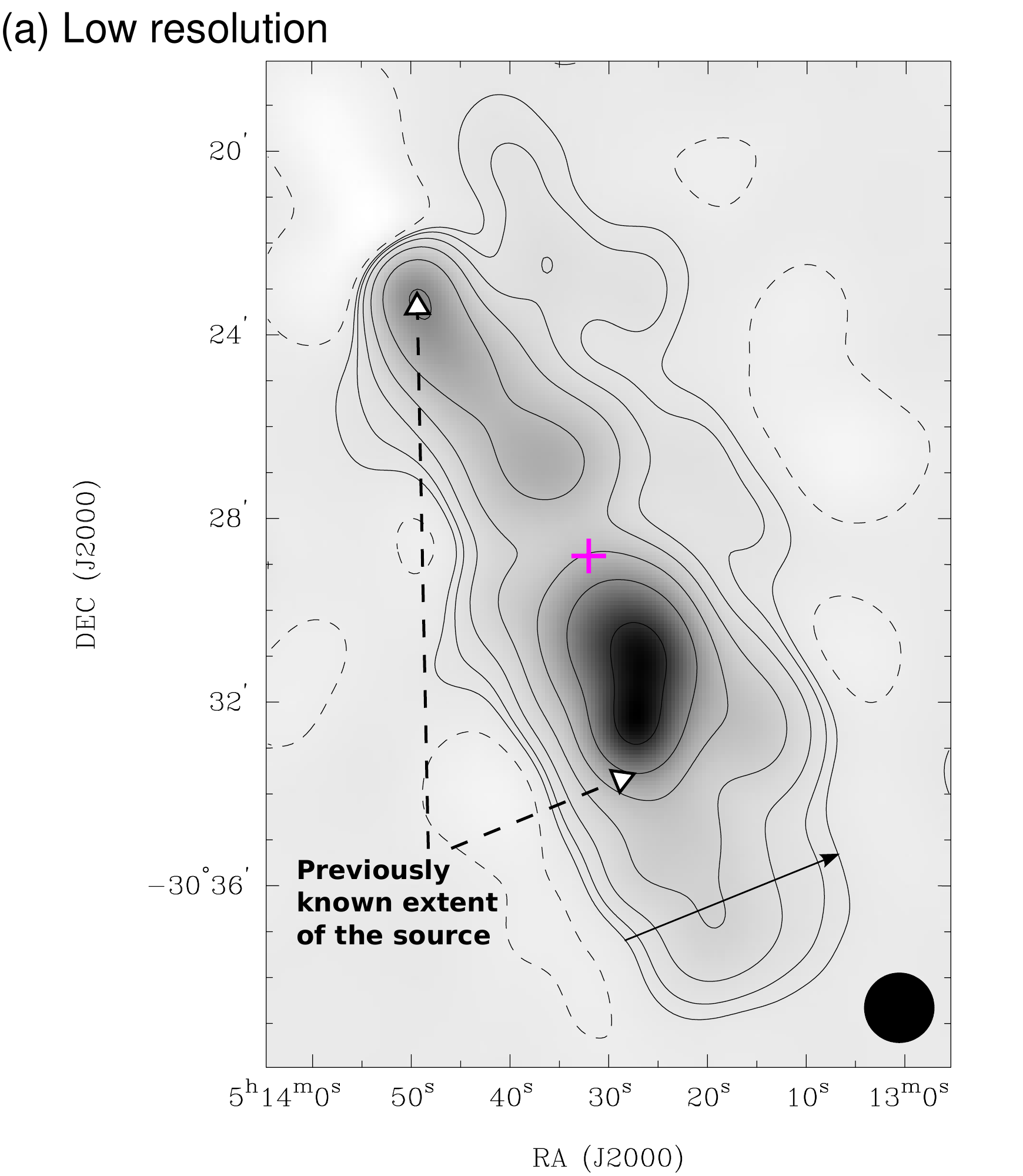} &
     \includegraphics[width=0.5\hsize]{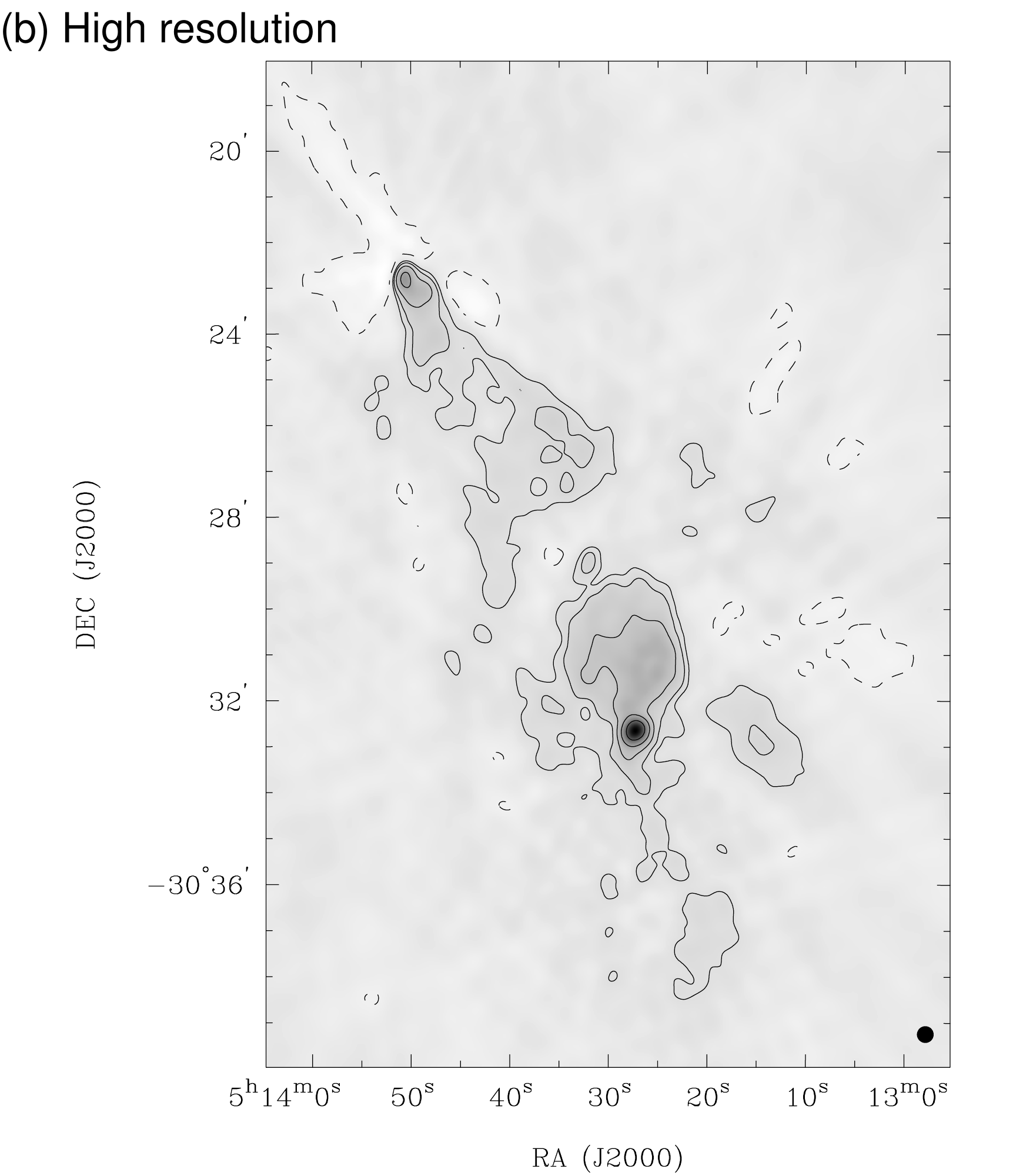} \\
     \includegraphics[width=0.5\hsize]{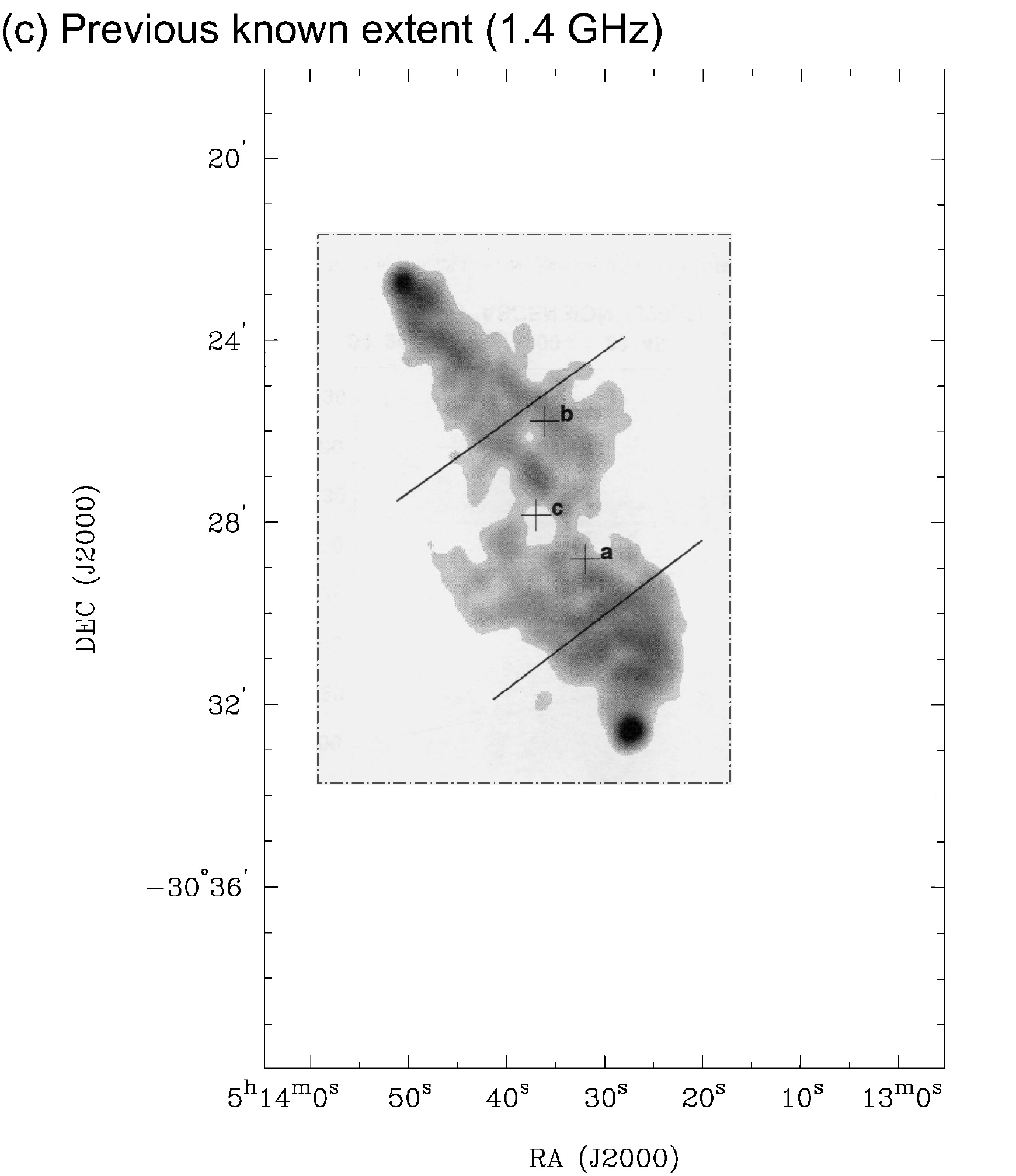} &
     \includegraphics[width=0.5\hsize]{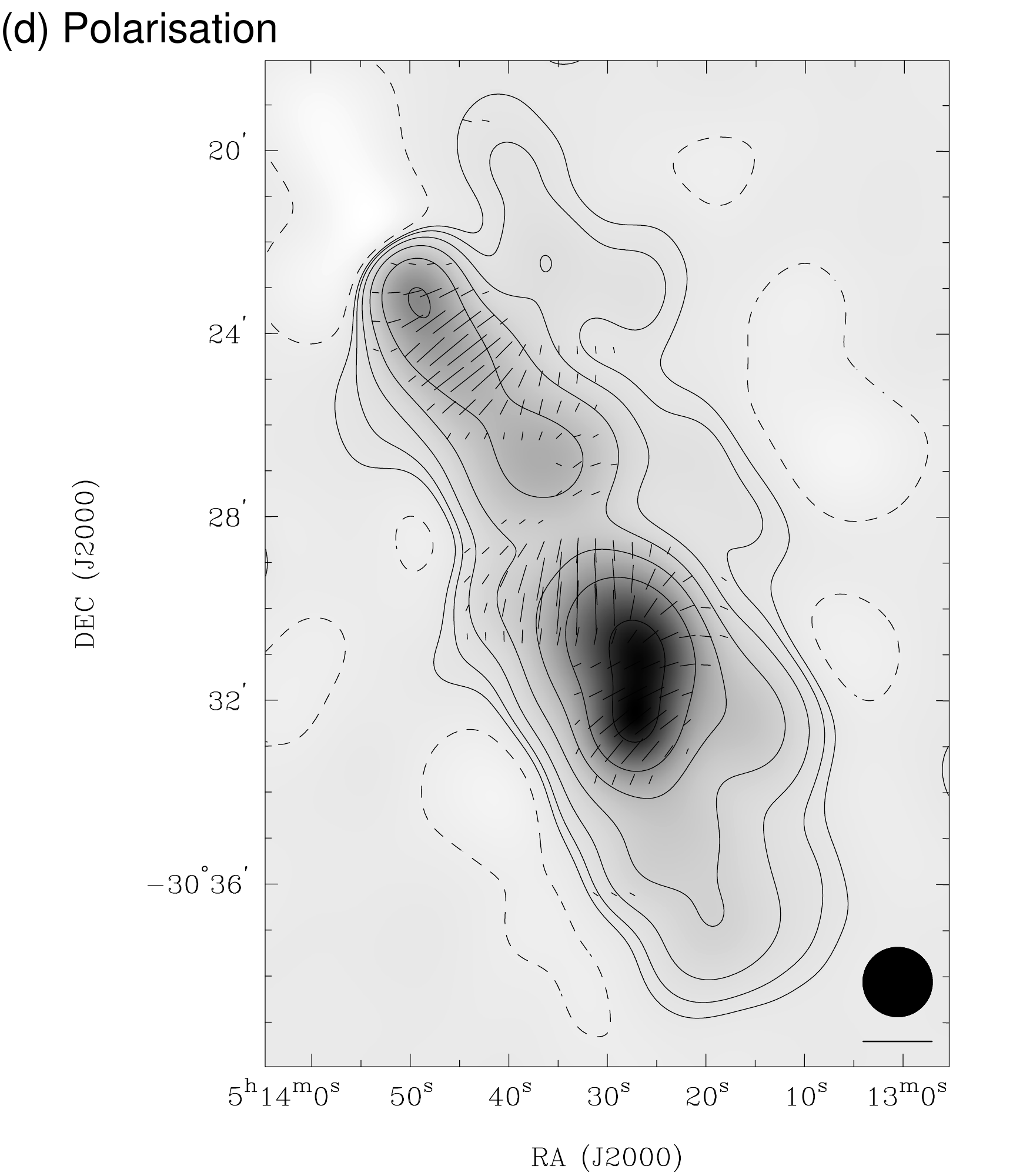}
\end{tabular}
\caption{(\subfigletter{a}) Low-resolution, wideband, total intensity image of B0511--305 at 2.1 GHz with contours at -3, 3, 6, 12, 24, 48, 96, and 192 $\times$ 1.25 mJy beam$^{-1}$ and a beam of FWHM 90\arcsec. (\subfigletter{b}) High resolution image of B0511--305 with a FWHM 20\arcsec beam and contour levels at -3, 3, 6, 12, 24, and 48 $\times$ 2 mJy beam$^{-1}$. (\subfigletter{c}) A previous 1.4 GHz ATCA image~\citep{Subrahmanyanetal1996} included for comparison. The dot-dashed line represents the boundary of the original greyscale image. (\subfigletter{d}) Distribution of polarised intensity at 2.8 GHz represented by the lengths of the overlaid Faraday rotation-corrected, electric field vectors. The scale-bar represents 25 mJy.}
\label{fig:B0511}
\end{figure*}

\begin{figure*}
  \centering
  \begin{tabular}{cc}
      \includegraphics[width=0.5\hsize]{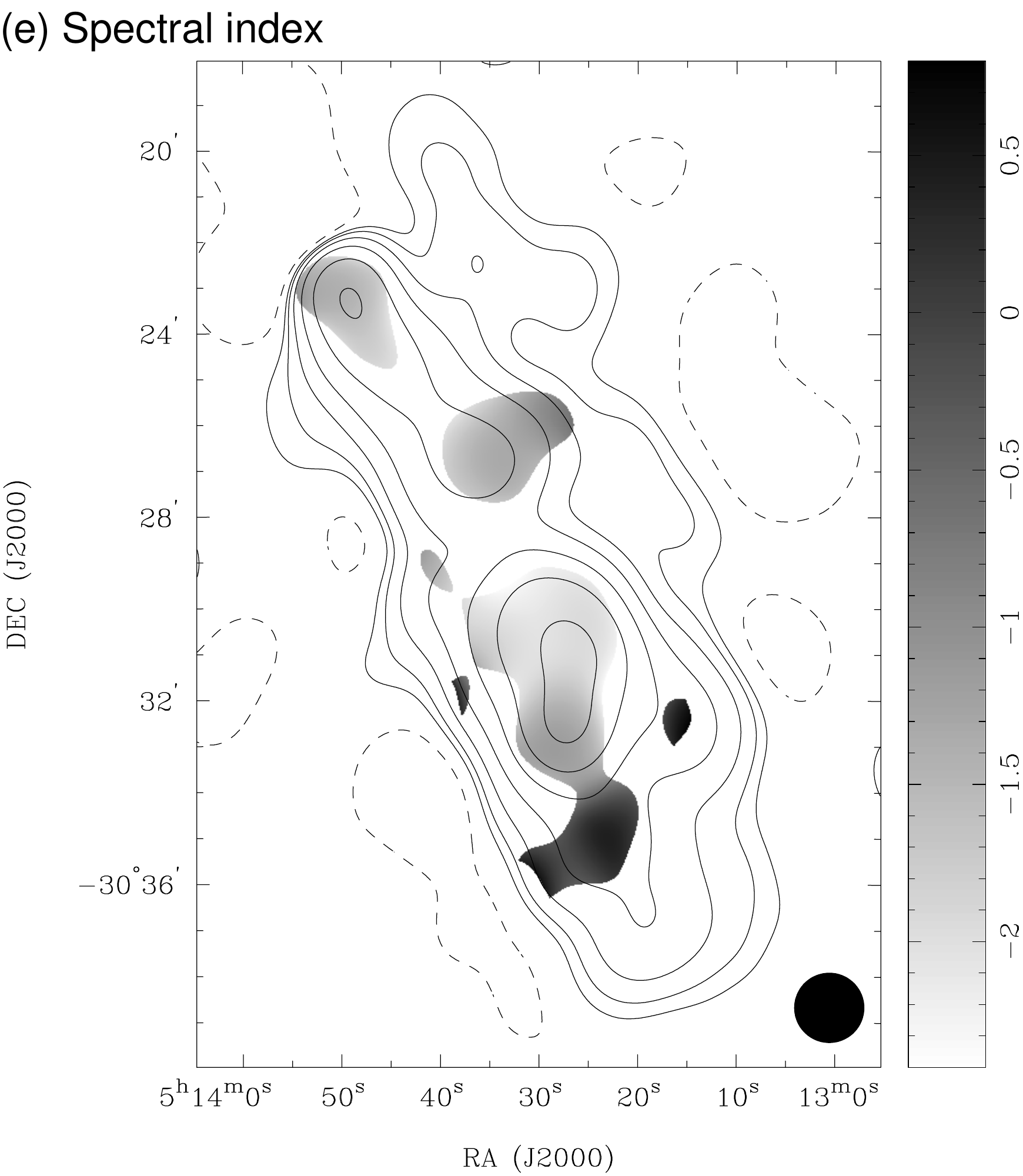} &
      \includegraphics[width=0.5\hsize]{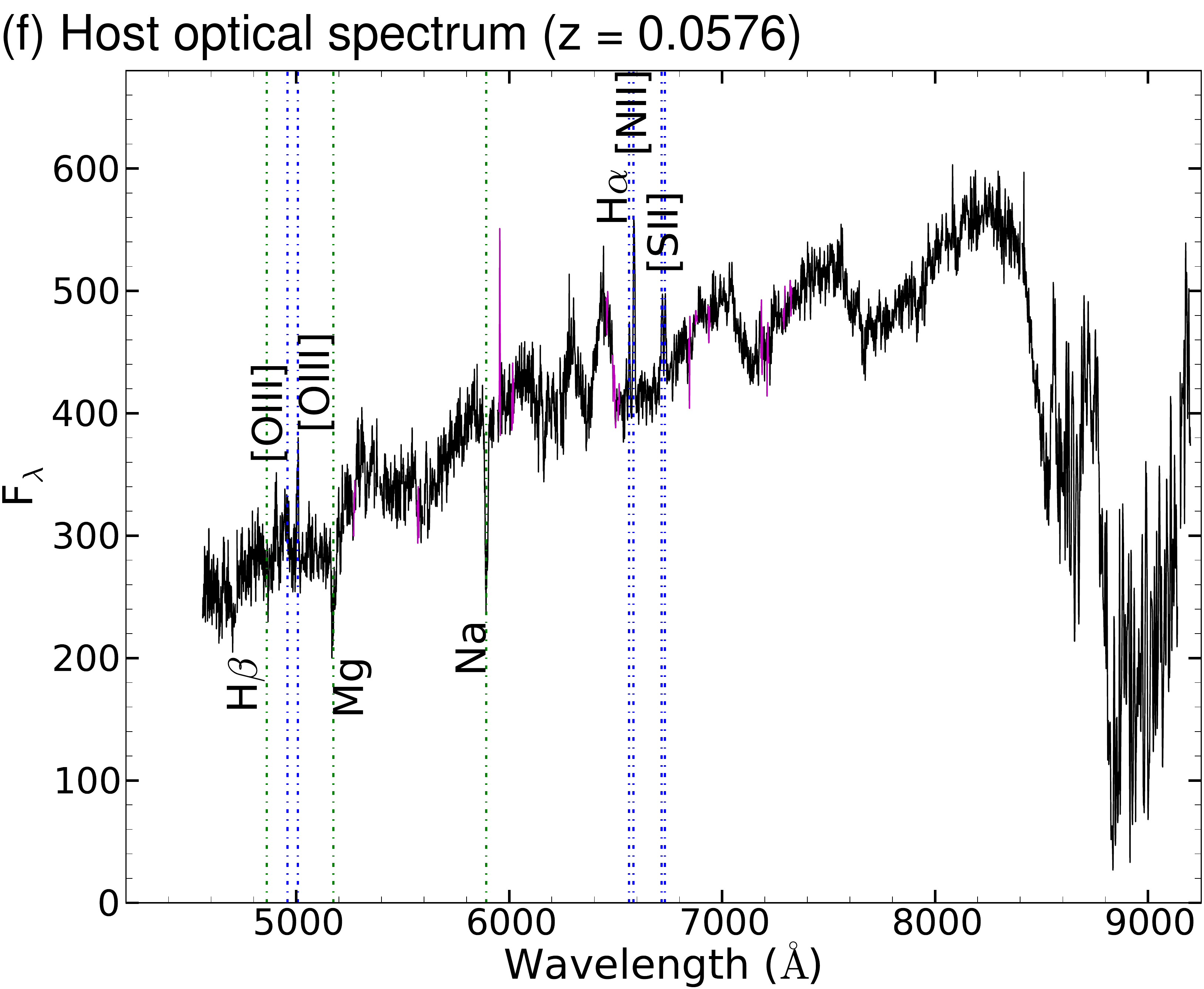}
  \end{tabular}
  \contcaption{(\subfigletter{e}) Distribution of spectral index computed from narrow band images at 843 MHz (SUMSS) and 2.8 GHz (ATCA) shown as greyscale in the range $-2.4$ to $0.8$ with a beam of FWHM 90\arcsec. The levels of overlaid total intensity contours are -3, 3, 6, 12, 24, 48, 96, and 192 $\times$ 1.25 mJy beam$^{-1}$. (\subfigletter{f}) An optical spectrum from AAOmega on the AAT of the host galaxy of B0511--305. The spectrum has been shifted to rest frame using the measured redshift of 0.0576. The undulating pattern in the continuum is due to fibre fringing.}
\end{figure*}

\subsection{B0703--451 (Fig.~\ref{fig:B0703})}
This giant source ($z=0.1242$) is one of the most asymmetric in terms of flux density, see Fig.~\ref{fig:B0703}. While previously the source identification was with an elliptical galaxy between a bright SW lobe and a faint NE lobe, our new image may suggest an entirely different picture. As seen in the ATCA images the brighter lobe has a bright `hotspot' at its inner end that was not previously known (Fig.~\ref{fig:B0703}\subfigletter{a}). This compact `hotspot' however is coincident with an optical object. Our AAOmega spectrum of this object (Fig.~\ref{fig:B0703}\subfigletter{e}) hints at broad emission lines and a redshift $z\sim 1.28$. Also present is a weaker unrelated point source to the west but aligned along the axis of the source as noted by \citet{JonesMcAdam1992}.

Our observations therefore reveal a different scenario for the source: a likely quasar associated with the bright core at RA: $07~05~17$ and Dec: $-45~14~01$ straddled to the NE by a bright lobe and to the SW by a compact hotspot. The two components of this source (also a giant radio galaxy) are quite different: to the NE although there is a bright lobe there is no prominent hotspot (although higher resolution observations may be needed to detect any compact structure) whereas to the SW there is no lobe and instead only a compact hotspot is seen. A faint bridge emission may also be present connecting the hotspot to the core (Fig.~\ref{fig:B0703}\subfigletter{a}).

In such a picture we need to understand the nature of the faint lobe extended along the source axis to the east. Its location, alignment, morphology as well as its relative faintness all suggest that it might represent relic emission from a past activity phase. We bring to attention a similar example of a giant quasar (J0031.8--6727) that has aligned relic emission observed as part of the ATLBS-ESS sample by~\citet{Saripallietal2012}.

In polarised intensity an edge is seen along the SE end of the bright lobe. Also, the integrated spectral index of the faint eastern lobe is steeper than that of the bright lobe. The optical spectrum of the host galaxy at the centroid is an ordinary early-type (Fig.~\ref{fig:B0703}\subfigletter{d}).

\begin{figure*}
  \centering
  \begin{tabular}{cc}
      \includegraphics[width=0.5\hsize]{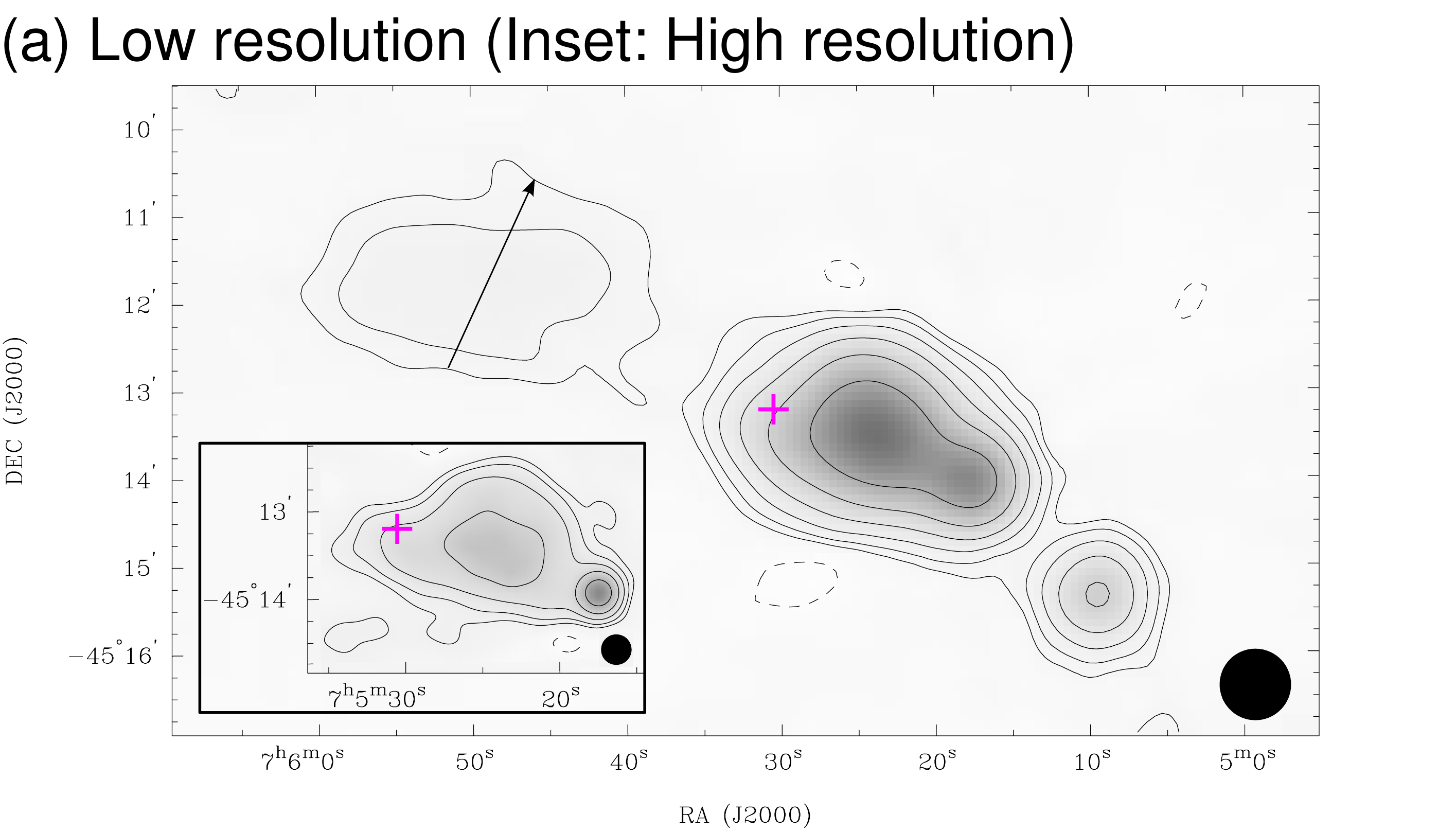} &
      \includegraphics[width=0.5\hsize]{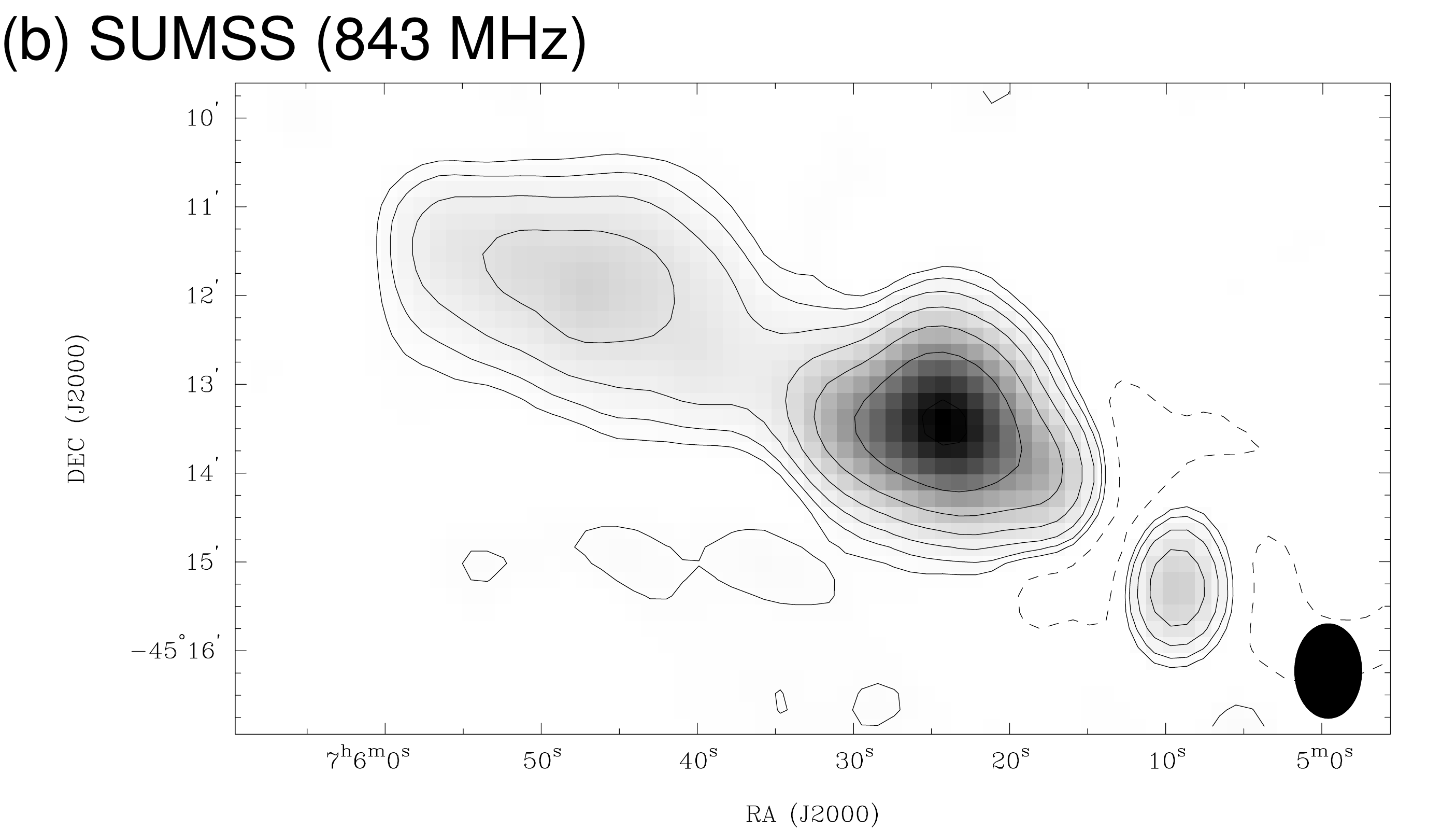} \\
      \includegraphics[width=0.5\hsize]{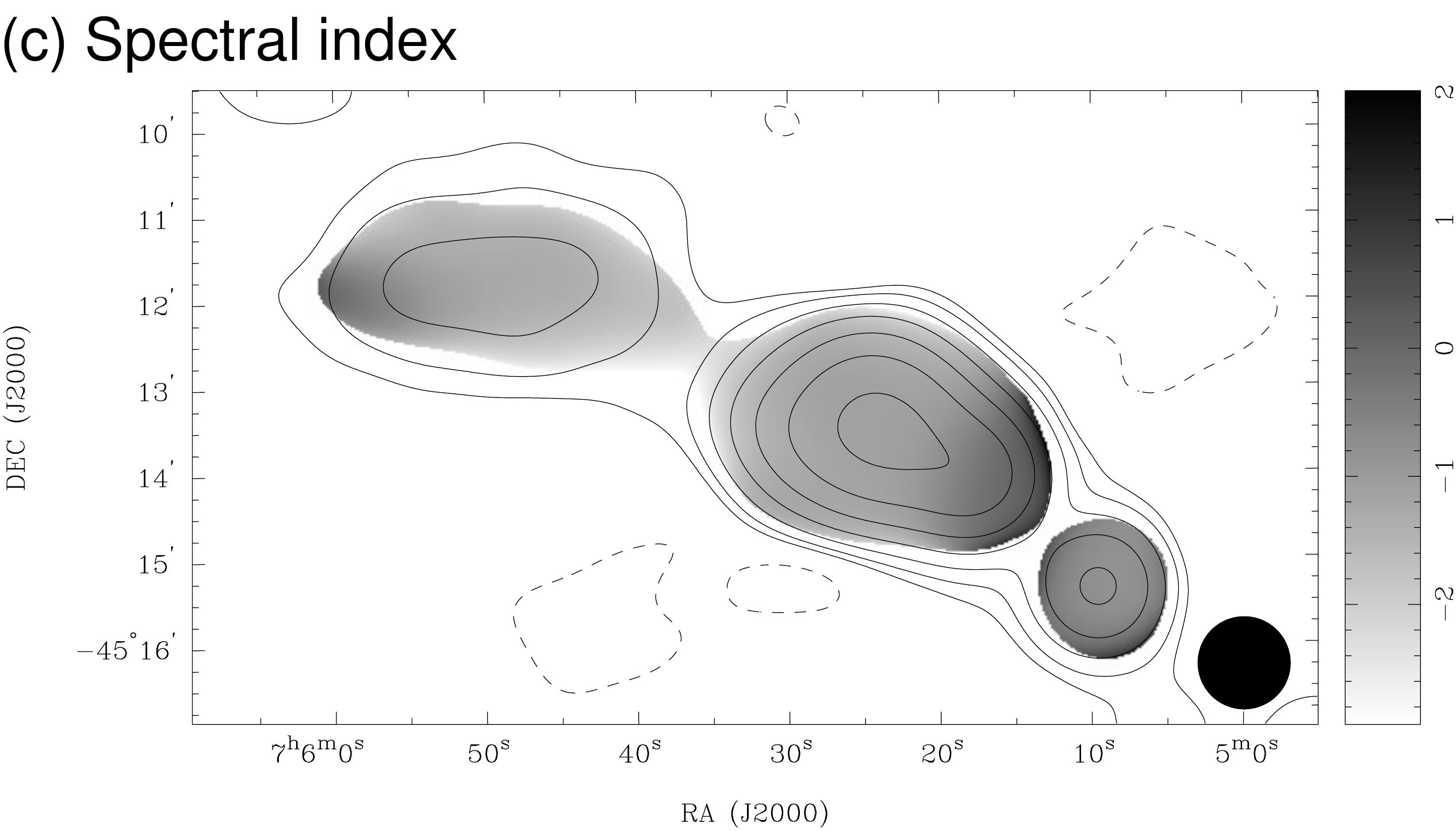} &
      \includegraphics[width=0.5\hsize]{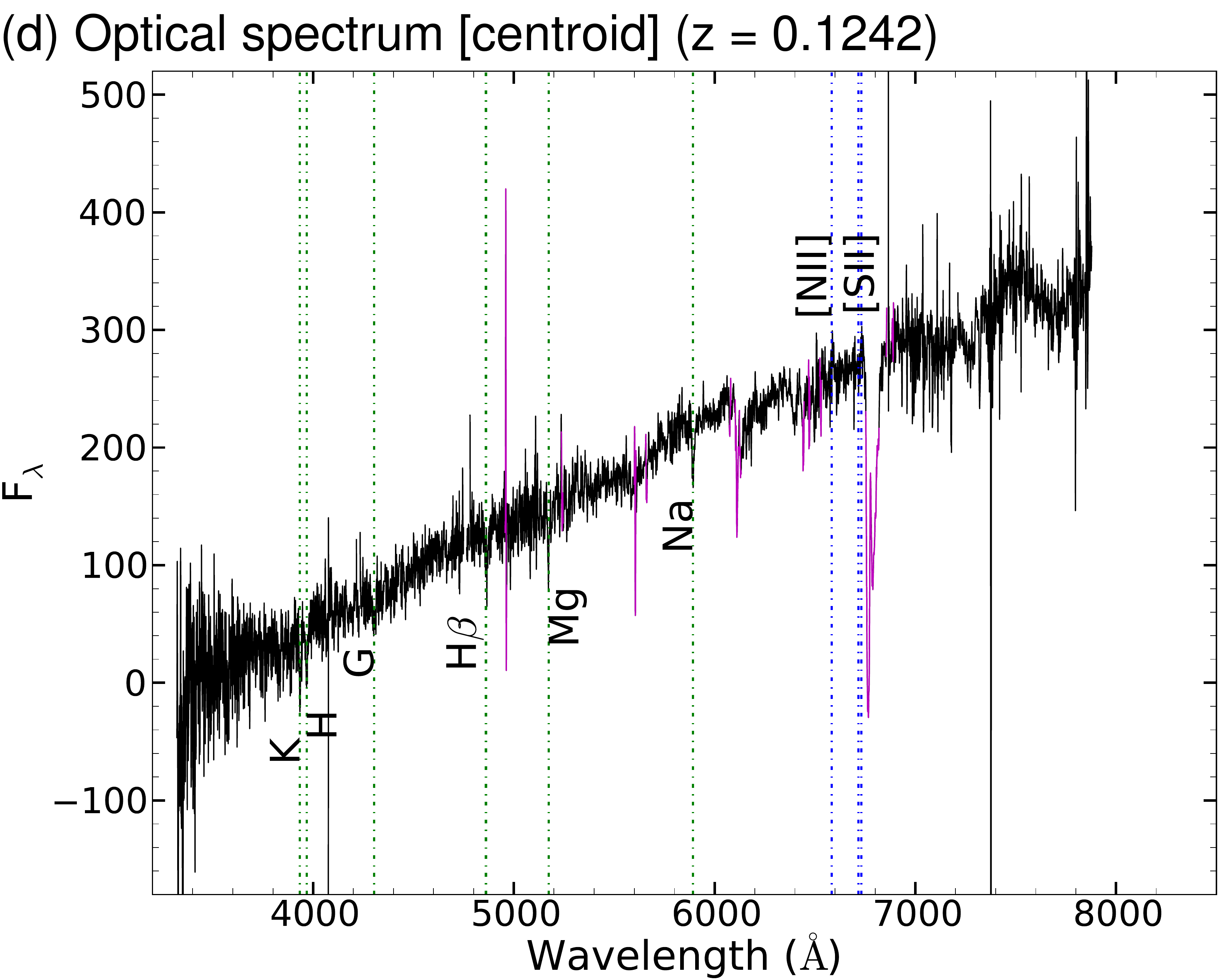}
\end{tabular}
\caption{(\subfigletter{a}) Low-resolution, wideband, total intensity image of B0703--451 at 2.1 GHz with contours at -3, 3, 6, 12, 24, 48, and 96 $\times$ 0.25 mJy beam$^{-1}$ and a beam of FWHM 48\arcsec. The inset shows the high resolution ATCA image of the SW lobe, including a resolved hotspot, with a beam of 20\arcsec and contour levels at -3, 3, 6, 12, 24, and 48 $\times$ 0.35 mJy beam$^{-1}$. (\subfigletter{b}) A 843 MHz SUMSS image included for comparison with a beam of FWHM 45\arcsec$\times$63\arcsec, total intensity contours at -3, 3, 6, 12, 24, 48, and 96 $\times$ 1.6 mJy beam$^{-1}$. (\subfigletter{c}) Distribution of spectral index computed from narrow band images at 843 MHz (SUMSS) and 1.55 GHz (ATCA), due to a primary beam offset that precluded the use of a 2.8 GHz ATCA subband, shown as greyscale in the range $-2.9$ to $2$ with a beam of FWHM 64\arcsec. The levels of overlaid total intensity contours are -3, 3, 6, 12, 24, 48, 96, and 192 $\times$ 0.25 mJy beam$^{-1}$. (\subfigletter{d}) An optical spectrum from AAOmega on the AAT of the host galaxy of B0703--451 at the centroid. The spectrum has been shifted to rest frame using the measured redshift of 0.1242.}
\label{fig:B0703}
\end{figure*}

\begin{figure}
  \centering
  \includegraphics[width=\hsize]{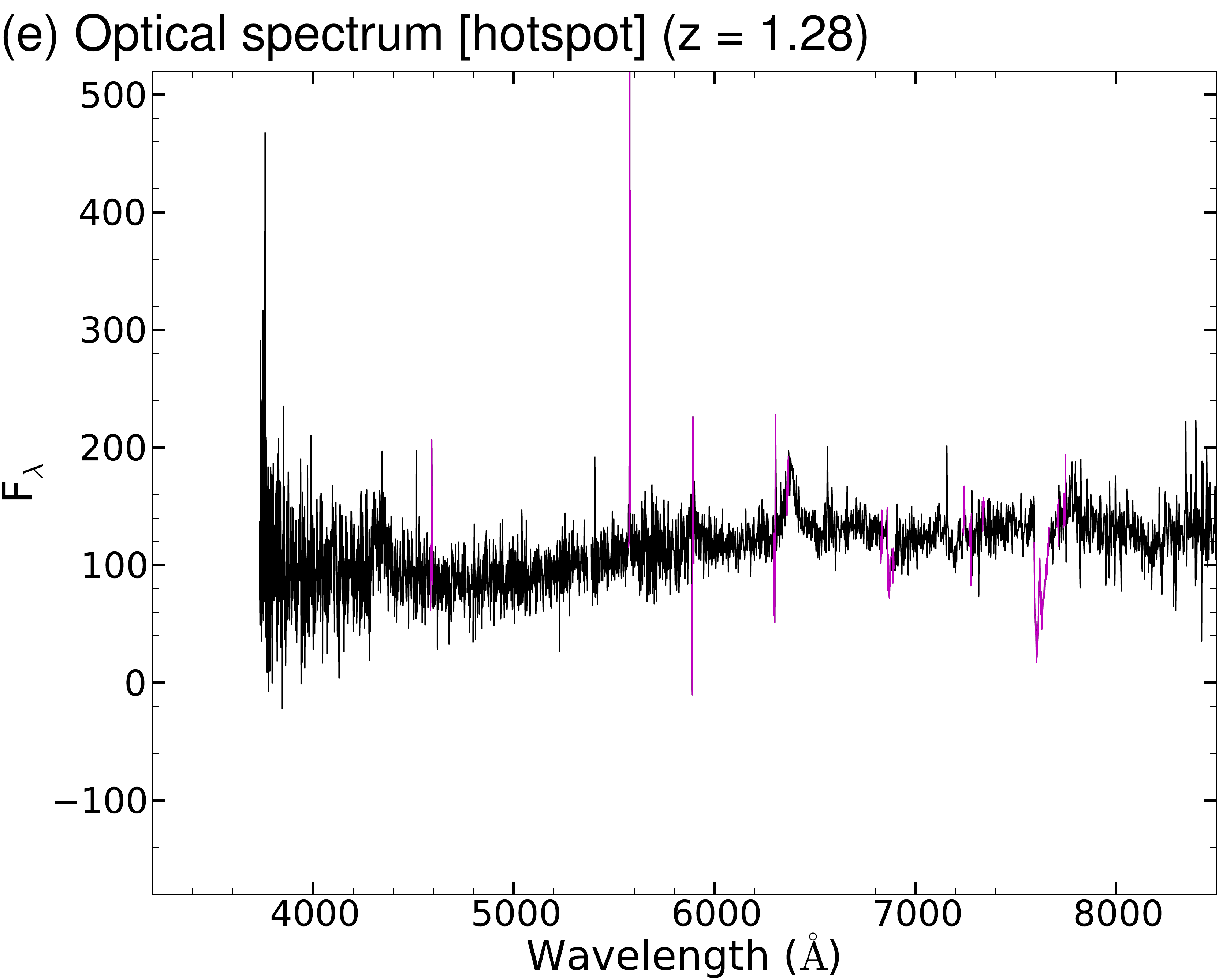} 
  \contcaption{(\subfigletter{e}) An optical spectrum from AAOmega on the AAT of the optical counterpart of the hotspot in the observed frame.}
\end{figure}

\subsection{J0746--5702 (Fig.~\ref{fig:J0746})}
The ATCA image of this GRG ($z=0.1301$) shows edge-brightened lobes with only weak emission peaks towards the lobe ends (Fig.~\ref{fig:J0746}). The previous high resolution ATCA image of this source~\citep{Saripallietal2005} does not reveal any compact hotspots within the lobes. Unusually for edge-brightened radio galaxies, the radio core is the brightest component in the source and as noted earlier it is accompanied by a partial, jet-like feature to the west (Fig.~\ref{fig:J0746}\subfigletter{b}). The ratio of core flux density to total flux density, the core fraction, is almost 15 per cent compared to the less than a few per cent expected for FRII sources. These characteristics of the source suggest that the AGN may have undergone a renewed jet activity. 

The axis formed by the core and the jet knot is skewed with respect to the line joining the hotspots in the lobes. The closer NE lobe is also the brighter. In this giant radio galaxy, the lack of a jet component towards the NE lobe will need to be understood. In particular we will need to determine if the asymmetry is related to asymmetries in the environment or projection effects. The optical spectrum of the host in Fig.~\ref{fig:J0746}\subfigletter{e} shows that the source is not a quasar. The spectrum is early-type with some weak emission features superposed. There is also the suggestion of line broadening in some of these features although the low signal-to-noise does not make this clear.

\begin{figure*}
  \centering
  \begin{tabular}{cc}
     \includegraphics[width=0.5\hsize]{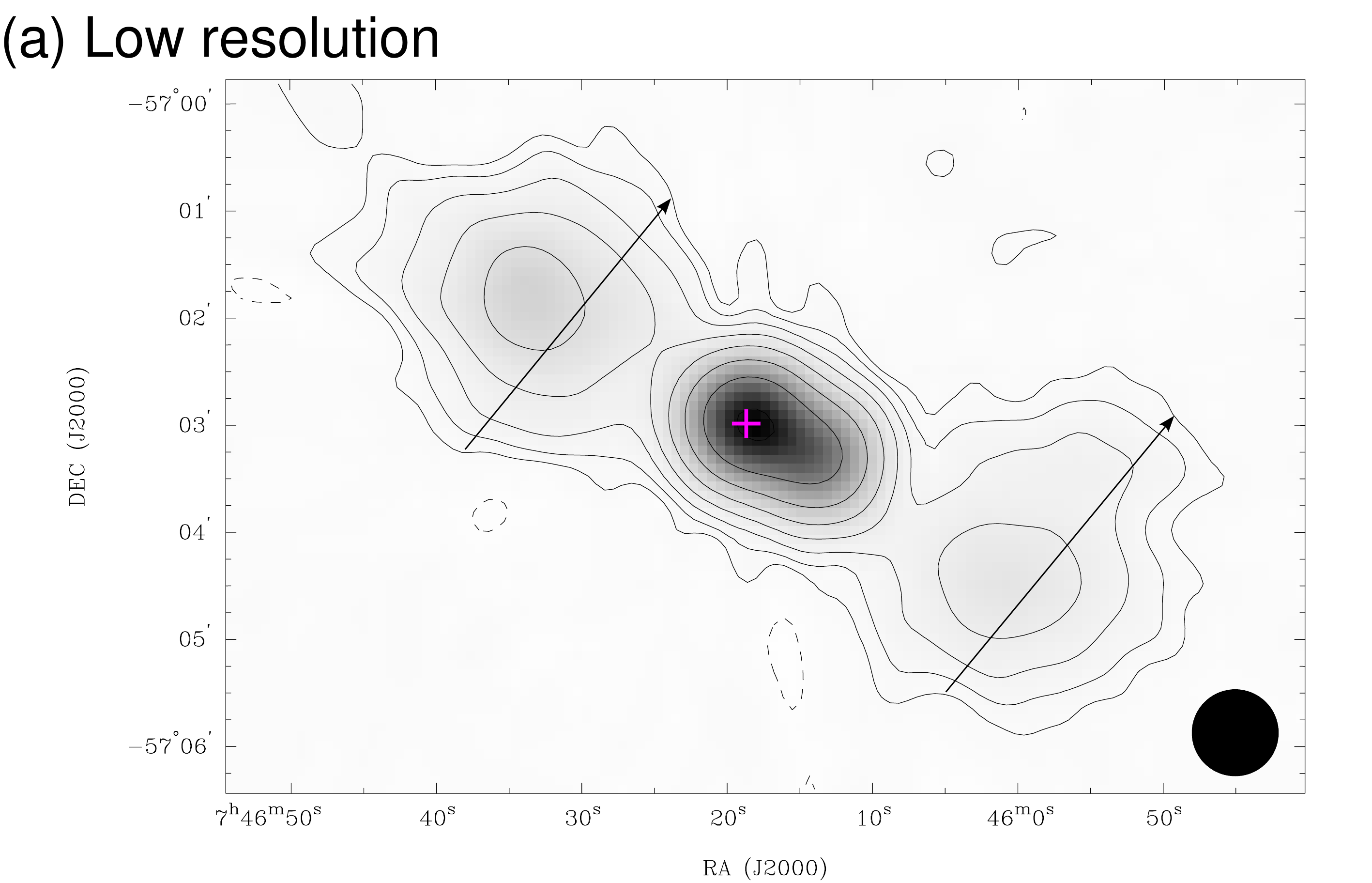} &
     \includegraphics[width=0.5\hsize]{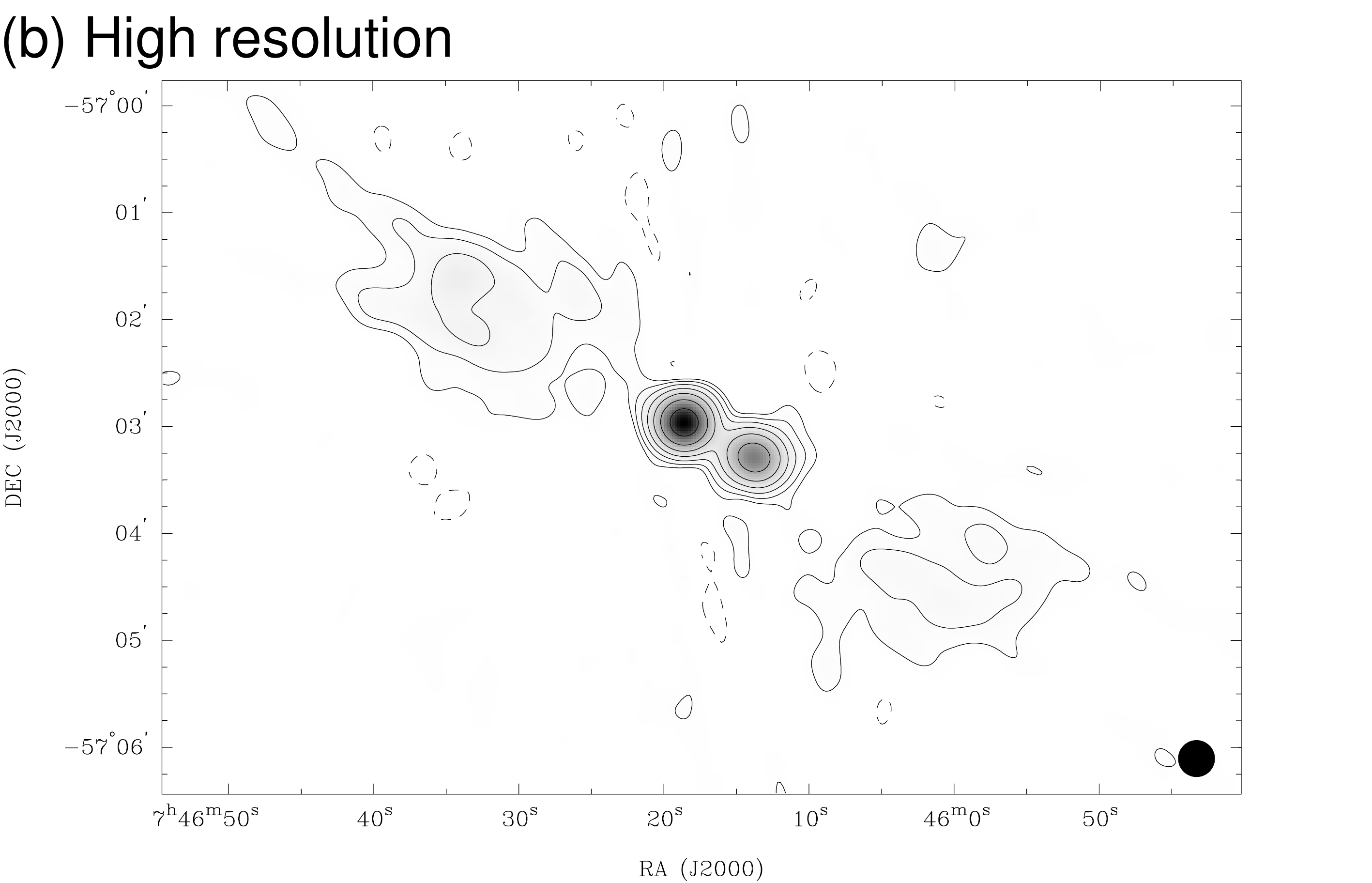} \\
     \includegraphics[width=0.5\hsize]{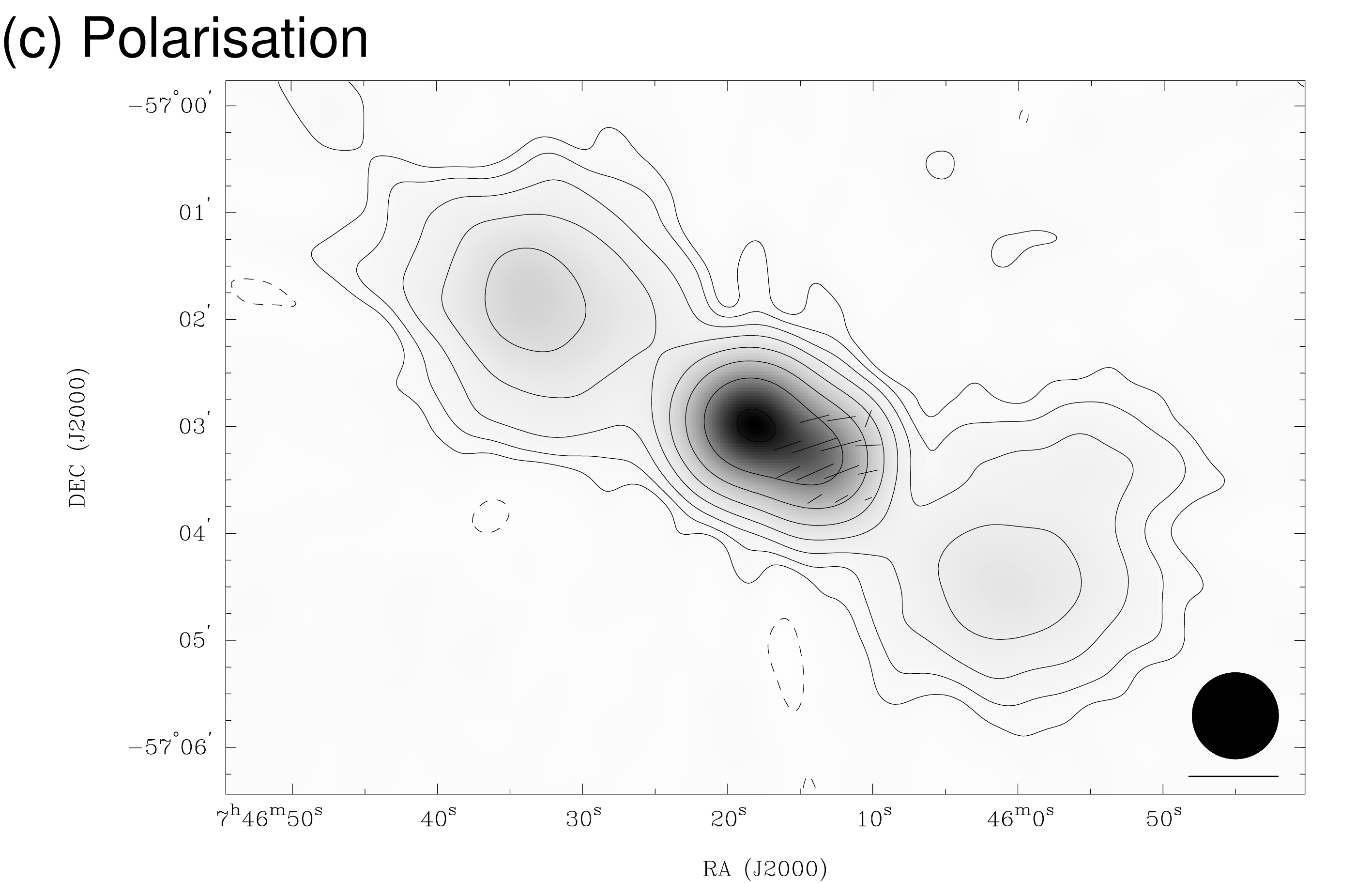} &
     \includegraphics[width=0.5\hsize]{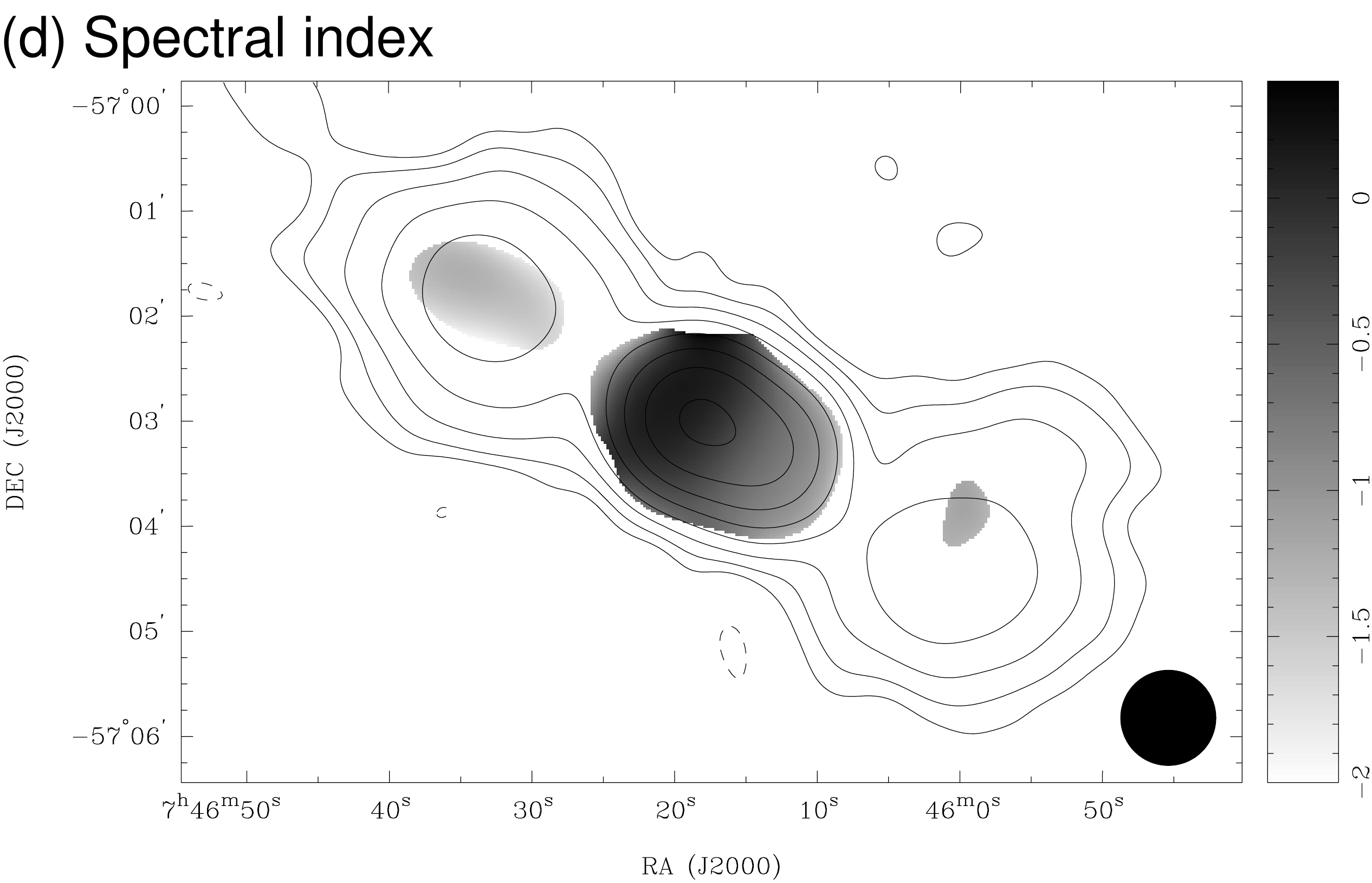}
\end{tabular}
\caption{(\subfigletter{a}) Low-resolution, wideband, total intensity image of J0746--5702 at 2.1 GHz with contours at -3, 3, 6, 12, 24, 48, 96, 192, and 384 $\times$ 0.12 mJy beam$^{-1}$ and a beam of FWHM 48\arcsec. (\subfigletter{b}) High resolution image of J0746--5702 with a beam of 20\arcsec and contour levels at -3, 3, 6, 12, 24, 48, 96, and 192 $\times$ 0.16 mJy beam$^{-1}$. (\subfigletter{c}) Distribution of polarised intensity at 2.8 GHz represented by the lengths of the overlaid Faraday rotation-corrected, electric field vectors. The scale-bar represents 4 mJy. (\subfigletter{d}) Distribution of spectral index between 843 MHz and 2.8 GHz shown as greyscale in the range $-2$ to $0.4$ with a beam of FWHM 54\arcsec~and total intensity contours at -3, 3, 6, 12, 24, 48, 96, 192, and 384 $\times$ 0.12 mJy beam$^{-1}$.}
\label{fig:J0746}
\end{figure*}

\begin{figure}
  \centering
      \includegraphics[width=\hsize]{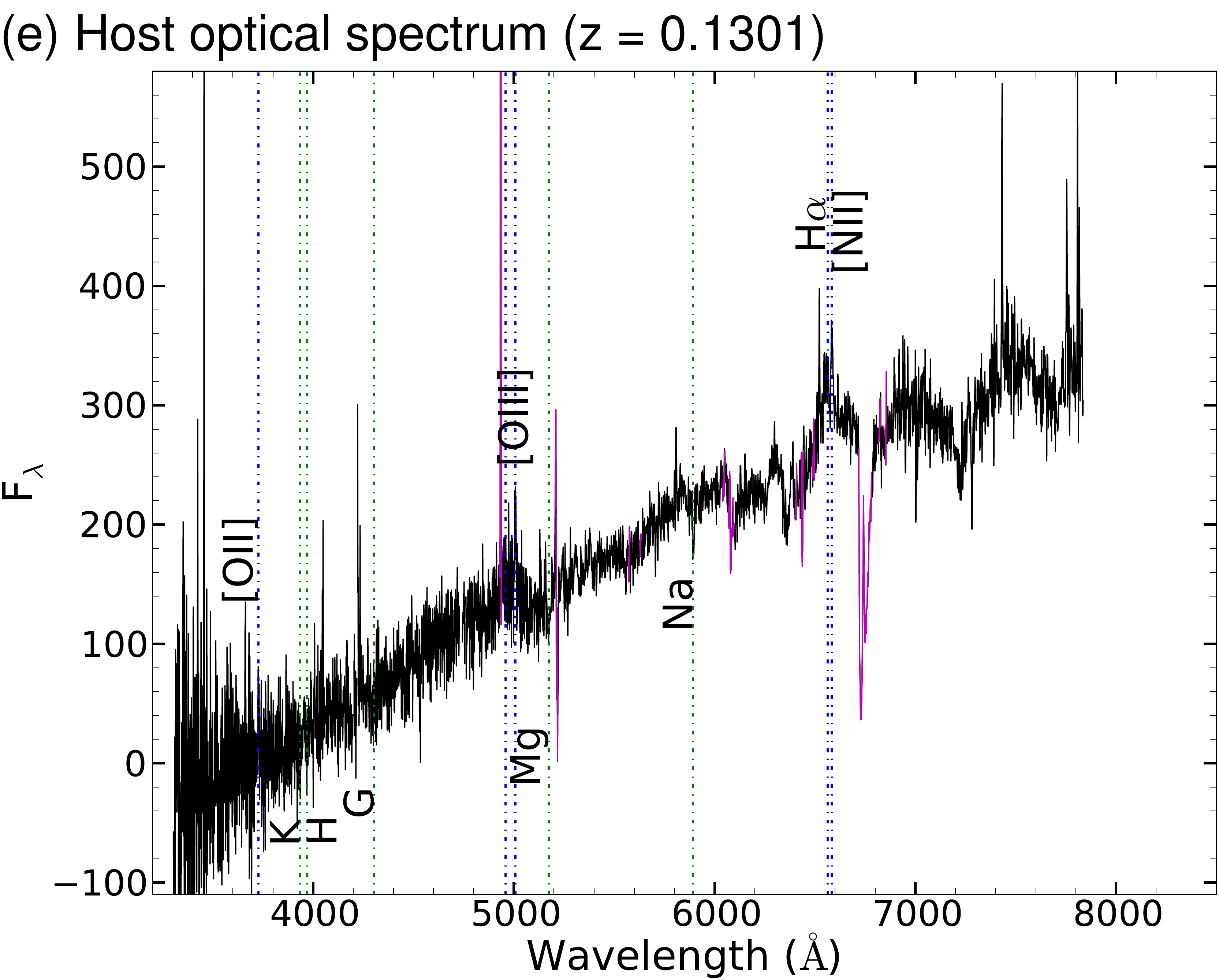}
  \contcaption{(\subfigletter{e}) An optical spectrum from AAOmega on the AAT of the host galaxy of J0746--5702. The spectrum has been shifted to rest frame using the measured redshift of 0.1301.}
\end{figure}

\subsection{J0843--7007 (Fig.~\ref{fig:J0843})}
This giant radio galaxy ($z=0.1390$) was imaged by~\citet{Saripallietal2005} as part of the SUMSS sample. They noted its similarity with another giant, J0746--5702, in that both showed weak lobes with a bright core. However a significant difference is that J0843--7007 has a bright, extended core with possible twin jets (Fig.~\ref{fig:J0843}\subfigletter{b}). The ATCA image has revealed that the lobes are edge-brightened with weak hotspots at their ends. Neither of these hotspots is detected at high resolution in the previous ATCA image. Our image has also revealed an inversion-symmetry in this source where the EW jets bend in opposite directions towards the two respective lobes. The jets and the lobes are connected by a continuous bridge of emission. Once again the fractional core flux is too large for an FRII radio galaxy (Fig.~\ref{fig:J0843}\subfigletter{c}) and together with the twin jets and lack of compact hotspots likely indicate a restarted jet activity in this giant.

The optical spectrum of the host indicates that it is not a quasar and in fact typical of an ordinary early-type spectrum (Fig.~\ref{fig:J0843}\subfigletter{e}). The peaks of emission to the east are likely an unrelated background source. This pair of point sources with a bridge of emission and located along the jet axis (to the east) of the GRG have no optical counterpart.

\begin{figure*}
  \centering
  \begin{tabular}{cc}
      \includegraphics[width=0.5\hsize]{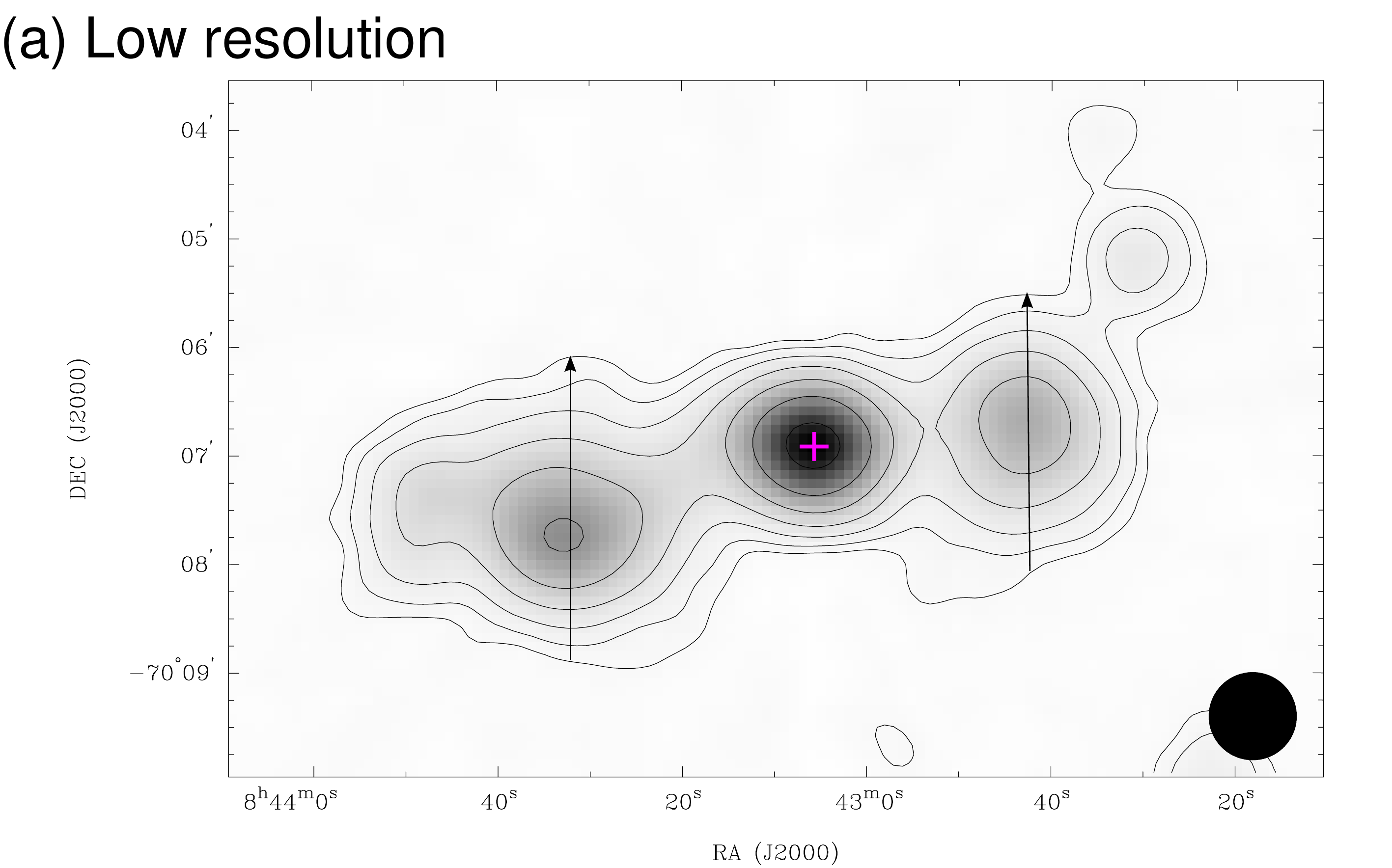} &
      \includegraphics[width=0.5\hsize]{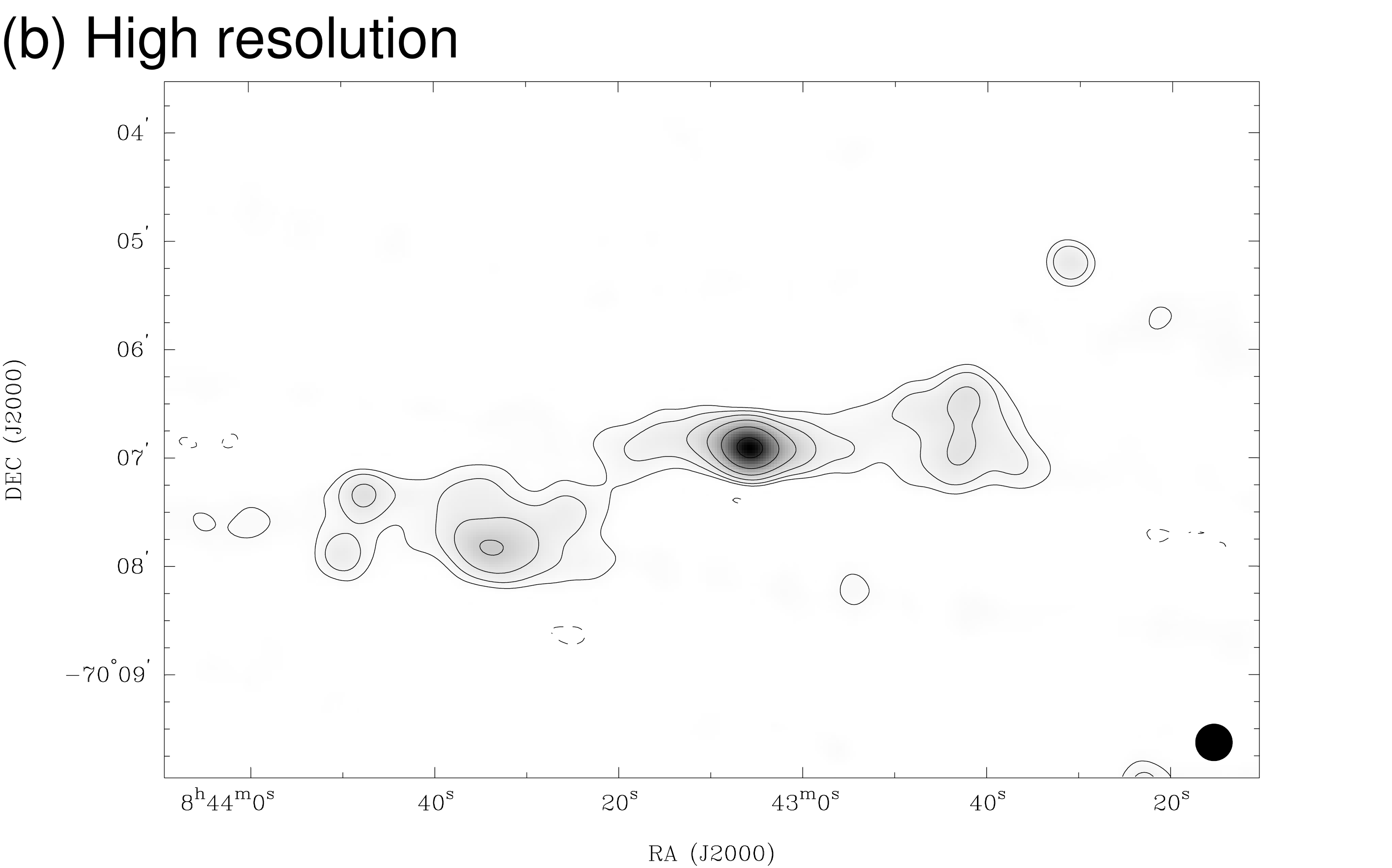} \\
      \includegraphics[width=0.5\hsize]{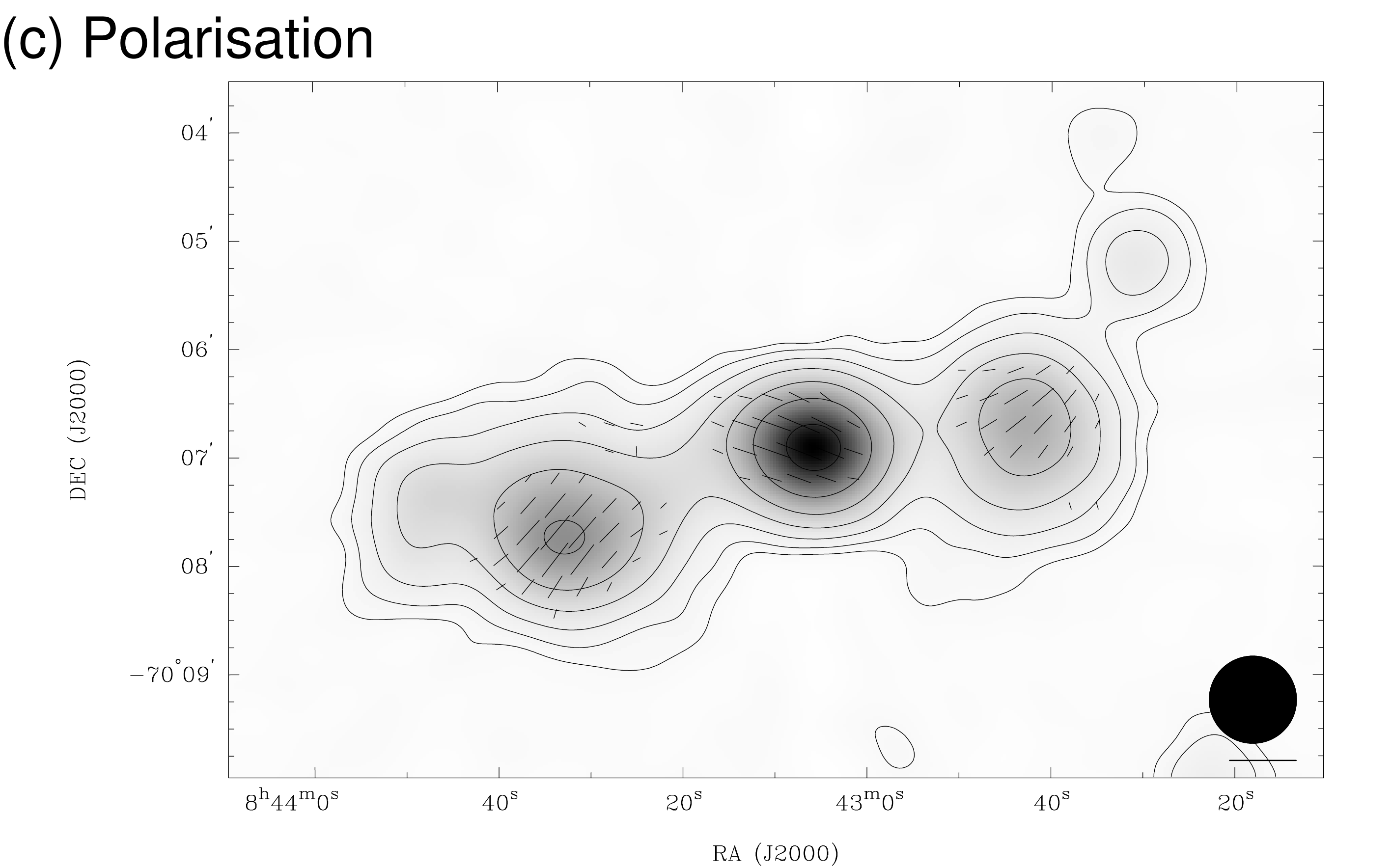} &
      \includegraphics[width=0.5\hsize]{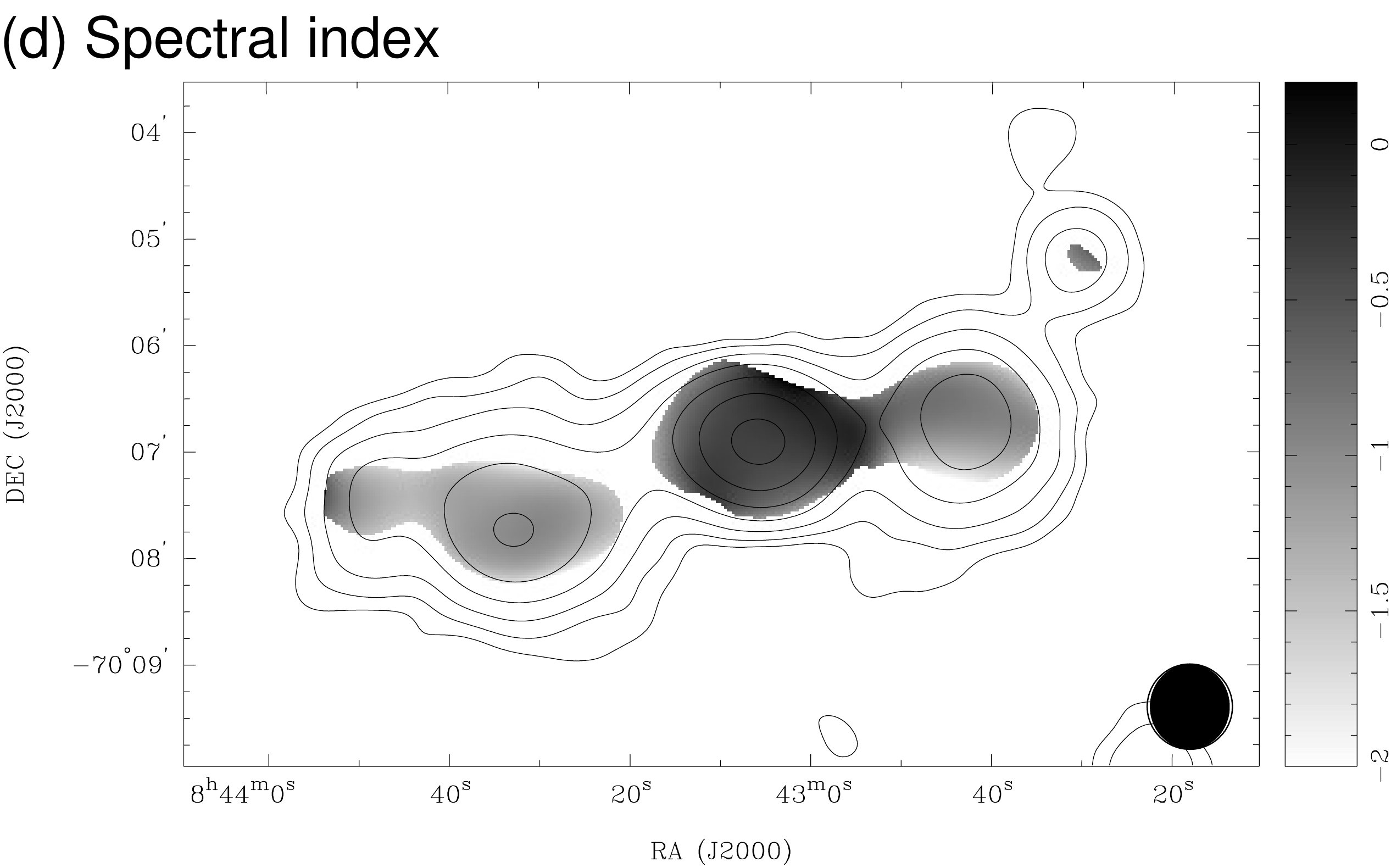}
\end{tabular}
\caption{(\subfigletter{a}) Low-resolution, wideband, total intensity image of J0843--7007 at 2.1 GHz with contours at -3, 3, 6, 12, 24, 48, 96, and 192 $\times$ 95 $\mu$Jy beam$^{-1}$ and a beam of FWHM 48\arcsec. (\subfigletter{b}) High resolution image of J0843--7007 with a beam of 20\arcsec and contour levels at -3, 3, 6, 12, 24, 48, and 96 $\times$ 0.15 mJy beam$^{-1}$. (\subfigletter{c}) Distribution of polarised intensity at 2.8 GHz represented by the lengths of the overlaid Faraday rotation-corrected, electric field vectors. The scale-bar represents 2 mJy. (\subfigletter{d}) Distribution of spectral index between 843 MHz and 2.8 GHz shown as greyscale in the range $-2$ to $0.2$ with a beam of FWHM 48\arcsec~and total intensity contours at -3, 3, 6, 12, 24, 48, 96, and 192 $\times$ 95 $\mu$Jy beam$^{-1}$.}
\label{fig:J0843}
\end{figure*}

\begin{figure}
  \centering
      \includegraphics[width=\hsize]{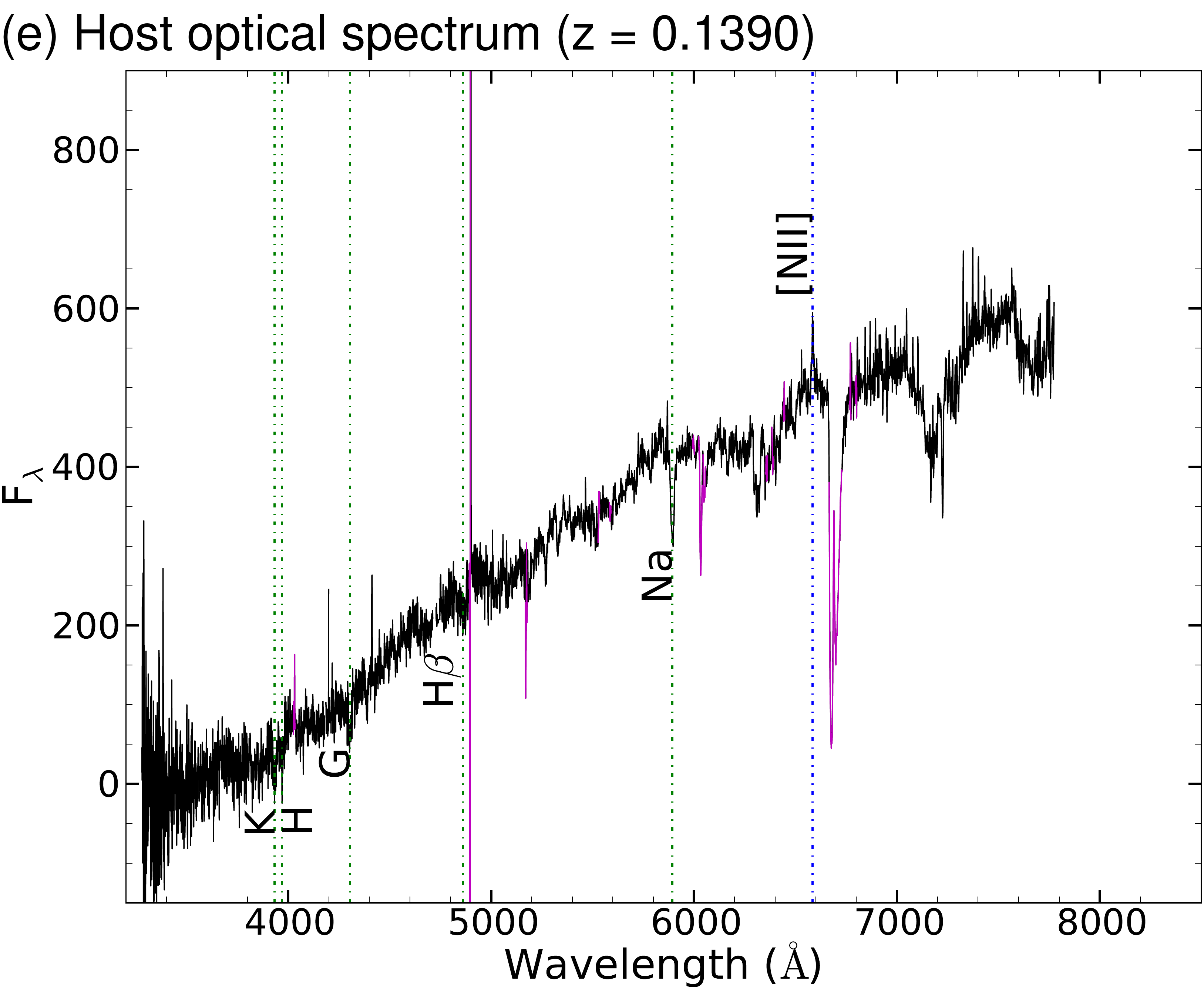}
  \contcaption{(\subfigletter{e}) An optical spectrum from AAOmega on the AAT of the host galaxy of J0843--7007. The spectrum has been shifted to rest frame using the measured redshift of 0.1390.}
\end{figure}

\subsection{B1302--325 (Fig.~\ref{fig:B1302})}
This radio galaxy ($z=0.1528$) has been previously imaged by~\citet{JonesMcAdam1992} at 843 MHz. As noted previously the source has edge-brightened lobes and both lobes have diffuse emission regions extending in opposite directions. The SW lobe has a compact and strong hotspot unlike the NE lobe (Fig. \ref{fig:B1302}\subfigletter{b}). In the optical spectrum, emission-lines feature prominently in what is otherwise an ordinary early-type spectrum (Fig. \ref{fig:B1302}\subfigletter{e}).

\begin{figure*}
  \centering
    \begin{tabular}{cc}
      \includegraphics[width=0.5\hsize]{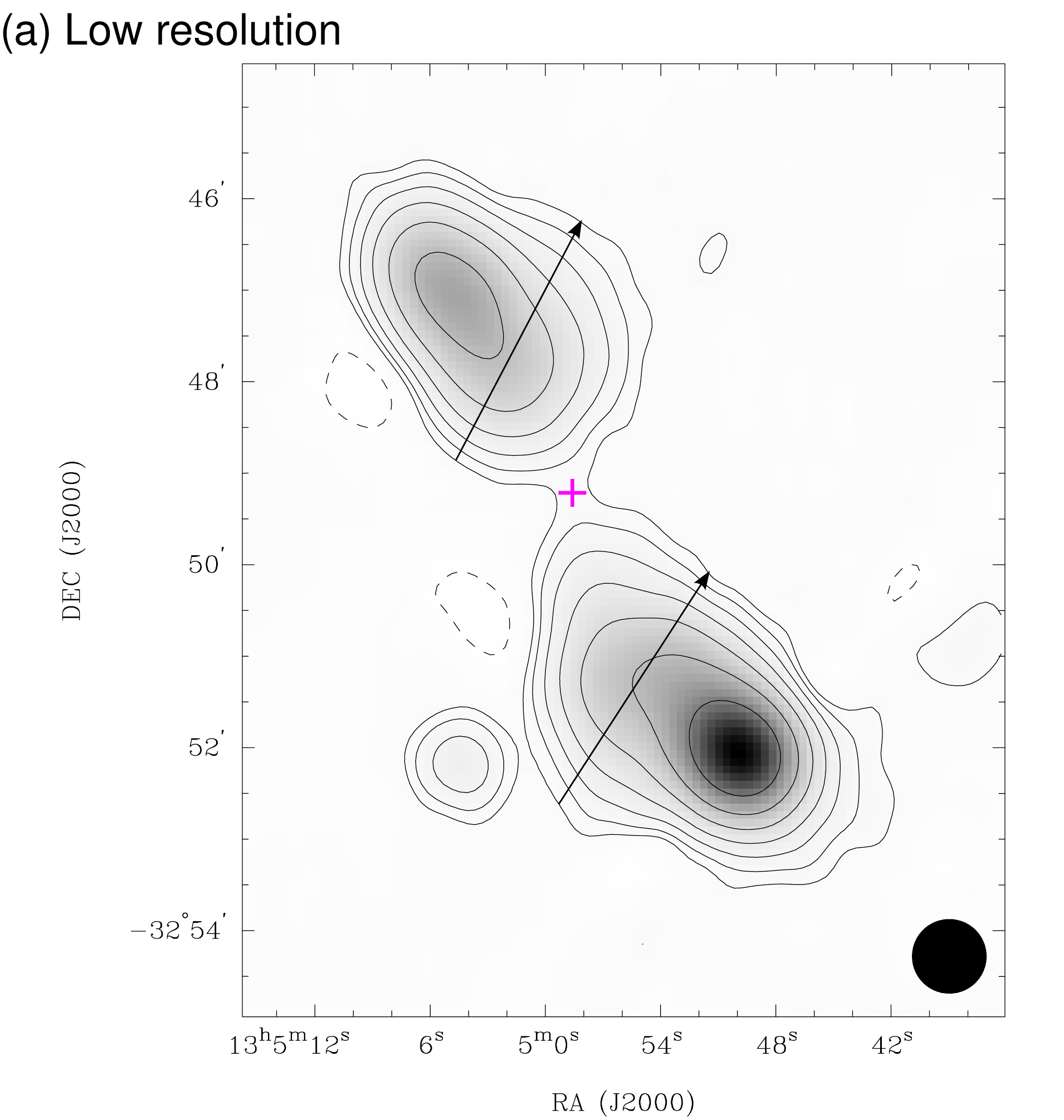} &
      \includegraphics[width=0.5\hsize]{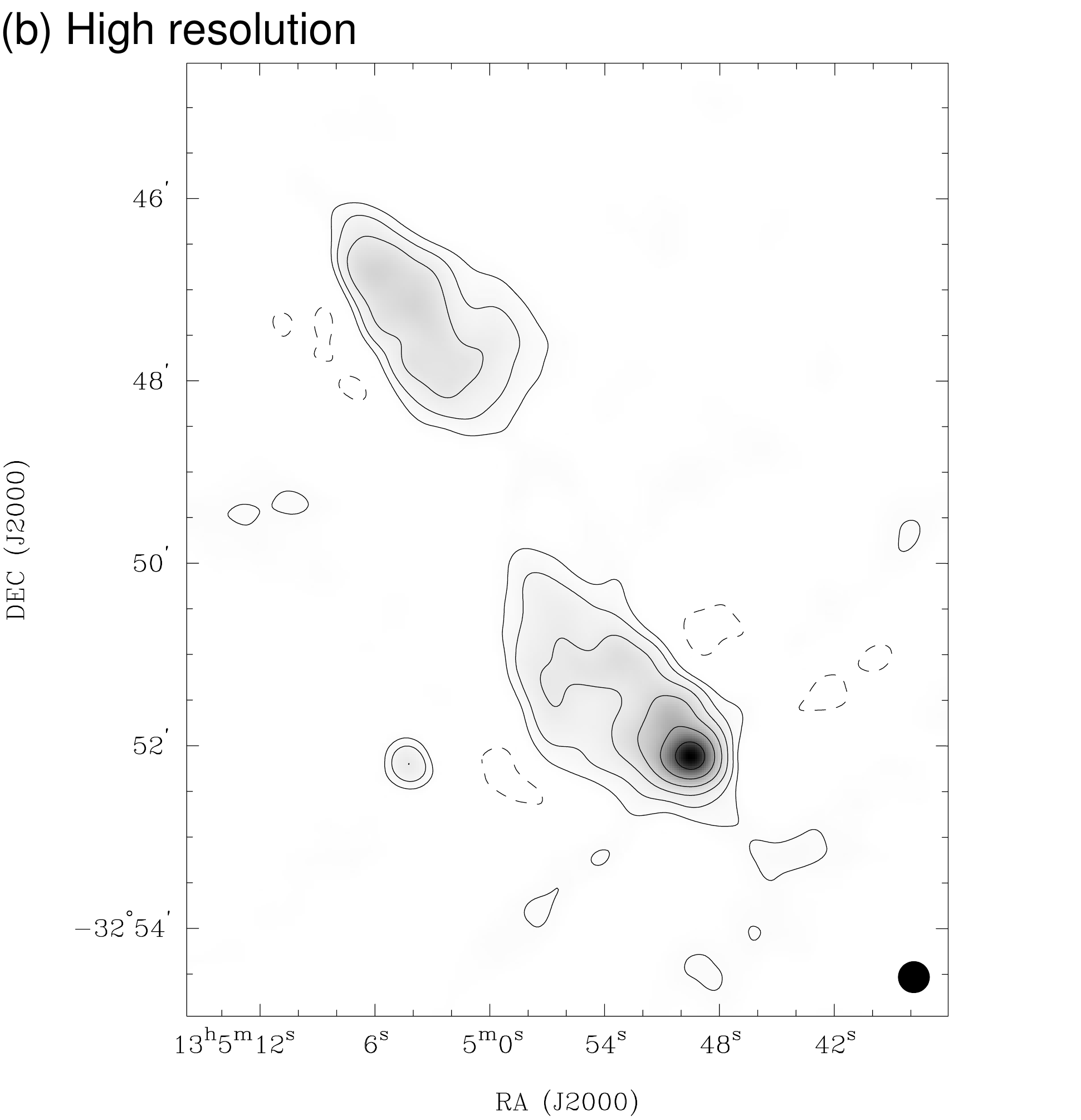} \\
      \includegraphics[width=0.5\hsize]{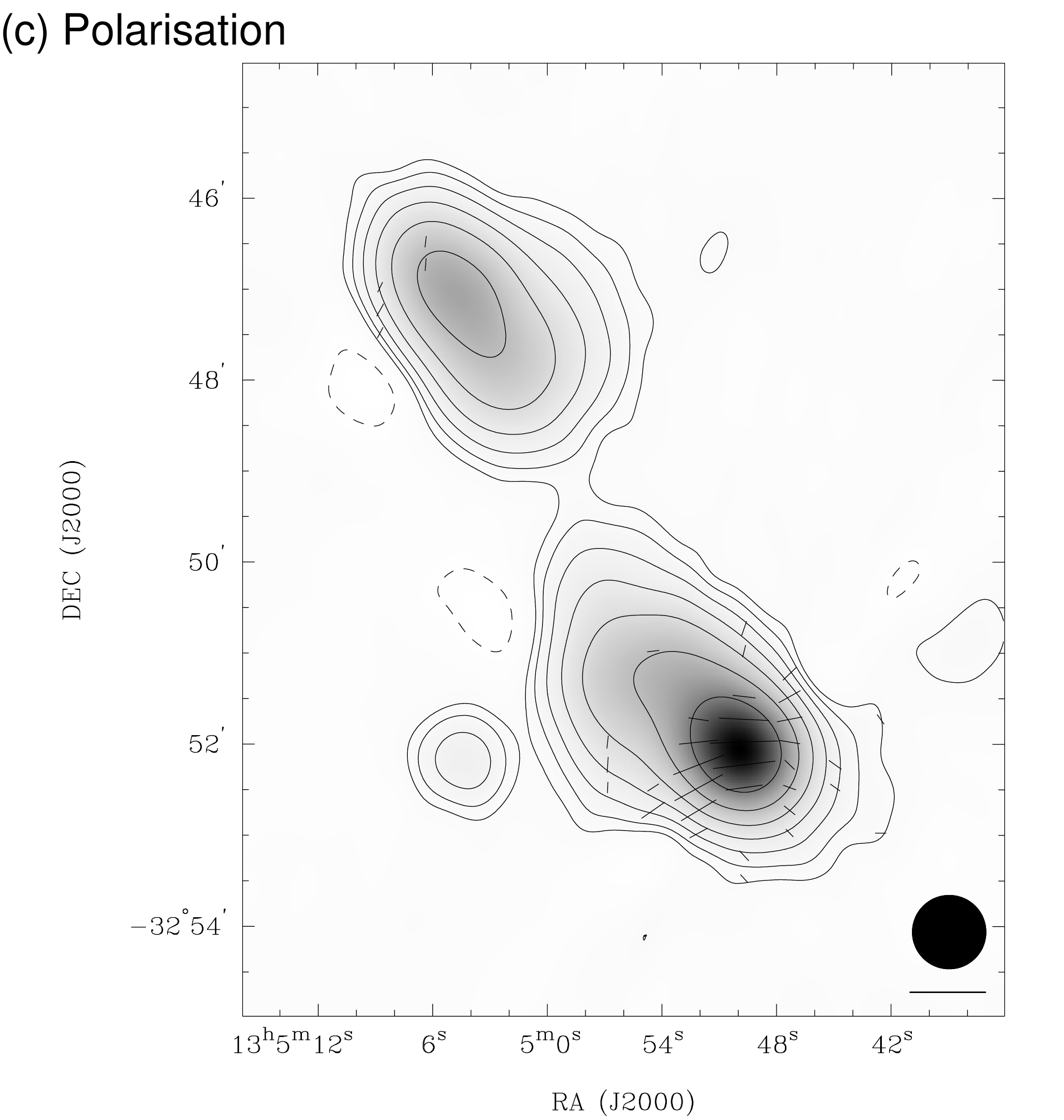} &
      \includegraphics[width=0.5\hsize]{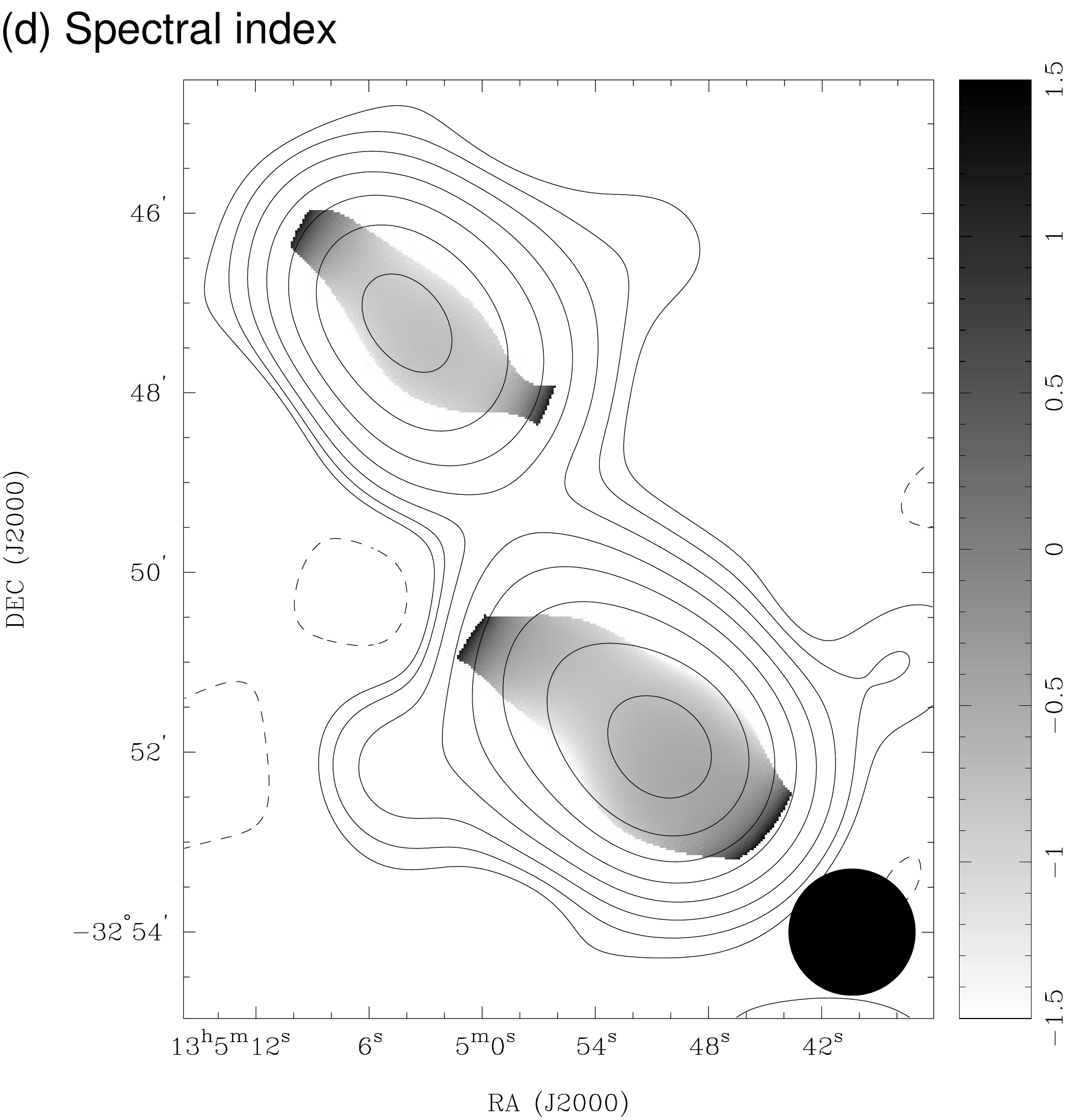}
\end{tabular}
\caption{(\subfigletter{a}) Low-resolution, wideband, total intensity image of B1302--325 at 2.1 GHz with contours at -3, 3, 6, 12, 24, 48, 96, and 192 $\times$ 0.8 mJy beam$^{-1}$ and a beam of FWHM 48\arcsec. (\subfigletter{b}) High resolution image of B1302--325 with a beam of 20\arcsec and contour levels at -3, 3, 6, 12, 24, 48, and 96 $\times$ 1.3 mJy beam$^{-1}$. (\subfigletter{c}) Distribution of polarised intensity at 2.8 GHz represented by the lengths of the overlaid Faraday rotation-corrected, electric field vectors. The scale-bar represents 40 mJy. (\subfigletter{d}) Distribution of spectral index between 843 MHz and 2.8 GHz shown as greyscale in the range $-1.5$ to $1.5$ with a beam of FWHM 84\arcsec~and total intensity contours at -3, 3, 6, 12, 24, 48, 96, 192, and 384 $\times$ 0.8 mJy beam$^{-1}$.}
\label{fig:B1302}
\end{figure*}

\begin{figure}
  \centering
      \includegraphics[width=\hsize]{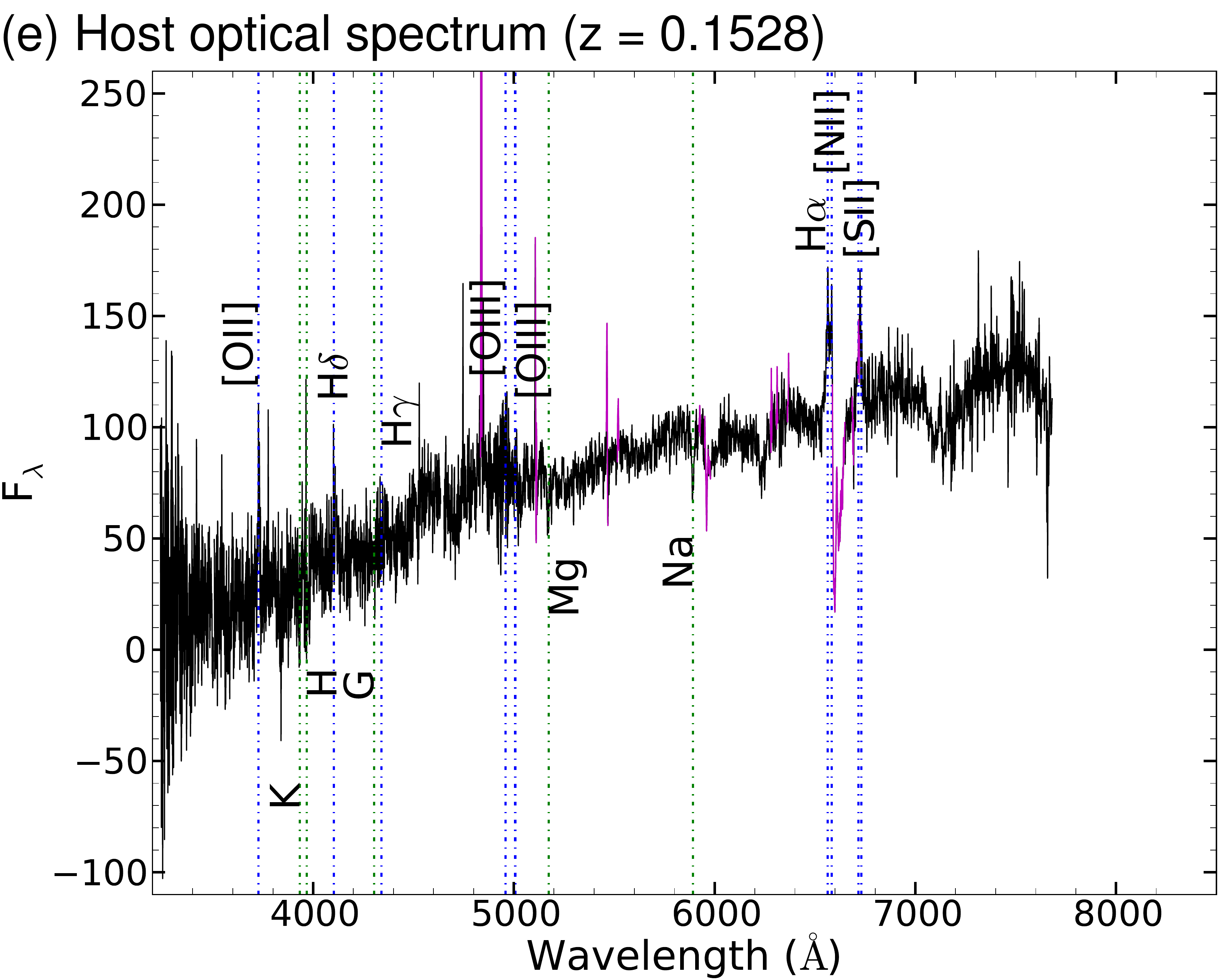}
  \contcaption{(\subfigletter{e}) An optical spectrum from AAOmega on the AAT of the host galaxy of B1302--325. The spectrum has been shifted to rest frame using the measured redshift of 0.1528.}
\end{figure}

\begin{figure*}
  \centering
  \begin{tabular}{cc}
      \includegraphics[width=0.5\hsize]{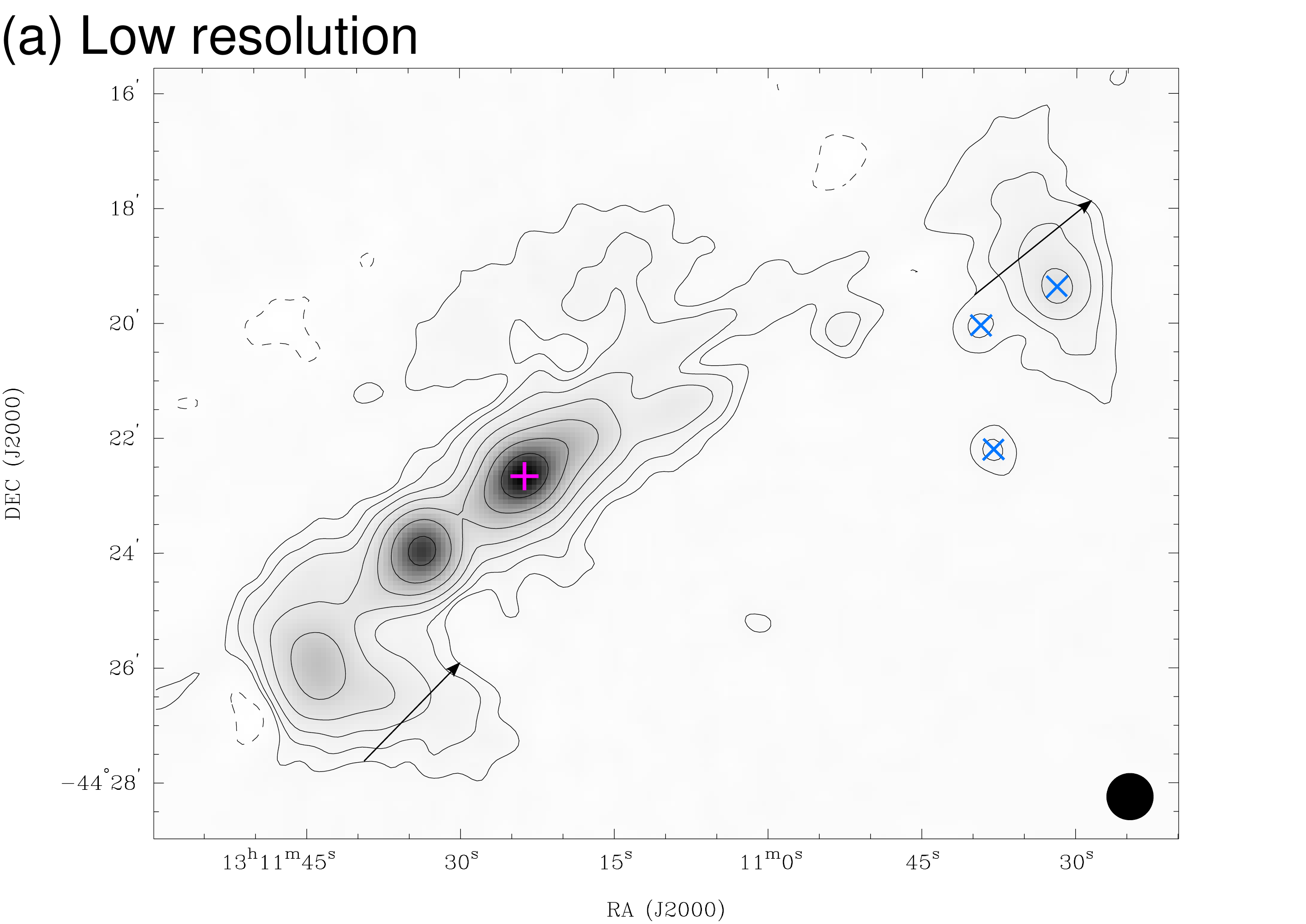} &
      \includegraphics[width=0.5\hsize]{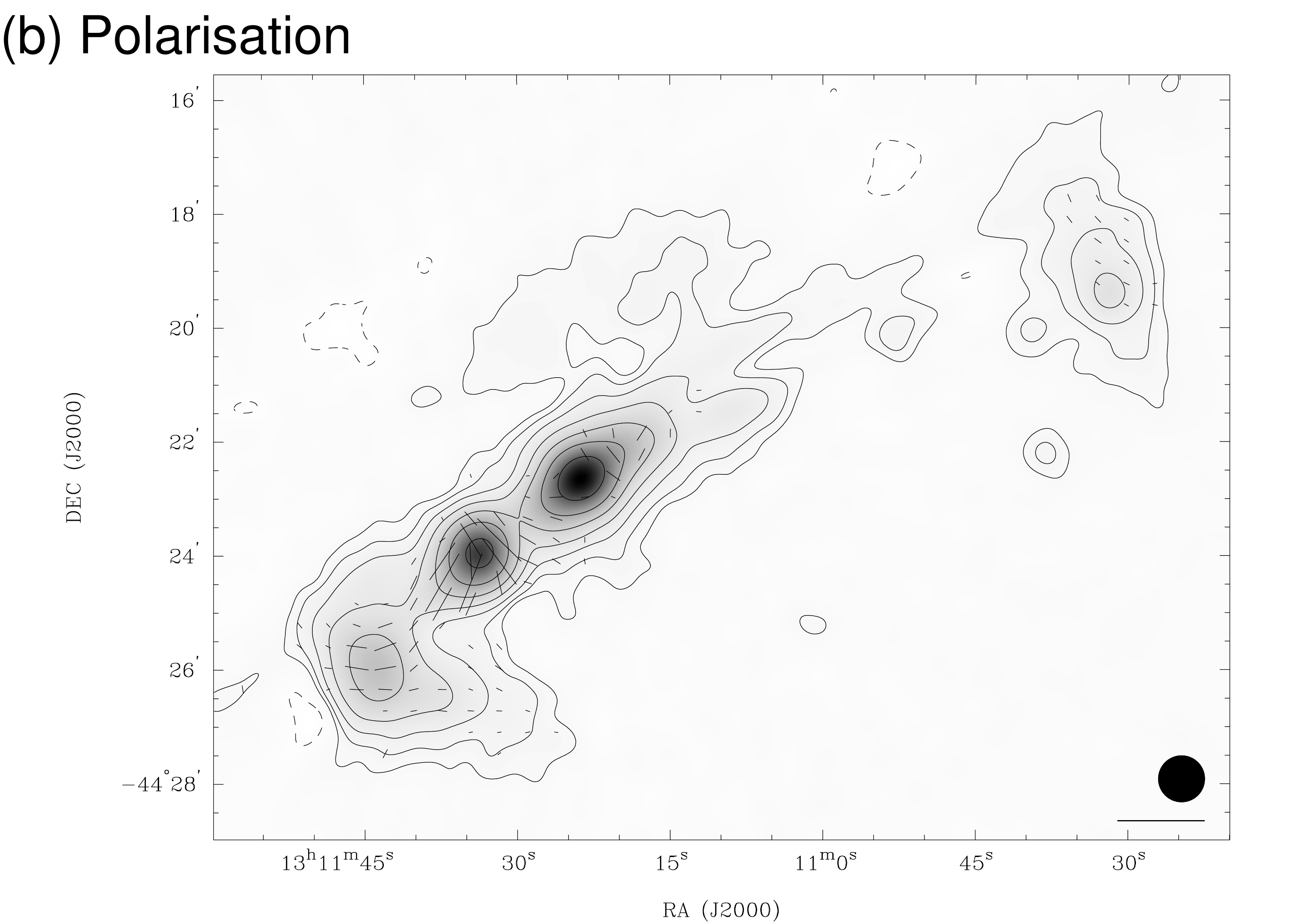} \\
      \includegraphics[width=0.5\hsize]{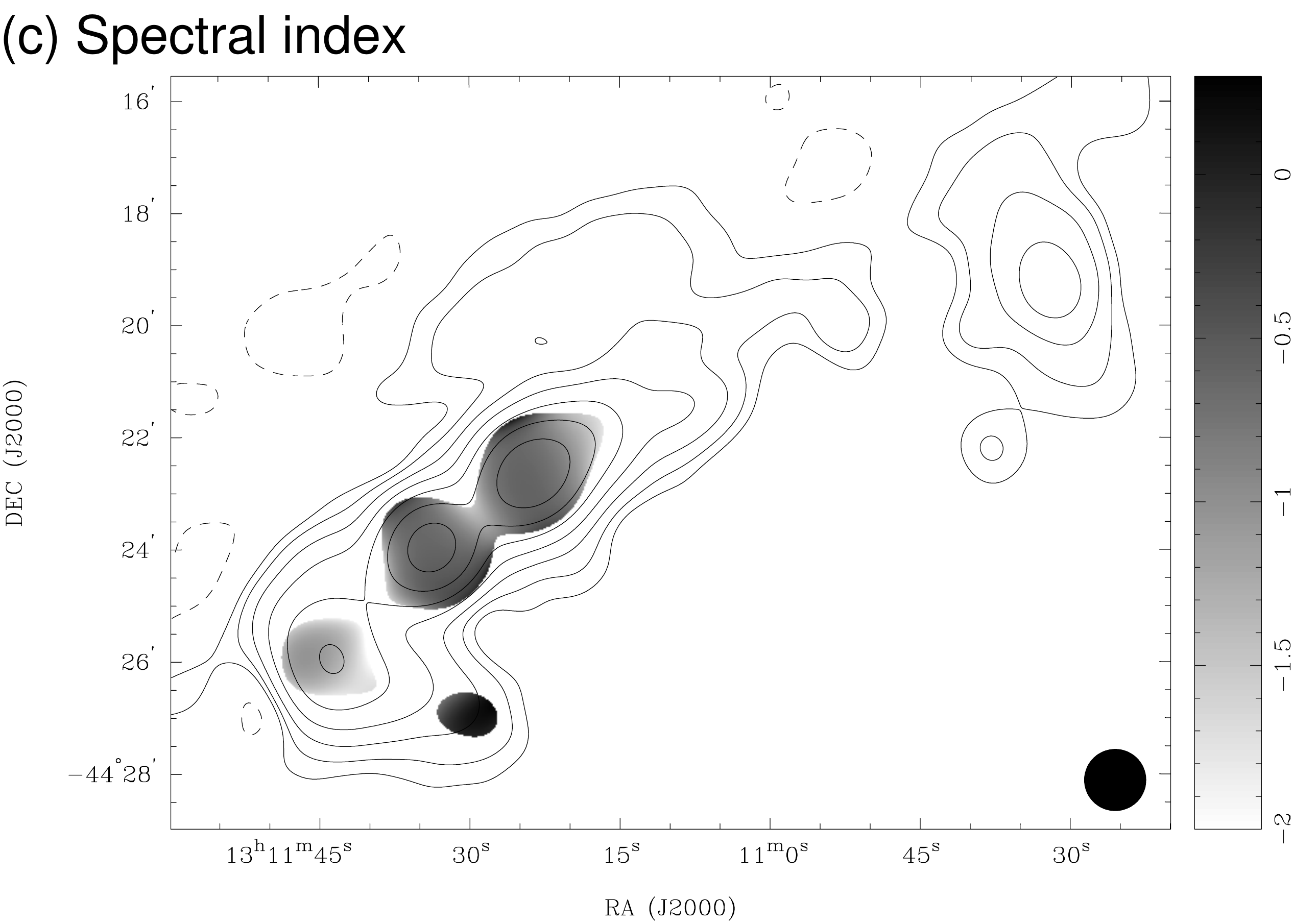} &
      \includegraphics[width=0.5\hsize]{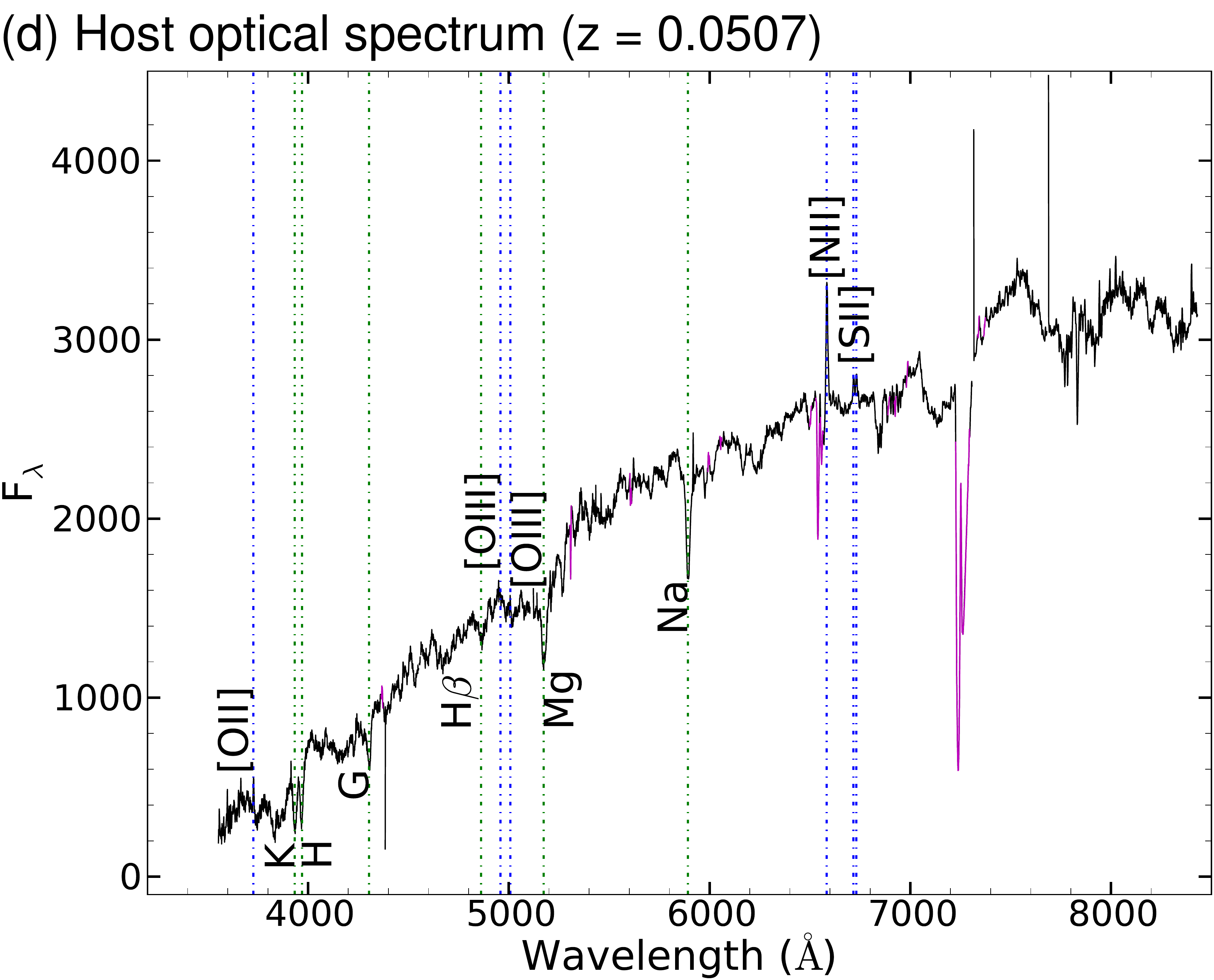}
\end{tabular}
\caption{(\subfigletter{a}) Low-resolution, wideband, total intensity image of B1308--441 at 2.1 GHz with contours at -3, 3, 6, 12, 24, 48, 96, and 192 $\times$ 0.4 mJy beam$^{-1}$ and a beam of FWHM 48\arcsec. Background point sources are marked with crosses. In the online version these unrelated objects are shown in blue (\textcolor{cyan}{$\times$}). (\subfigletter{b}) Distribution of polarised intensity at 2.8 GHz represented by the lengths of the overlaid Faraday rotation-corrected, electric field vectors. The scale-bar represents 10 mJy. (\subfigletter{c}) Distribution of spectral index between 843 MHz and 2.8 GHz shown as greyscale in the range $-2$ to $0.3$ with a beam of FWHM 65\arcsec~and total intensity contours at -3, 3, 6, 12, 24, 48, 96, and 192 $\times$ 0.4 mJy beam$^{-1}$. (\subfigletter{d}) An optical spectrum from AAOmega on the AAT of the host galaxy of B1308--441. The spectrum has been shifted to rest frame using the measured redshift of 0.0507.}
\label{fig:B1308}
\end{figure*}

\subsection{B1308--441 (Fig.~\ref{fig:B1308})}
The total intensity image (Fig.~\ref{fig:B1308}\subfigletter{a}) has revealed that there is no broad plume of emission in the NW as seen in the SE as suggested by the~\citet{JonesMcAdam1992} image. The extended emission to the south of the host in the previous 843 MHz image has been detected with the ATCA. Our high resolution image (not presented) shows that there is jet-like collimated emission present even at large distances from the core. The broad lobe seen in the 843 MHz image is resolved into three background point sources (positions marked on the low resolution map) and a linear jet feature in the same position angle as the jet closer to the core. This FRI giant radio galaxy ($z=0.0507$) has very asymmetric lobes: to the NW is seen a collimated jet that remains narrow even to 150 kpc from the core whereas to the SE is seen a jet to a much smaller distance that exhibits a strong knot and ends in a broad plume of emission. The very high signal-to-noise optical spectrum in Fig.~\ref{fig:B1308}\subfigletter{d} is a classic early-type spectrum with major absorption features present. The SuperCOSMOS catalogue designated two positions for one object in B1308--441 and so we chose the candidate three magnitudes brighter as the host.

\subsection{J2159--7219 (Fig.~\ref{fig:J2159})}
The ATCA image of this giant radio galaxy (Fig. \ref{fig:J2159}) shows a remarkable structure where a bright core is connected to two edge-brightened lobes by continuous and narrow jets. This source may be classified as an FRII radio galaxy ($z=0.0967$) as it is edge brightened yet the bright core and continuous jets, each of which can be traced to the lobes, are characteristic of FRI sources. The jets were not detected in~\citet{Saripallietal2005}. The high resolution ATCA image (Fig. \ref{fig:J2159}\subfigletter{b}) shows an emission peak at the end of the NW lobe whereas in the SE the lobe is seen to have recessed peaks following the position angle of the southern jet. Although edge brightened, previous high-resolution ATCA images failed to reveal compact hotspots in the two lobes. There are two prominent knots in the jets symmetrically located on either side of the core, which may represent a new epoch of activity.
The host galaxy is bright and has several neighbours. There is also a background source to the south of the NW lobe. The optical spectrum of the host is clearly early-type.

\begin{figure*}
  \centering
  \begin{tabular}{cc}
      \includegraphics[width=0.5\hsize]{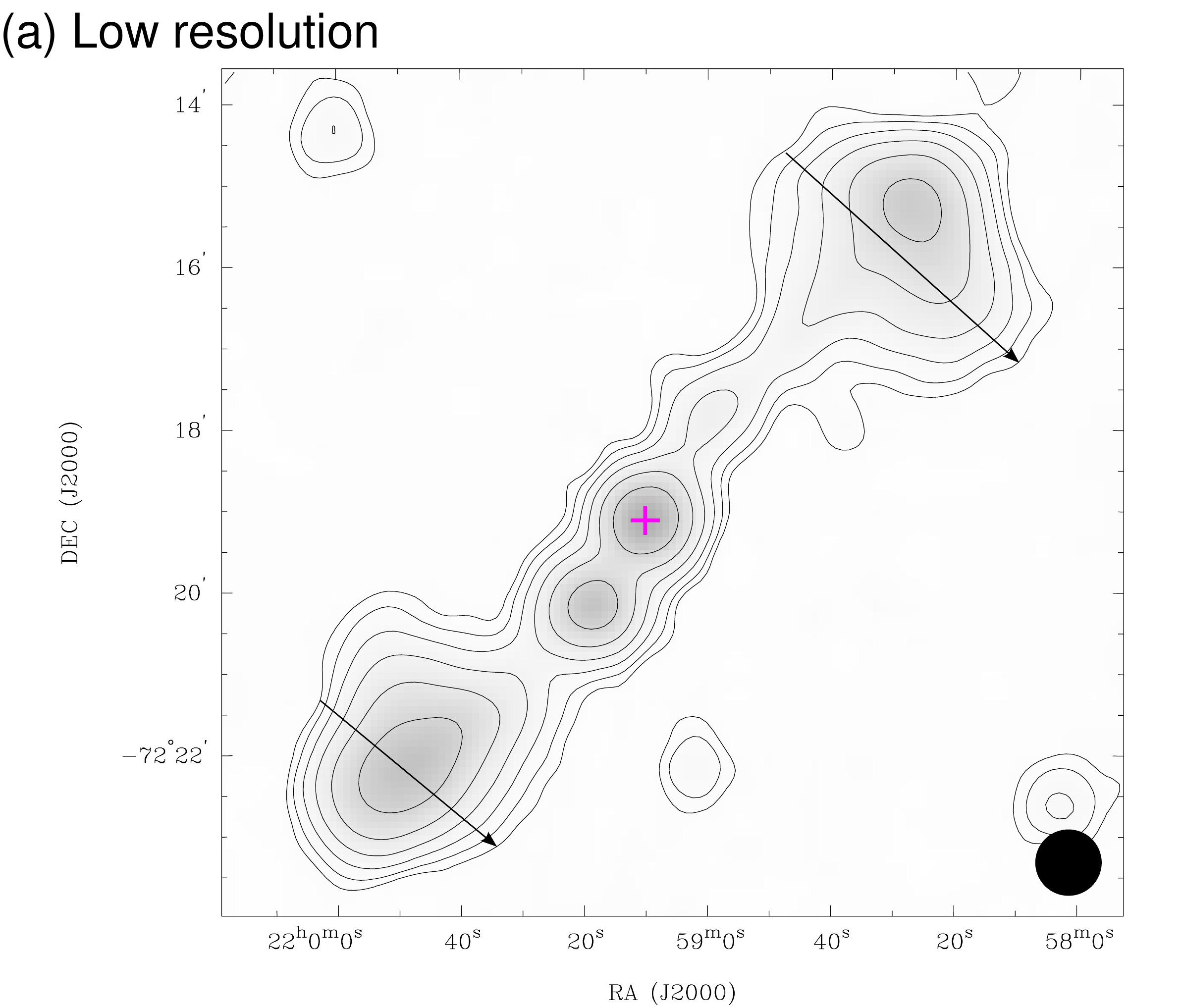}&
      \includegraphics[width=0.5\hsize]{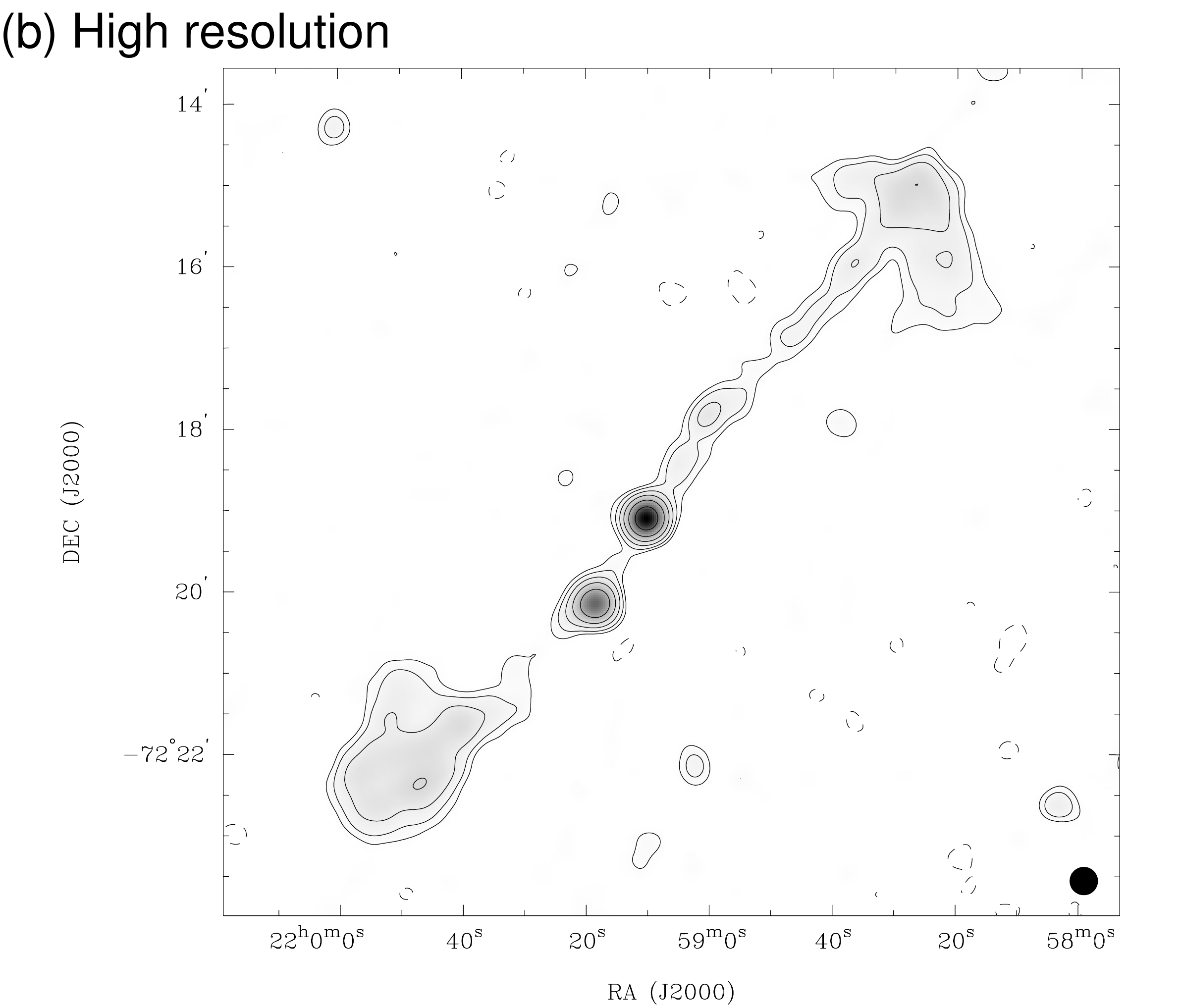} \\
      \includegraphics[width=0.5\hsize]{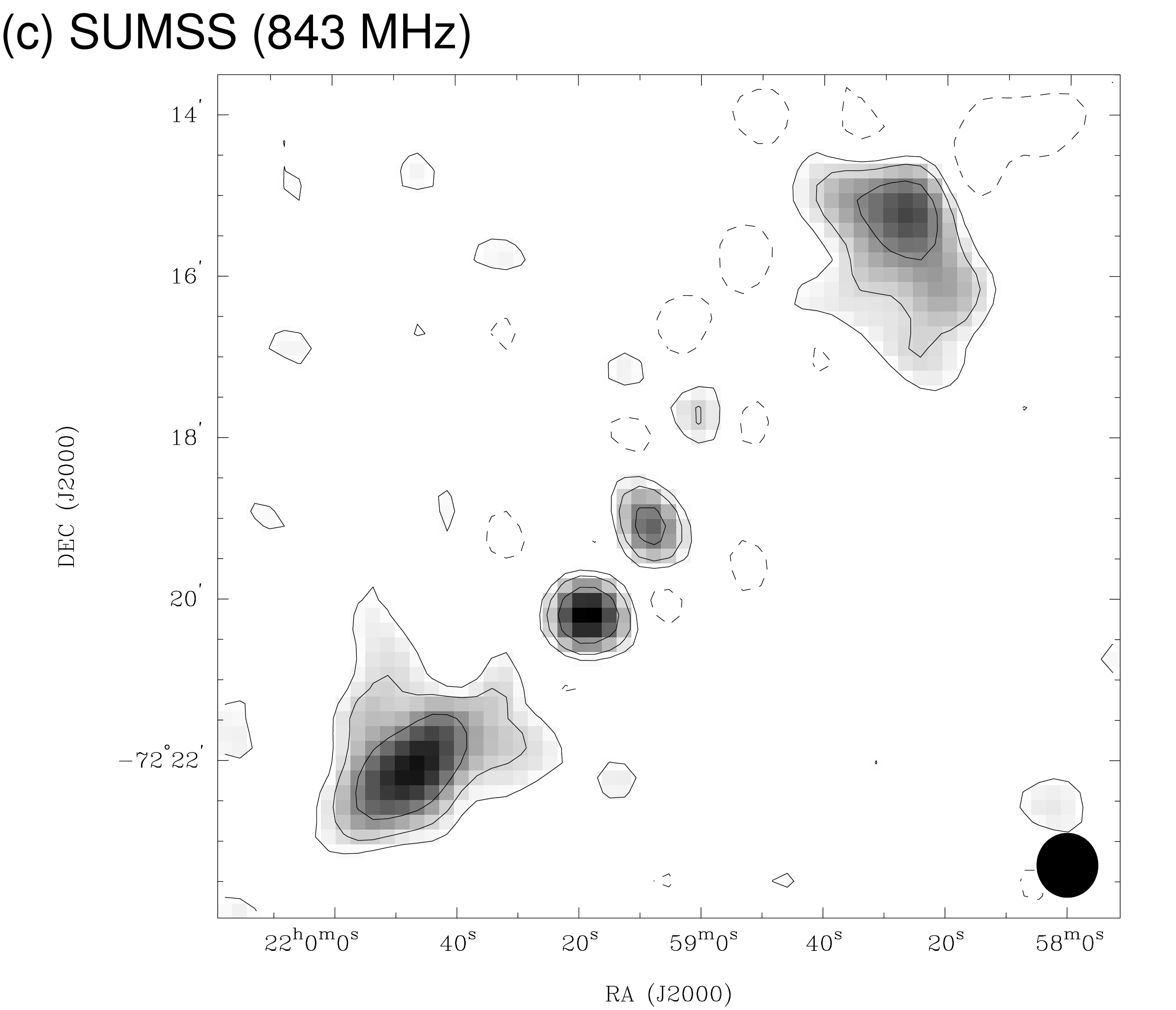} &
      \includegraphics[width=0.5\hsize]{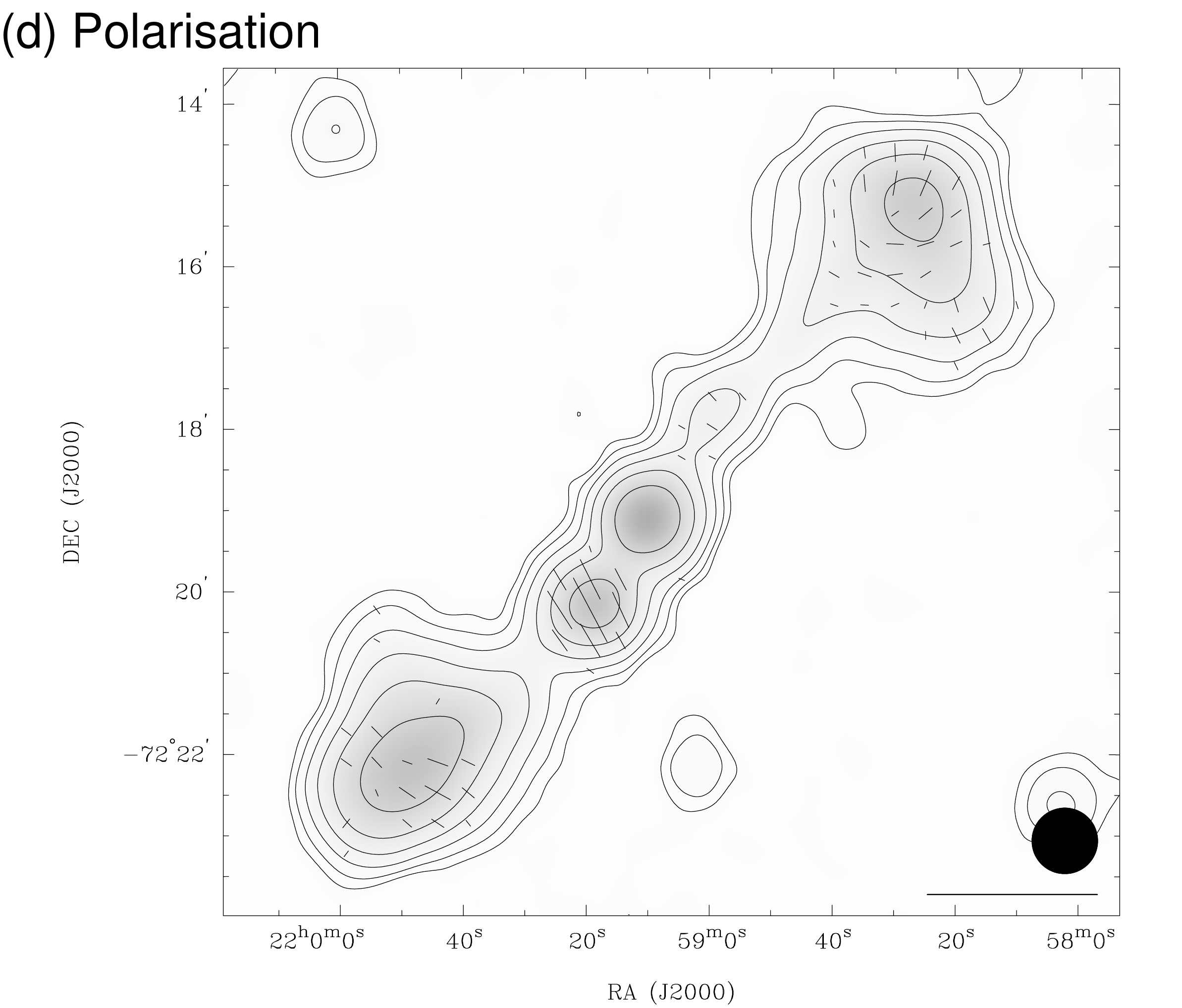}
\end{tabular}
\caption{(\subfigletter{a}) Low-resolution, wideband, total intensity image of J2159--7219 at 2.1 GHz with contours at -3, 3, 6, 12, 24, 48, and 96 $\times$ 75 $\mu$Jy beam$^{-1}$ and a beam of FWHM 48\arcsec. (\subfigletter{b}) High-resolution, total intensity ATCA image of J2159--7219 at 2.1 GHz with a beam of FWHM 20\arcsec~and contours at -3, 3, 6, 12, 24, 48, and 96 $\times$ 90 $\mu$Jy beam$^{-1}$. (\subfigletter{c}) The 843 MHz SUMSS image of J2159--7219 included for comparison with a beam of FWHM 45\arcsec$\times$47\arcsec,  total intensity contours at -1, 1, 2, and 4 $\times$ 2.5 mJy beam$^{-1}$ and greyscale in the range 2.5--19.0 mJy beam$^{-1}$. (\subfigletter{d}) Distribution of polarised intensity at 2.8 GHz represented by the lengths of the overlaid Faraday rotation-corrected, electric field vectors. The scale-bar represents 5 mJy.}
\label{fig:J2159}
\end{figure*}

\begin{figure*}
  \centering      
  \begin{tabular}{cc}
      \includegraphics[width=0.5\hsize]{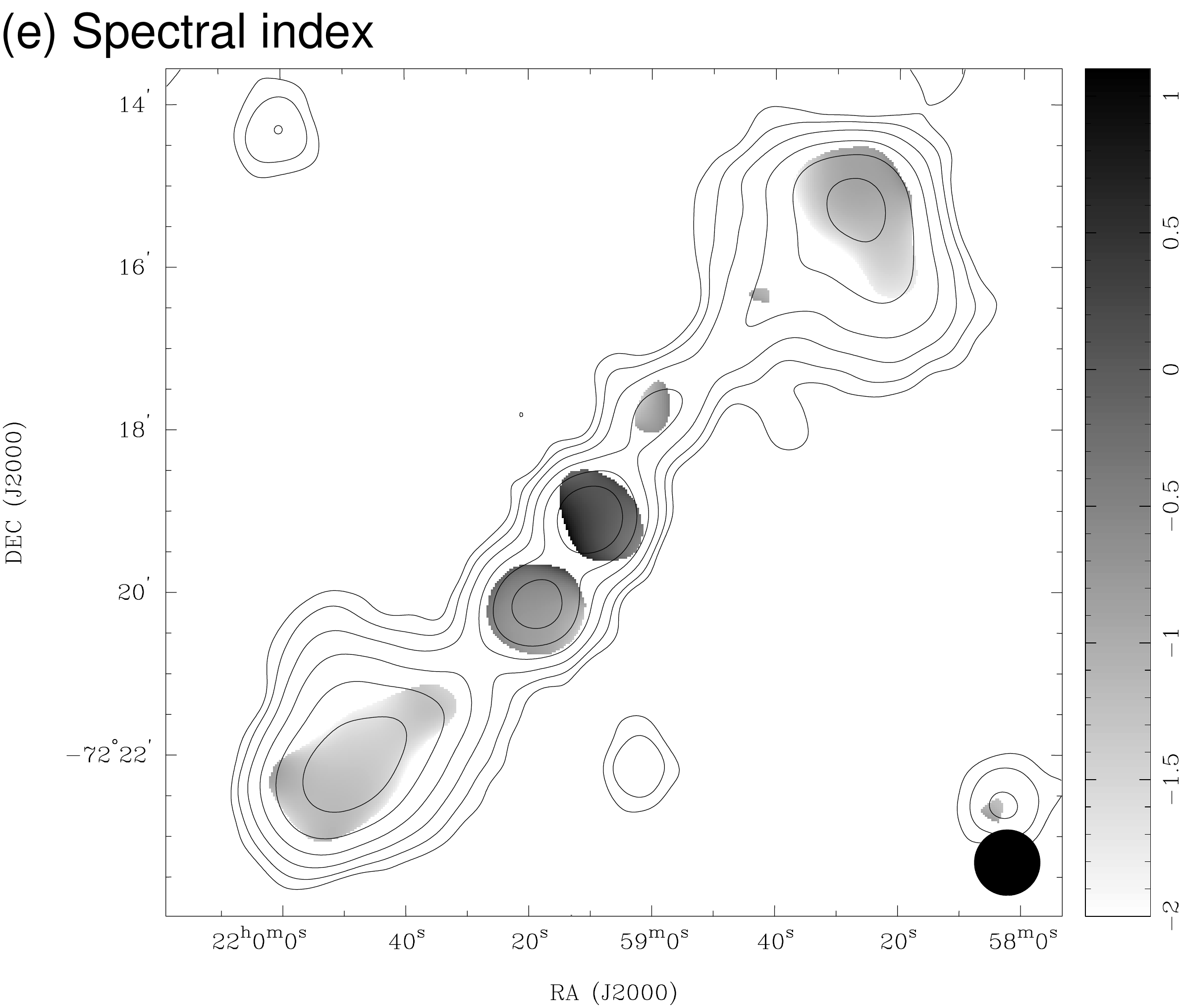} &
      \includegraphics[width=0.5\hsize]{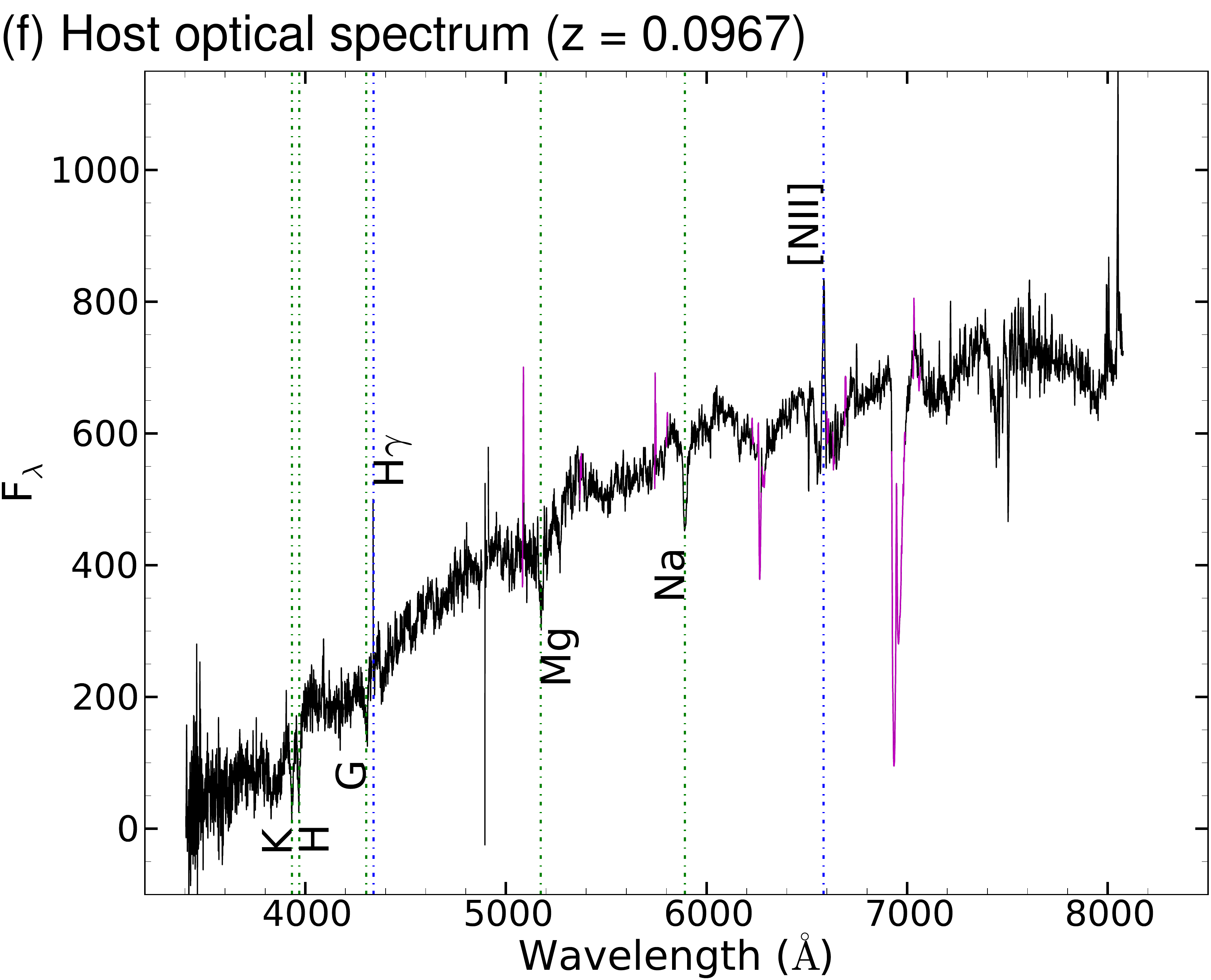}
  \end{tabular}
  \contcaption{(\subfigletter{e}) Distribution of spectral index between 843 MHz and 2.8 GHz shown as greyscale in the range $-2$ to $1.1$ with a beam of FWHM 48\arcsec~and total intensity contours at -3, 3, 6, 12, 24, 48, and 96 $\times$ 75 $\mu$Jy beam$^{-1}$. (\subfigletter{f}) An optical spectrum from AAOmega on the AAT of the host galaxy of J2159--7219. The spectrum has been shifted to rest frame using the measured redshift of 0.0967.}
\label{fig:J2159spin}
\end{figure*}

\begin{figure*}
  \centering
  \begin{tabular}{cc}
      \includegraphics[width=0.5\hsize]{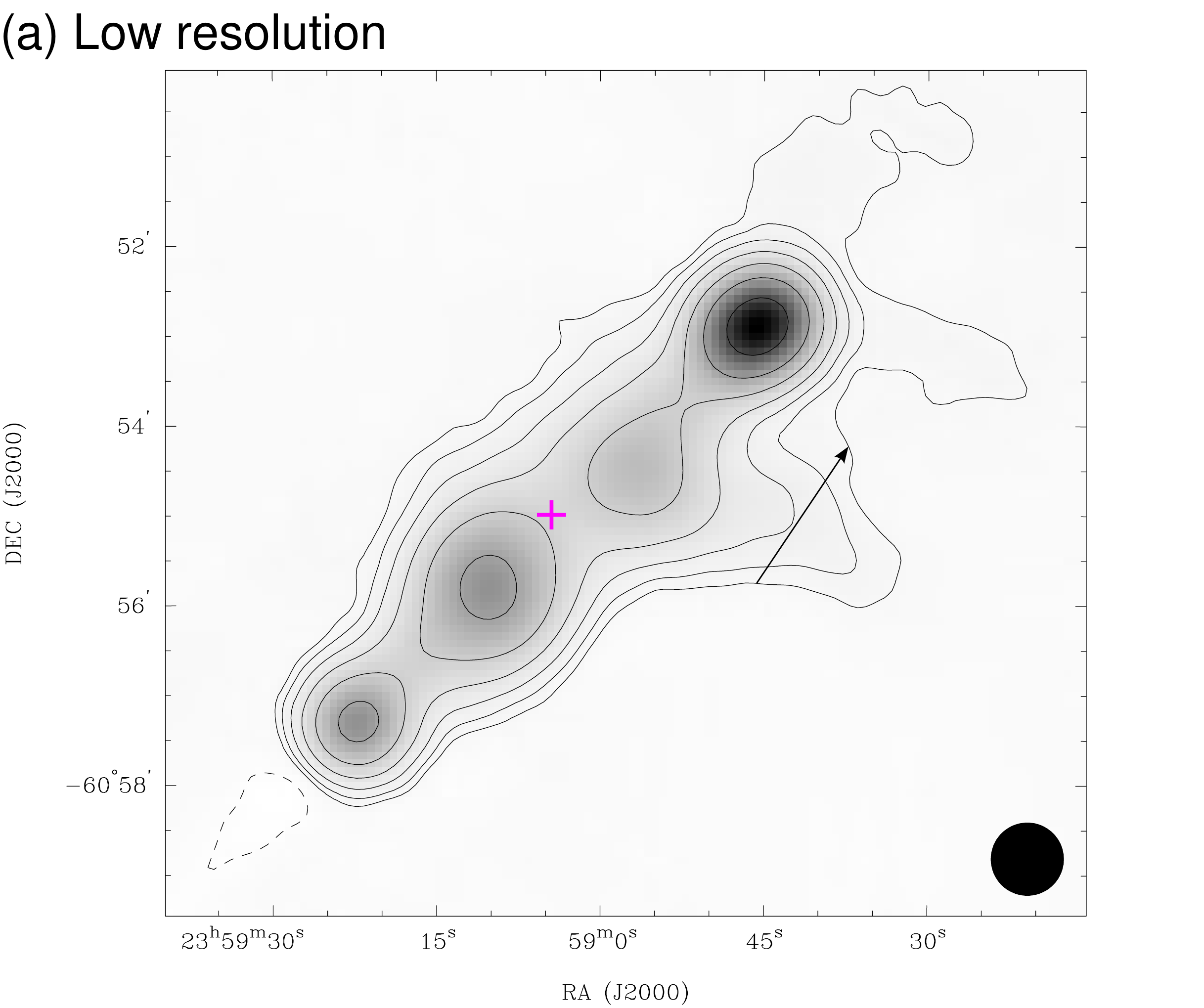} &
      \includegraphics[width=0.5\hsize]{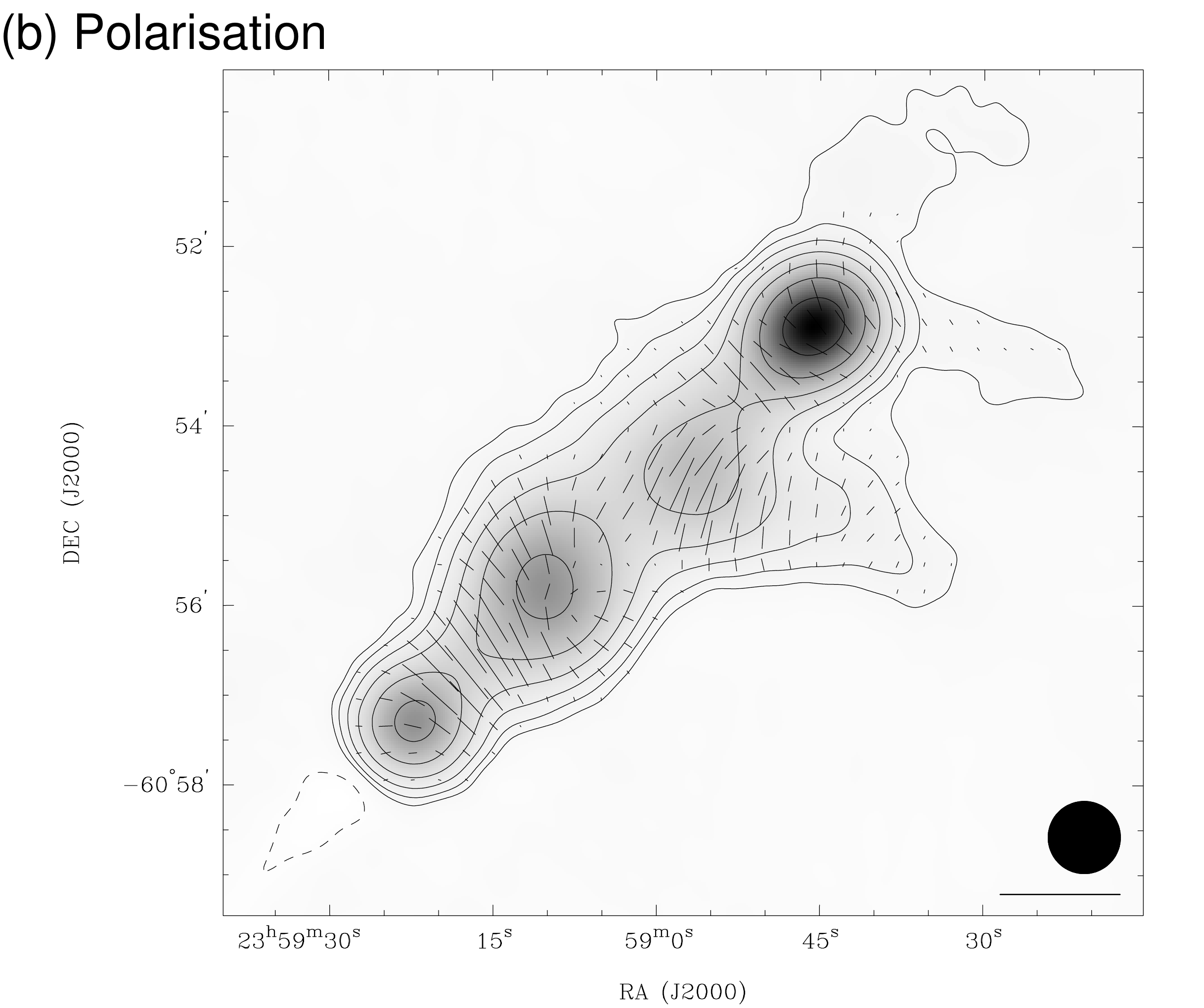} \\
      \includegraphics[width=0.5\hsize]{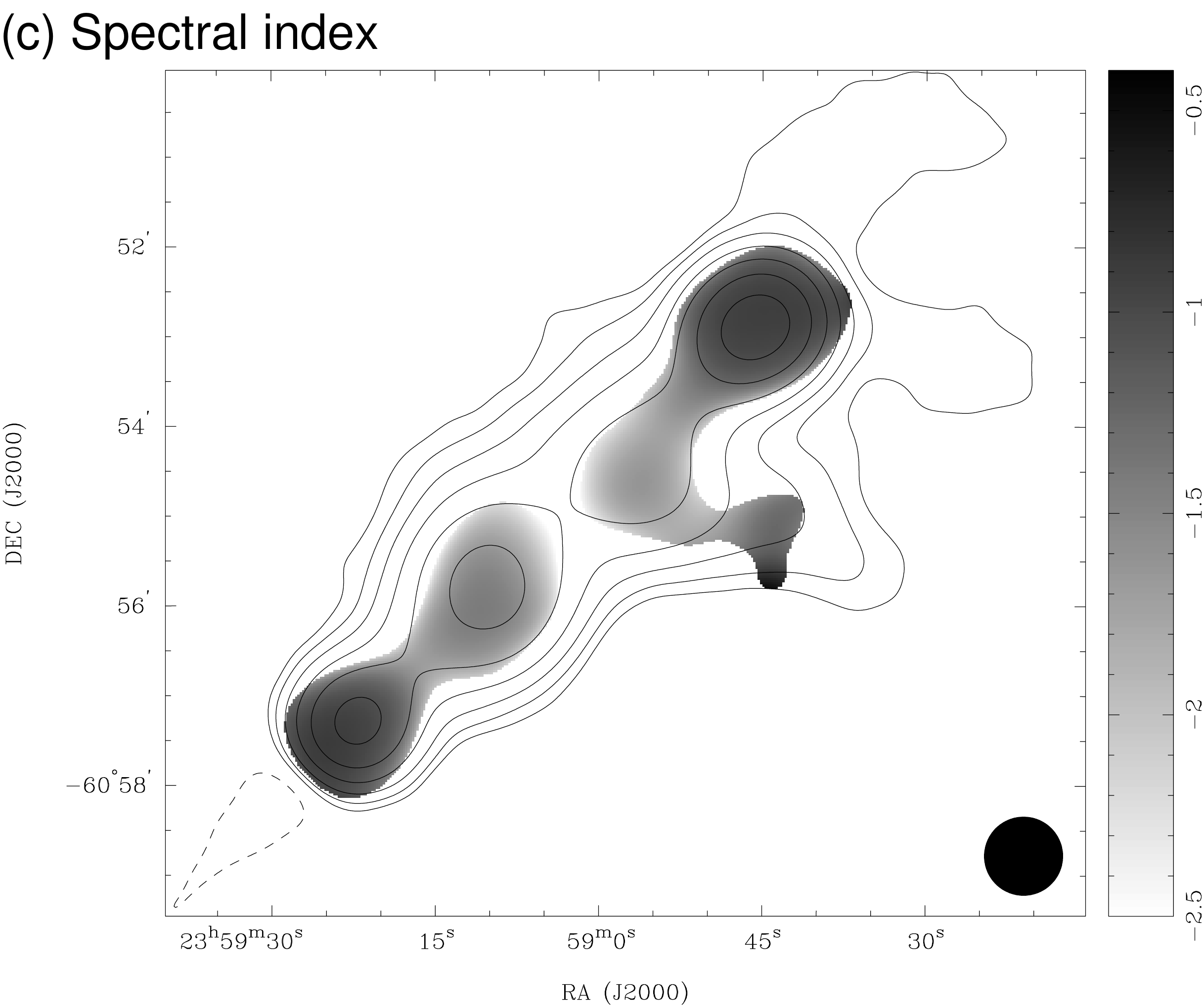} &
      \includegraphics[width=0.5\hsize]{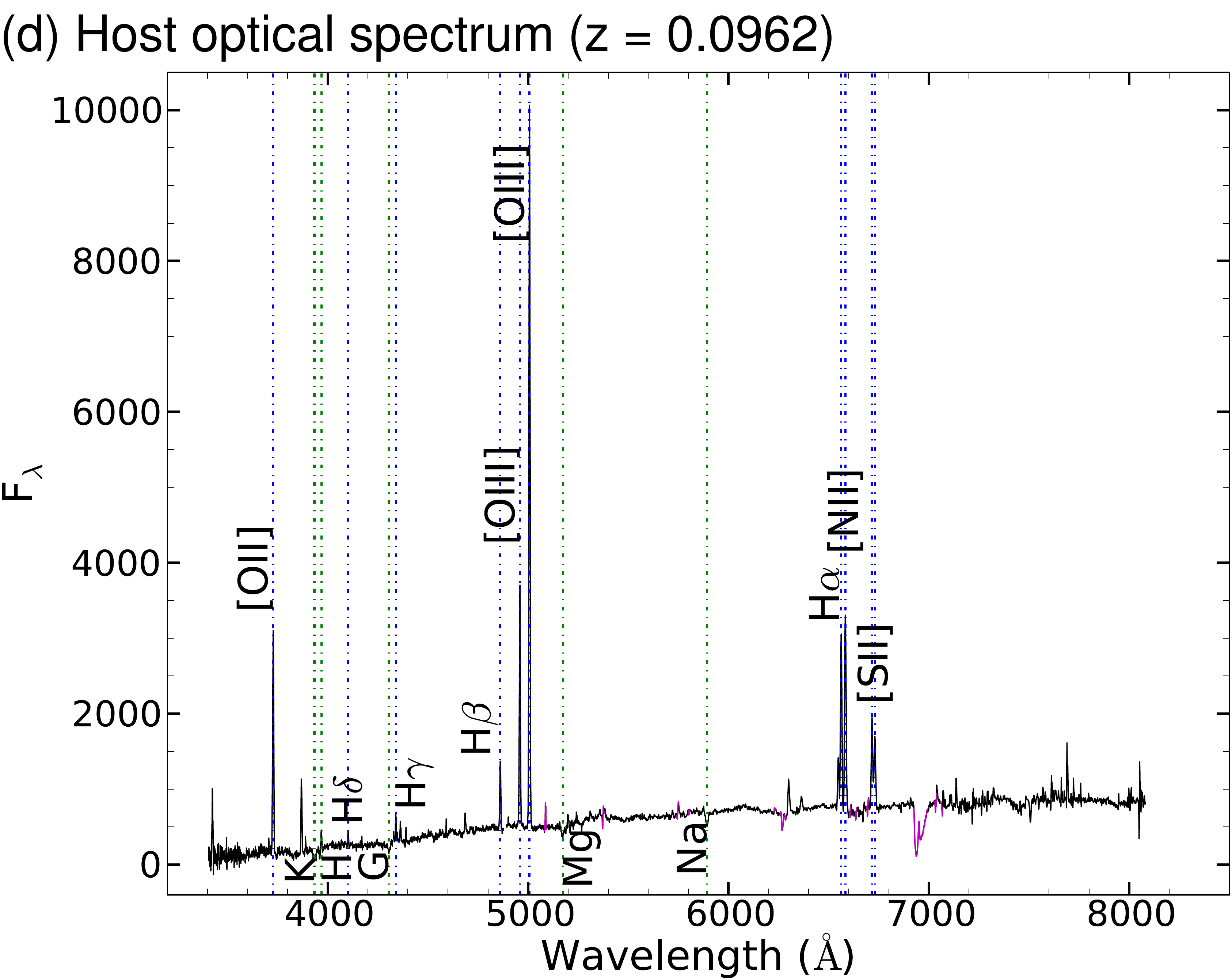}
\end{tabular}
\caption{(\subfigletter{a}) Low-resolution, wideband, total intensity image of B2356--611 at 2.1 GHz with contours at -3, 3, 6, 12, 24, 48, 96, and 192 $\times$ 15 mJy beam$^{-1}$ and a beam of FWHM 48\arcsec. (\subfigletter{b}) Distribution of polarised intensity at 2.8 GHz represented by the lengths of the overlaid Faraday rotation-corrected, electric field vectors. The scale-bar represents 0.25 Jy. (\subfigletter{c}) Distribution of spectral index between 843 MHz and 2.8 GHz shown as greyscale in the range $-2.5$ to $-0.4$ with a beam of FWHM 52\arcsec~and total intensity contours at -3, 3, 6, 12, 24, 48, 96, and 192 $\times$ 15 mJy beam$^{-1}$. (\subfigletter{d}) An optical spectrum from AAOmega on the AAT of the host galaxy of B2356--611. The spectrum has been shifted to rest frame using the measured redshift of 0.0962.}
\label{fig:B2356}
\end{figure*}

\subsection{B2356--611 (Fig.~\ref{fig:B2356})}
Our ATCA image of this bright giant radio galaxy ($z=0.0962$) shows the emission wing to the west more prominently than seen previously~\citep{Subrahmanyanetal1996,KoekemoerBicknell1998}. Our observations do not detect any counterpart for this feature (Fig. \ref{fig:B2356}\subfigletter{a}). We have detected highly polarised emission across the entire source as presented in Fig.~\ref{fig:B2356}\subfigletter{b} and find that the projected magnetic field is along the wing.

The optical spectrum of the host in Fig.~\ref{fig:B2356}\subfigletter{d} is unusual, in that it is characterised by numerous prominent emission lines, (including H$\alpha$, [NII], [SII], as well as the forbidden oxygen lines [OII] and [OIII]). The absence of any noticeable line-broadening suggests a spectrum more typical of a Type II Seyfert galaxy.

\begin{figure*}
  \centering
  \begin{tabular}{cc}
      \includegraphics[width=0.5\hsize]{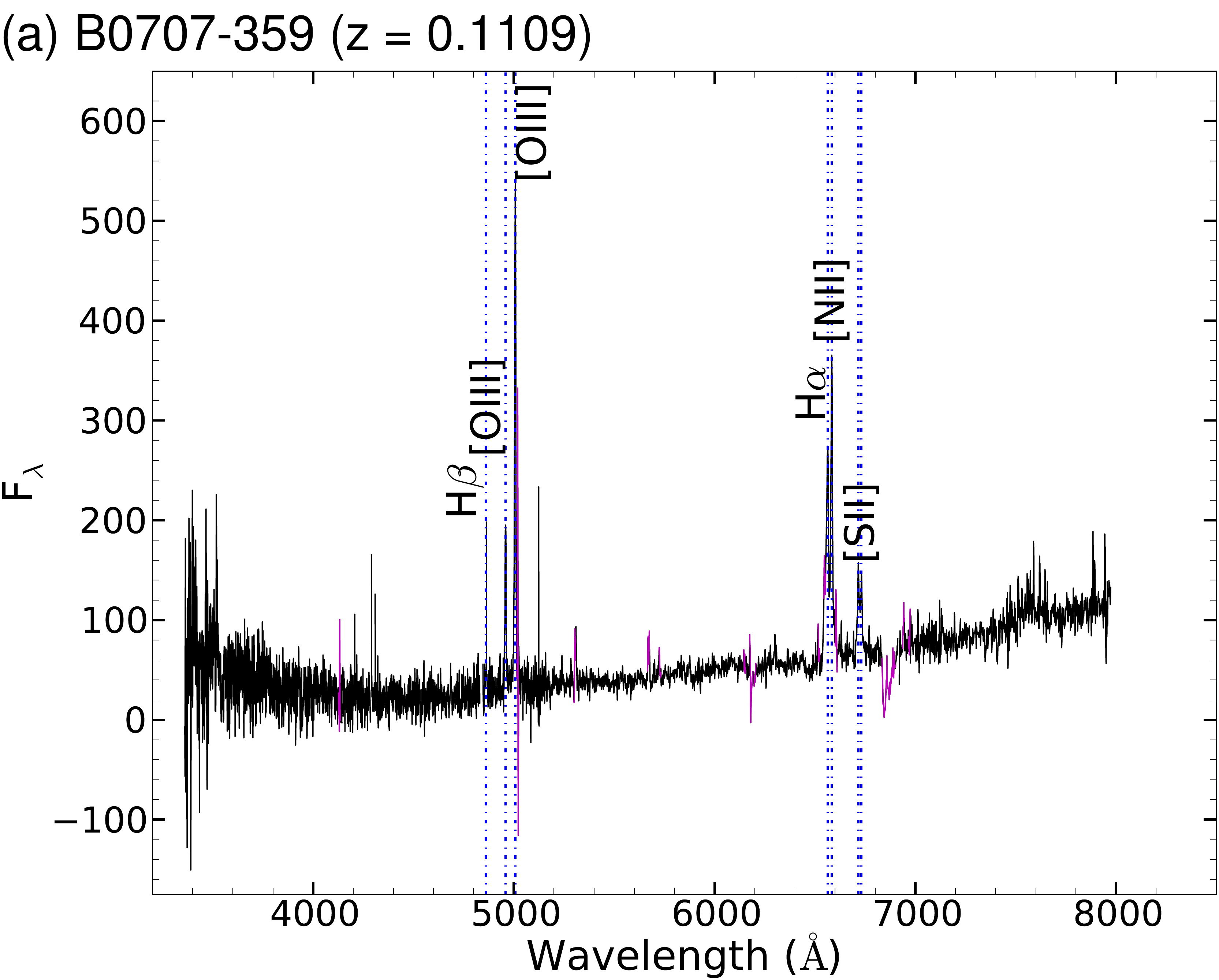} &
      \includegraphics[width=0.5\hsize]{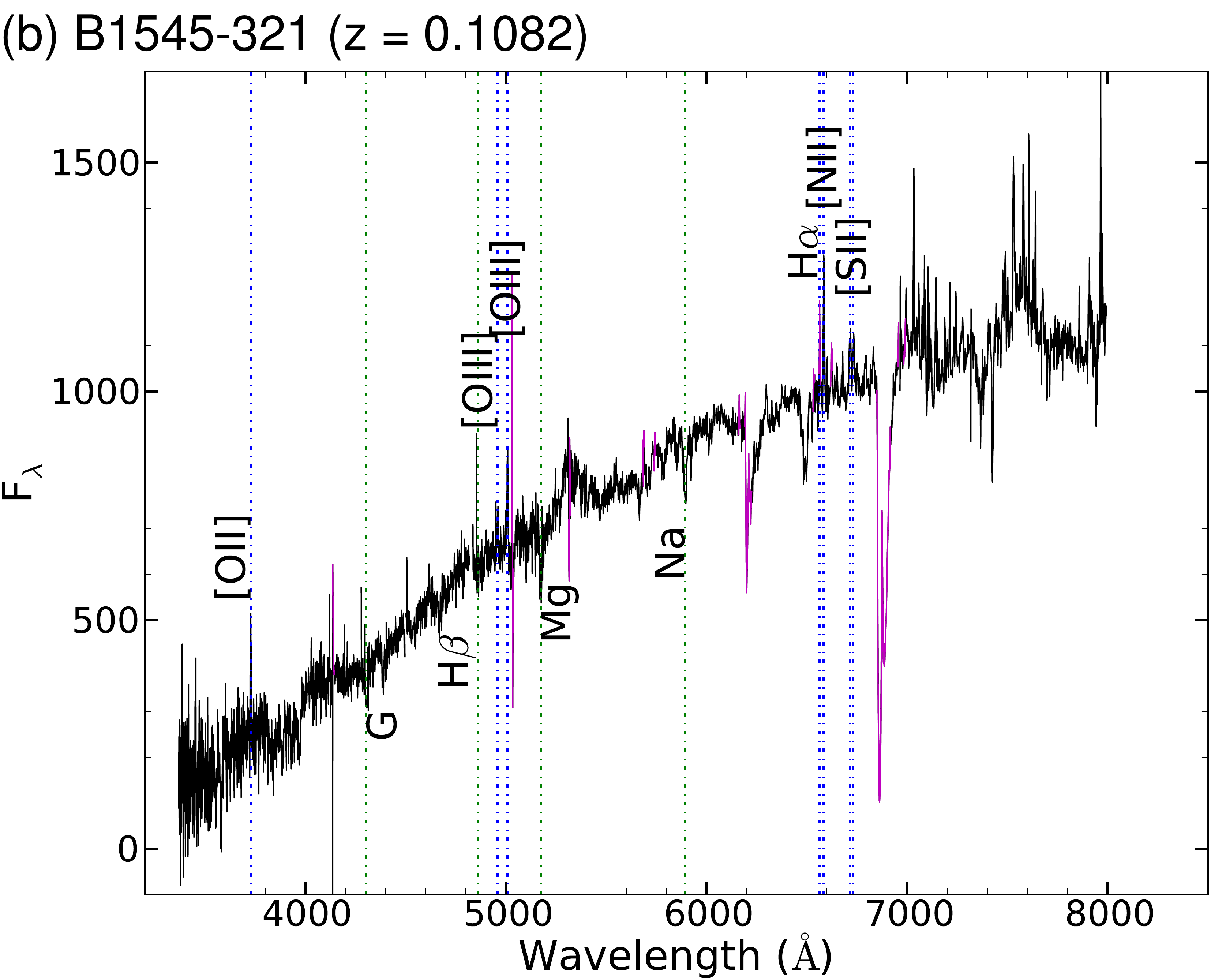} \\
      \includegraphics[width=0.5\hsize]{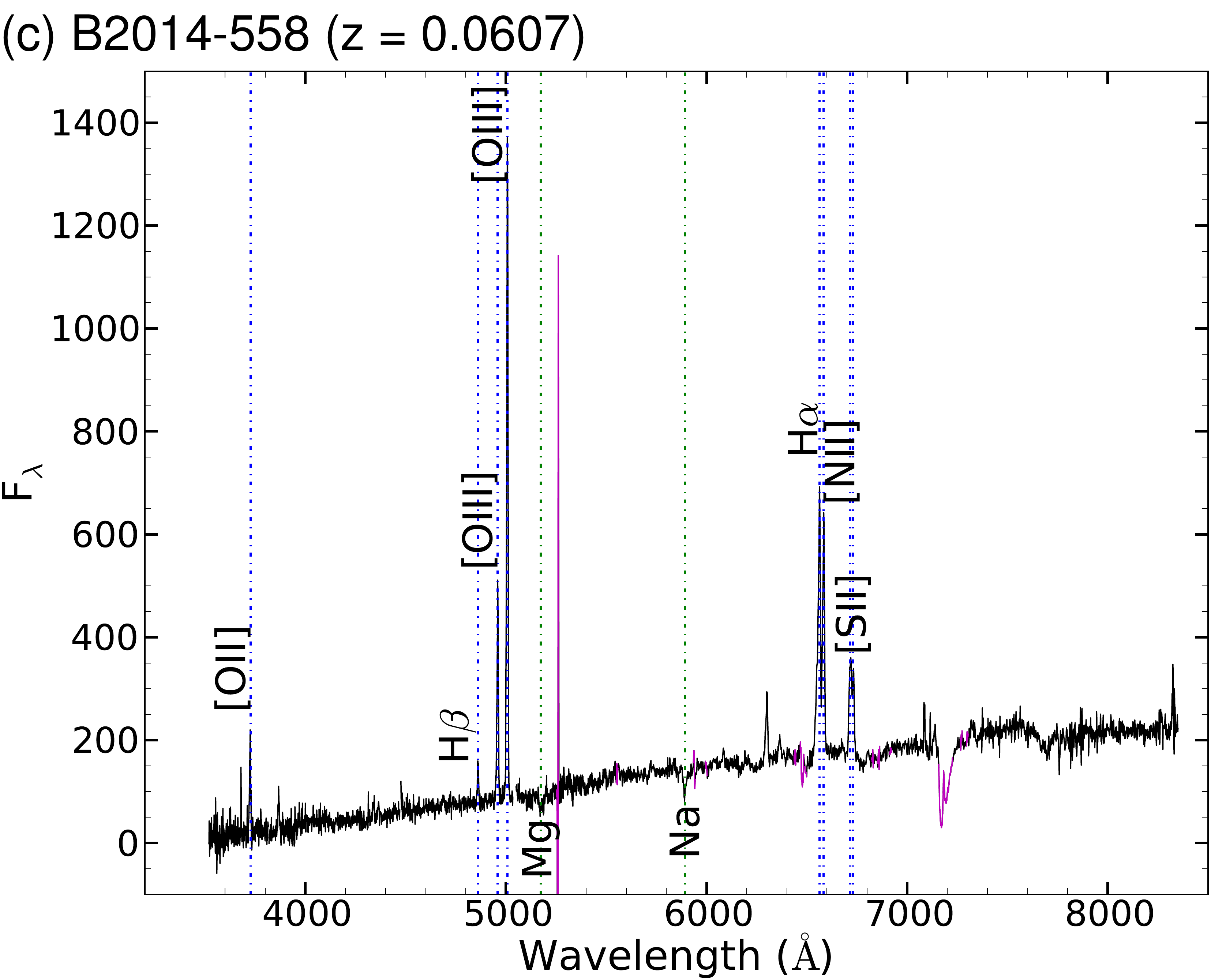}
\end{tabular}
\caption{Rest frame optical spectra of the remaining GRG hosts observed with the AAT using the AAOmega. Spectral regions traced in magenta indicate non-zero sky-subtraction residuals from telluric emission and absorption features. Common galaxy absorption (dotted green) and emission-like features (dotted blue) are shown by the vertical lines.}
\label{fig:optical}
\end{figure*}

\subsection{Additional optical observations (Fig.~\ref{fig:optical})}
We present the host optical spectra for the GRGs B0707--359, B1545--321 and B2014--558 observed with the AAOmega spectrograph on the AAT in Fig.~\ref{fig:optical}. The optical spectrum of the host galaxy of B0707--359 ($z=0.1109$) shown in Fig. \ref{fig:optical}\subfigletter{a} has prominent emission lines (including H$\alpha$, H$\beta$, [NII], [SII] and [OIII]) suggesting a Type II Seyfert galaxy. The optical spectrum of the host of B1545--321 ($z=0.1082$) in Fig. \ref{fig:optical}\subfigletter{b} shows both emission lines as well as the absorption line features typical of an ordinary early-type galaxy. The prominence of the H$\alpha$ lines suggests that the line emission is due to small amounts of star formation rather than nuclear activity. The optical spectrum of the host of B2014--558 ($z=0.0607$) shown in Fig. \ref{fig:optical}\subfigletter{c}, like that of B2356--611, is characterised by numerous prominent emission lines (including H$\alpha$, [NII], [SII], [OII] and [OIII]). The strong presence of the oxygen lines (in particular [OIII]) and the negligible line-broadening suggest a Type II Seyfert galaxy.

\section{Lobe pressures as a probe of the WHIM}
\label{sec:LobePressures}
Derivation of lobe pressures has been fraught with uncertainty. Assumption of a state of minimum energy for the radiating synchrotron plasma, where there is rough equipartition between the relativistic electrons and magnetic field, is commonly used to arrive at values for the lobe magnetic field and particle pressures. However, the very assumption of such a minimum energy condition has been a primary cause for skepticism in accepting the derived lobe physical parameters. In addition, accounting for the source emissivity, assumptions are made for the ratio of non-radiating to radiating particles (often assumed zero) and for the slope, shape and energy range of the electron energy distribution, which again lend uncertainty to the derived pressures.

X-ray observations of hotspots and lobes of radio galaxies have offered valuable insights into this subject. X-ray emission from radio lobes, attributed to the Inverse Compton (IC) up-scattering of cosmic microwave background photons by the lobe relativistic electrons, has provided an independent means of estimating the lobe energy densities.

However, as~\citet{ColafrancescoMarchegiani2011} point out, if there exists a low energy electron population with Lorentz factors $\gamma$ well below $10^3$, it is the Sunyaev-Zel'dovich effect (SZE) rather than the IC X-ray observations that would reveal their presence. Hence, the open problem of whether particles and fields in radio galaxy lobes are close to or depart significantly from the equipartition condition is difficult to answer based solely on comparison of energy densities estimated from radio emissivity with those using IC X-ray observations.

In the giant radio galaxy 6C 0905+3955,~\citet{Blundelletal2006} find strong evidence for a lower limit to the Lorentz factor $\gamma_{min}$ of the lobe electrons. This is based on the lack of X-ray emission from a hotspot together with strong IC X-ray emission from the corresponding lobe that is nearly 80 kpc downstream. They argue for a relatively substantial deficit or near absence of low energy electrons in the lobes of this GRG below $\gamma_{min}$ of $10^3$. If equipartition conditions are assumed for the lobes of this GRG, in which there is the aforementioned evidence of a cutoff in the electron spectrum at $\gamma_{min}$$\sim$$10^3$, then it turns out that there is a close agreement between the equipartition particle pressure derived from the emissivity and the independently derived value based on IC X-rays.

If this were also the case for other radio galaxies~\citep{Brunettietal2002,HarrisKrawczynski2002,Crostonetal2005} -- that there is a minimum to the Lorentz factor of lobe electrons -- then there is closer agreement with equipartition values compared to what was reported by these authors (who adopted low Lorentz factors $\gamma_{min}$ of a few tens).

It is not yet clear whether equipartition conditions are universal in radio galaxy lobes. We may only state that if it holds, and if low $\gamma$ electrons form an insignificant fraction, then it appears that the independent particle pressures obtained via IC X-ray observations (that suggest the absence of low Lorentz $\gamma$ electrons at least for one giant radio galaxy) are in close agreement with those obtained via the assumption of equipartition.

Our sample consists of giant radio galaxies. We do not have IC-X-ray observations available for our sources. Nevertheless we are encouraged (1) by the lower limit to the lobe electron Lorentz factors inferred for the case of the GRG 6C 0905+3955, (2) by the closeness of the independently estimated particle energy densities with the equipartition values, and (3) by the argument that since GRGs are amongst the largest of radio galaxies there is reason to assume that the GRGs are also older as compared to smaller sized radio galaxies and hence particles and fields in their lobes are more likely to have reached equilibrium.

For these reasons, we compute lobe pressures for our sample of GRGs by assuming equipartition conditions and using a $\gamma_{min}$ for electrons of $10^3$ (\mbox{$E_{min}=511$ MeV}). The procedure we adopt is as follows: we make slice profiles across the most relaxed components of the lobes (avoiding the jets and other peaks in total intensity) to find a peak flux density in Jy beam$^{-1}$ and a deconvolved width in arcsec for each lobe at the reference frequency $\nu^*$ in GHz~\citep[equation 1,][]{Parmaetal2007}. It is assumed that each lobe is cylindrically symmetric. Hence, the path length through the source in the line of sight is equal to its width in the plane of the sky in kpc, which is determined from its deconvolved width at the redshift z of the host galaxy. We then take the volume V of the emission region in kpc$^3$ to be a cylinder orientated through the source along the line of sight with a cross-section equal to the beam. The synchrotron emission may be attributed entirely to light relativistic particles and so we assume that the energy density ratio of all particles to relativistic particles $(1 + k)$ is unity~\citep{Subrahmanyanetal2008}. While there may be non-radiating population of particles within the lobes because of entrainment, their significance has been argued to be small considering the close agreement found between independently derived particle energy density and equipartition value where equipartition is assumed between the fields and radiating particles~\citep[see e.g.][]{Crostonetal2005}. A typical spectral index of $-1$ ($S_{\nu} \propto \nu^{\alpha}$) is assumed in the diffuse lobe regions over the electron energy range. The monochromatic luminosity $L(\nu$*$)$ in W Hz$^{-1}$ is determined from the peak flux of the lobe, and $F(\alpha)$, which is a quantity in the range 10$^{-39}$ to 10$^{-38}$ for $0.5 \leq \alpha \leq 1.0$, is taken to be 10$^{-38.5}$. The lobe pressure is computed as one third of the energy density given a relativistic gas.

There were five sources for which we did not have access to reduced data and so could not make slice profiles directly. However, slice profiles are available for J0116--473 and B0503--286~\citep[see][]{Subrahmanyanetal2008}. In the remaining three cases, we fitted slice profiles to existing maps (B0319--454:~\citet{Safourisetal2009}; J0515--800:~\citet{Subrahmanyanetal2006}; B1545--321:~\citet{Safourisetal2008}).

Table~\ref{tbl:FluxPressure} lists properties of the complete GRG sample including pressures and integrated flux densities. The integrated fluxes were measured for each lobe separately from the low resolution images, summing over multiple components where present. The flux density values in the literature have been converted to 2.1 GHz using an integrated spectral index of $\alpha=-0.8$ ($S_{\nu} \propto \nu^{\alpha}$), except J0116--473 for which the quoted value of $-0.93$ was used~\citep{Saripallietal2002}. The corresponding total power values are expressed in the rest frame of the respective source. Given the varied jet-axis position angles of the GRGs, we assign numbers to the lobes arbitrarily in the table and for each source give cardinal directions on the sky in columns 3 and 4 for lobes 1 and 2, respectively. The locations of the slice profiles used in determining the lobe pressures are overlaid as solid arrows on the low resolution radio maps in Fig.~\ref{fig:J0034}--\ref{fig:B2356}. In particular cases, we used other radio components in preference to more overpressured lobes for this analysis, including the extended emission regions of J0331--7710 and B0511--305, B0703--451's faint eastern lobe, and B2356--611's emission wing to the west.

\begin{table*}
 \caption{Derived properties of the complete GRG sample (2.1 GHz). The radio lobes are arbitrarily numbered but may be identified with the given cardinal directions. The flux density values in the literature have been converted to 2.1 GHz using an integrated spectral index of $\alpha=-0.8$ ($S_{\nu} \propto \nu^{\alpha}$), except J0116--473 for which the quoted value of $-0.93$ was used~\citep{Saripallietal2002}. Power values are expressed in the rest frame of the respective source. The low resolution, total intensity maps in Fig.~\ref{fig:J0034}--\ref{fig:B2356} are overlaid with solid arrows representing the locations of the slice profiles used to determine the lobe pressures.}\label{tbl:FluxPressure}
 \begin{tabular}{l@{\hskip 2.5mm}lccc@{\hskip 2.5mm}c@{\hskip 2.5mm}c@{\hskip 2.5mm}c@{\hskip 2.5mm}c@{\hskip 2.5mm}c@{\hskip 2.5mm}c}
  \hline
  Name              & Redshift      & \multicolumn{2}{c}{Cardinal direction}& \multicolumn{3}{c}{Integrated flux density}   & Power                        & \multicolumn{3}{c}{Pressure}                                       \\
                    &               & Lobe 1      & Lobe 2                  & Lobe 1 & Lobe 2 & Total                       &                              &  Lobe 1              & Lobe 2               & Mean                 \\  
                    &               &             &                         & (mJy)  & (mJy)  & (mJy)                       & (W Hz$^{-1}$)                & (Pa)                 & (Pa)                 & (Pa)                 \\
  \hline
  J0034--6639       & $0.1103$      & N           & S                       & $29.8$ & $39.1$ & $73.8$                      & $2.45\times10^{24}$          & $1.44\times10^{-15}$ & $1.91\times10^{-15}$ & $1.68\times10^{-15}$ \\
  J0116--473$^a$    & $0.146^{*a}$  & N           & S                       & -      & -      & $1.96\times10^{3}$          & $1.25\times10^{26}$          & $8.00\times10^{-15}$ & $8.82\times10^{-15}$ & $8.41\times10^{-15}$ \\
  B0319--454$^b$    & $0.0622^{*b}$ & SW          & -                       & -      & -      & $2.76\times10^{3}$          & $2.62\times10^{25}$          & $1.84\times10^{-15}$ & -                    & -                    \\
  J0331--7710       & $0.1446$      & N           & S                       & $183$  & $40.1$ & $223$                       & $1.36\times10^{25}$          & $1.11\times10^{-15}$ & $1.57\times10^{-15}$ & $1.34\times10^{-15}$ \\
  J0400--8456       & $0.1033$      & N           & S                       & $44.8$ & $46.2$ & $120$                       & $3.45\times10^{24}$          & $3.08\times10^{-15}$ & $2.43\times10^{-15}$ & $2.76\times10^{-15}$ \\
  J0459--528        & $0.0957$      & E           & W                       & $14.5$ & $18.3$ & $51.0$                      & $1.24\times10^{24}$          & $2.89\times10^{-15}$ & $3.58\times10^{-15}$ & $3.23\times10^{-15}$ \\
  B0503--286$^c$    & $0.0383^{*c}$ & N           & S                       & -      & -      & $2.28\times10^{3}$          & $7.79\times10^{24}$          & $4.74\times10^{-15}$ & $3.90\times10^{-15}$ & $4.32\times10^{-15}$ \\
  B0511--305        & $0.0576$      & S           & -                       & $1.39\times10^{3}$ & - & $1.97\times10^{3}$   & $1.60\times10^{25}$          & $3.30\times10^{-15}$ & -                    & -                    \\
  J0515--800$^{de}$ & $0.1052^{*e}$ & N           & S                       & $35.4$ & $87.7$ & $125$                       & $3.72\times10^{24}$          & $2.05\times10^{-15}$ & $2.60\times10^{-15}$ & $2.33\times10^{-15}$ \\
  B0703--451        & $0.1242$      & E           & -                       & $16.3$ & -      & $175$                       & $7.56\times10^{24}$          & $1.97\times10^{-15}$ & -                    & -                    \\
  B0707--359$^f$    & $0.1109$      & N           & S                       & -      & -      & $1.39\times10^{3}$          & $4.69\times10^{25}$          & $2.03\times10^{-14}$ & $1.79\times10^{-14}$ & $1.91\times10^{-14}$ \\
  J0746--5702       & $0.1301$      & E           & W                       & $26.5$ & $18.0$ & $123$                       & $5.94\times10^{24}$          & $3.21\times10^{-15}$ & $2.27\times10^{-15}$ & $2.74\times10^{-15}$ \\
  J0843--7007       & $0.1390$      & E           & W                       & $22.5$ & $13.9$ & $64.7$                      & $3.62\times10^{24}$          & $3.82\times10^{-15}$ & $3.30\times10^{-15}$ & $3.56\times10^{-15}$ \\
  B1302--325        & $0.1528$      & NE          & SW                      & $331$  & $697$  & $1.03\times10^{3}$          & $7.14\times10^{25}$          & $1.24\times10^{-14}$ & $1.32\times10^{-14}$ & $1.28\times10^{-14}$ \\
  B1308--441        & $0.0507$      & SE          & NW                      & $119$  & $43.5$ & $626$                       & $3.86\times10^{24}$          & $4.16\times10^{-15}$ & $3.48\times10^{-15}$ & $3.82\times10^{-15}$ \\
  B1545--321$^g$    & $0.1082$      & N           & S                       & -      & -      & $1.22\times10^{3}$$\dagger$ & $3.87\times10^{25}$$\dagger$ & $1.35\times10^{-14}$ & $1.68\times10^{-14}$ & $1.51\times10^{-14}$ \\
  B2014--558$^h$    & $0.0607$      & N           & S                       & -      & -      & $1.17\times10^{3}$          & $1.05\times10^{25}$          & $3.75\times10^{-15}$ & $3.80\times10^{-15}$ & $3.78\times10^{-15}$ \\
  J2159--7219       & $0.0967$      & NW          & SE                      & $31.7$ & $37.2$ & $99.2$                      & $2.46\times10^{24}$          & $3.31\times10^{-15}$ & $4.11\times10^{-15}$ & $3.71\times10^{-15}$ \\
  B2356--611        & $0.0962$      & W$\ddagger$ & -                       & $484$  & -      & $1.65\times10^{4}$          & $4.04\times10^{26}$          & $1.96\times10^{-14}$ & -                    & -                    \\
  \hline
 \end{tabular}

\medskip
Notes -- references for sources that were not part of the wideband ATCA subset: $^a$\citet{Saripallietal2002}, $^b$\citet{Safourisetal2009}, $^c$\citet{Subrahmanyanetal2008}, $^d$\citet{Saripallietal2005}, $^e$\citet{Subrahmanyanetal2006}, $^f$Malarecki et al. (in preparation), $^g$\citet{Safourisetal2008}, and $^h$\citet{Saripallietal2008}. *literature redshifts for sources that we have not observed with the AAT. $\dagger$B1545--321's \emph{outer} double radio source. $\ddagger$B2356--611's emission wing to the west.
\end{table*}

A number of studies have estimated electron density associated with potential detections of X-ray emission from hot gas in filaments connecting galaxy clusters~\citep{BrielandHenry1995, Werneretal2008, FraserMcKelvieetal2011}. We converted the baryon density reported by \citet{Werneretal2008} to electron density assuming gas particles in a primordial composition of 10 H to 1 He nuclei, where each ionised H and He atom contributed 1 and 2 electrons respectively. These X-ray emission studies are most sensitive to the denser component of the WHIM gas in galaxy filaments, particularly filaments aligned close to the line of sight~\citep[e.g.][]{Werneretal2008,Dietrichetal2012}. We computed pressures from these electron densities, which are presented in Table~\ref{tbl:xraystudies}, as estimates of the WHIM for comparison with the GRG lobe pressures.

We also estimate the particle number densities, assuming the same primordial composition, for each of the X-ray emission studies, which are presented in Table~\ref{tbl:xraystudies} as overdensities and later used in comparison with our GRG sample. Particle overdensity is defined as $n/\bar{n}$, where $n$ is the estimated particle number density and $\bar{n}$ is the mean particle number density in the IGM. We take the baryon density in the Universe to be $\Omega_b=0.0455$~\citep{Larsonetal2011} and assume that half of the baryons are in the gas phase at low redshifts giving a mean IGM gas density of $2.15\times10^{-28}\mbox{ kg m}^{-3}$. For a fully ionised gas of primordial composition, this corresponds to a mean particle number density of $\bar{n} = 0.21\mbox{ m}^{-3}$.

\begin{table}
 \caption{Estimates of WHIM properties from X-ray observations of galaxy filaments.}\label{tbl:xraystudies}
 \begin{tabular}{@{}l@{}c@{}c@{ }c@{ }c@{}}
  \hline
                                                           & $n_{e}$     & Temperature         & Pressure                        & Particle \\
  Reference                                      & (m$^{-3}$)& ($\times10^7$ K)  & ($\times10^{-15}$ Pa) & overdensity \\
  \hline
  \citet{BrielandHenry1995}          & 62.4           & $1.04$                    & $<$$8.99$                     & 565 \\
  \citet{Werneretal2008}                & 37.1           & $1.06$                    & $5.41\pm3.55$              & 336 \\
  \citet{FraserMcKelvieetal2011} & 396            & $1.04$                    & $57.1\pm2.43$             & 3589 \\
  \citet{Dietrichetal2012}*              & 38.1           & $1.06$                    & $5.55\pm3.65$             & 345 \\
  \hline
 \end{tabular}
 
\medskip
Notes -- Particle overdensity is defined as $n/\bar{n}$, where $n$ is the estimated particle density and $\bar{n} = 0.21\mbox{ m}^{-3}$ is the mean particle number density in the IGM. * from the work of \citet{Werneretal2008}.
\end{table}

\begin{figure}
  \centering
  \includegraphics[width=\hsize]{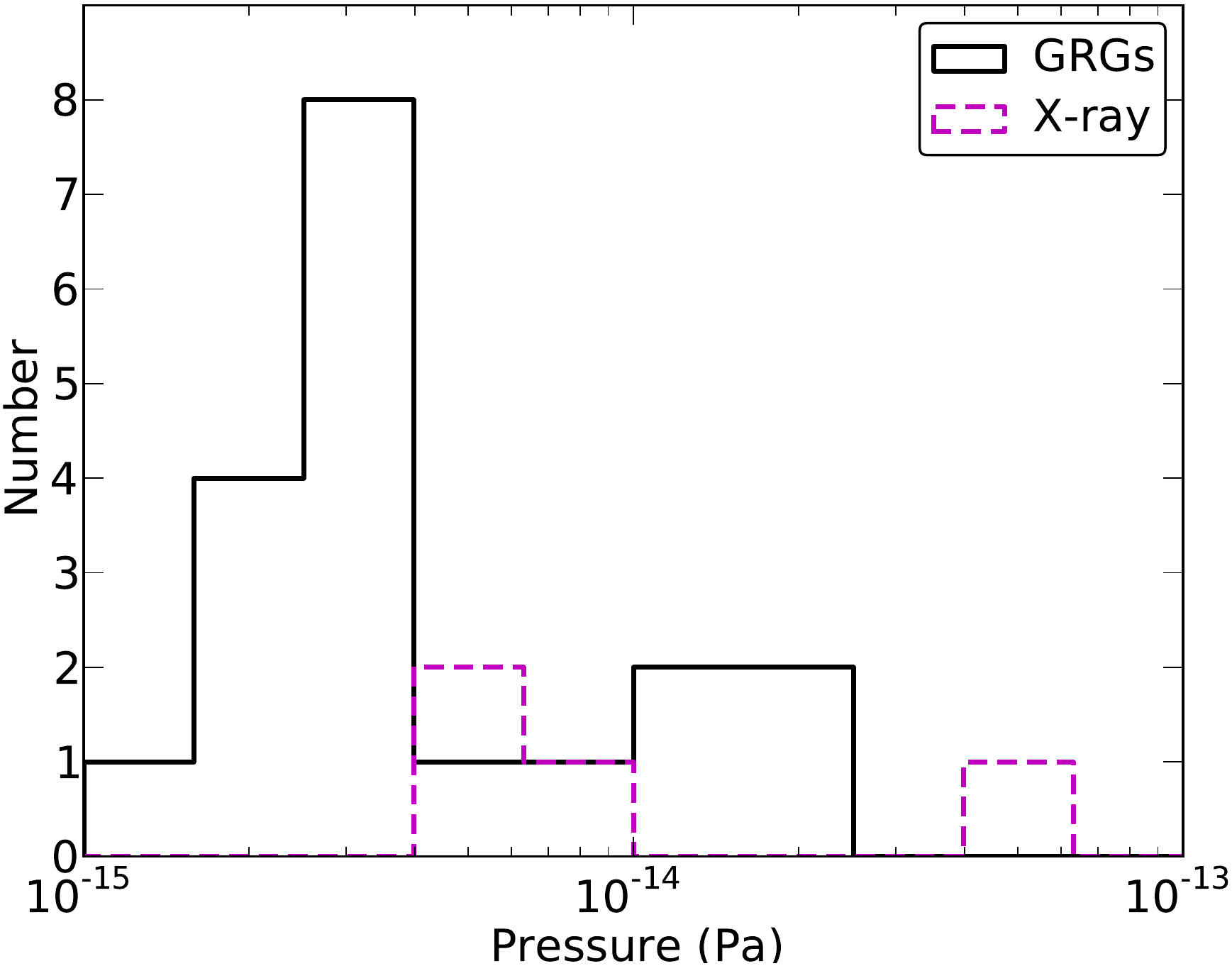}
  \caption{Complete GRG sample averaged radio lobe pressures (black bins) overlaid with pressure estimates from X-ray studies (magenta dashed bins; see also Table~\ref{tbl:xraystudies}).}
  \label{fig:pressurehistogram}
\end{figure}

In Figure~\ref{fig:pressurehistogram} we plot a histogram showing the distribution of derived GRG lobe pressures overlaid with the set of X-ray observations of filaments listed in Table~\ref{tbl:xraystudies}. We compare these samples using the nonparametric Kolmogorov-Smirnov test and find that these samples do not come from a common distribution at a 95 per cent level of confidence (p-value = 0.03). This result suggests that the radio lobes may be probing pressures slightly less than inferred by X-ray observations, such that they may be in balance with a lower-pressure component of the WHIM gas. In contrast, more typical kpc-scale FRII sources (at redshifts $z \leq0.15$) do not sample the WHIM, and these sources have pressures an order of magnitude greater than our GRG sample for the same values of $\alpha$ and $k$~\citep{HardcastleWorrall2000}.

\begin{figure*}
  \centering
  \begin{tabular}{cc}
  \includegraphics[width=0.5\hsize]{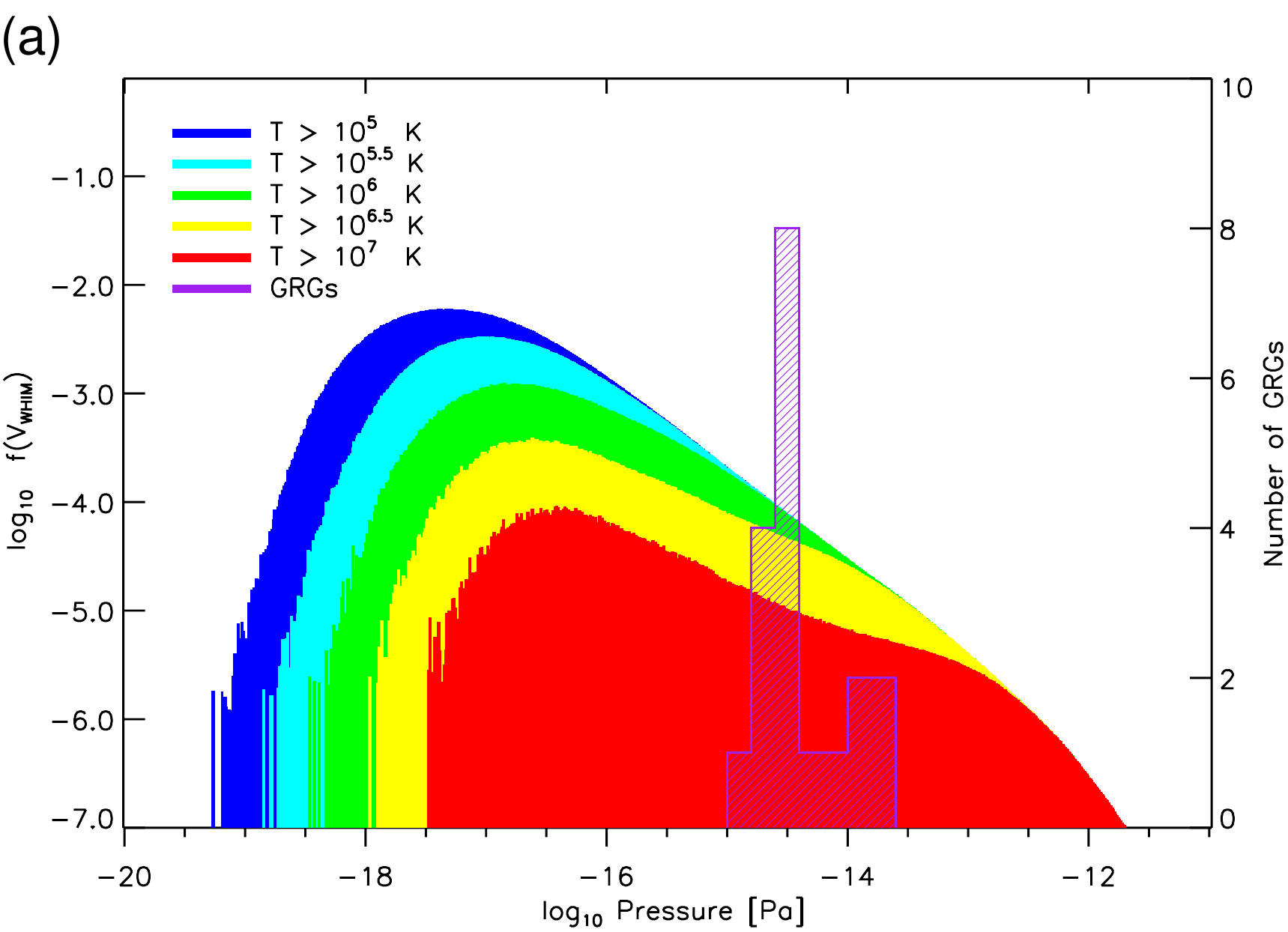} &
  \includegraphics[width=0.5\hsize]{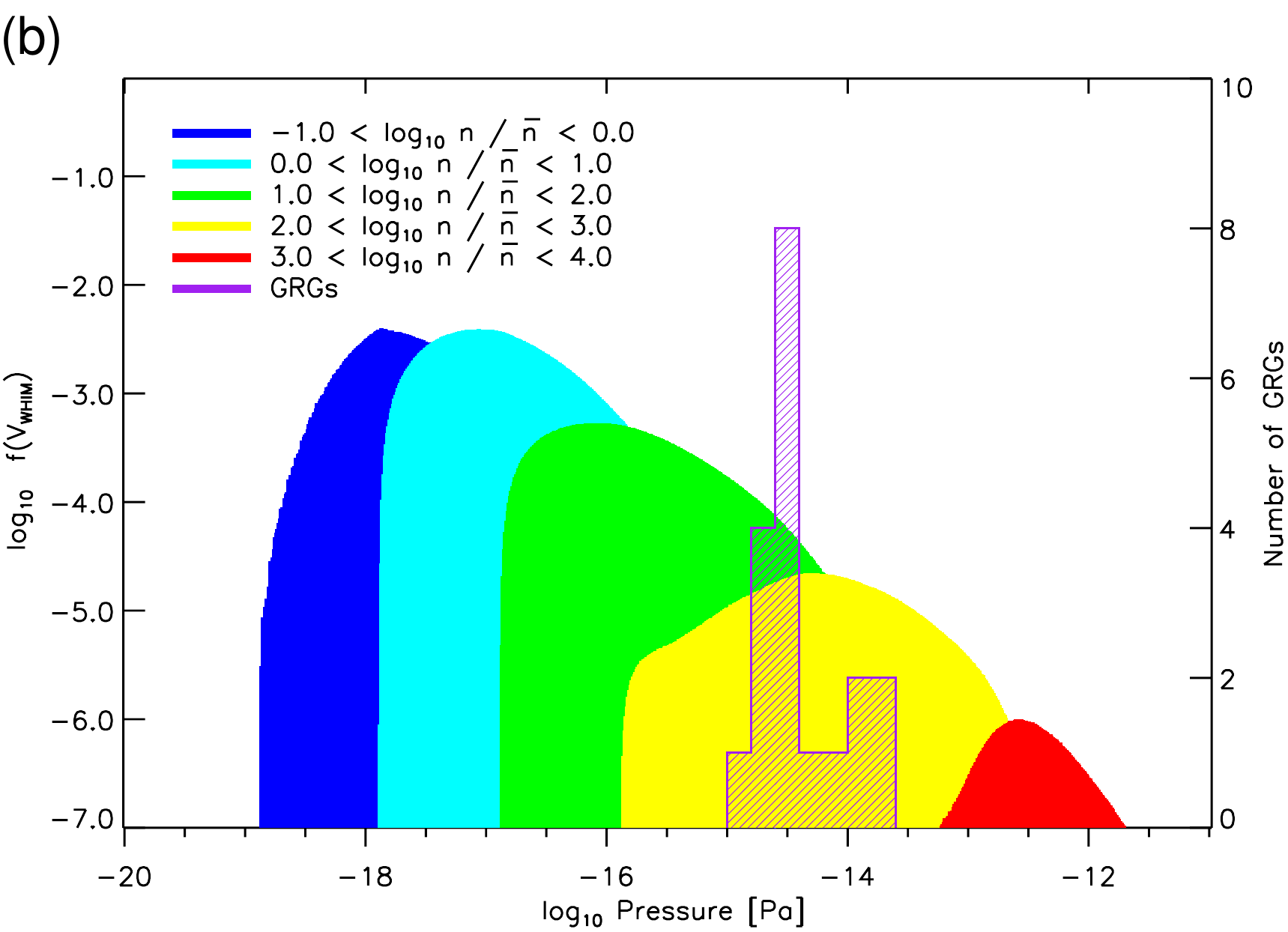}
  \end{tabular}
  \caption{The probability distribution functions for the WHIM from the OverWhelmingly Large Simulations project (and the complete sample of averaged GRG radio lobes in hatched purple) as a function of pressure for different cuts in (\subfigletter{a}) temperature and (\subfigletter{b}) particle overdensity as given in the legend. The values plotted for the simulations are volume weighted, normalised by the entire volume of the WHIM and divided by the bin size, which is uniform logarithmic steps of $\Delta \log_{10}$Pressure [Pa] $= 0.01$ and $0.2$ for the OWLS and GRG distributions, respectively. This simulation is known as ``AGN\textunderscore L100N512'' in~\citet{Schayeetal2010}, which includes feedback from AGN and supernovae as well as cooling from metal-line emission.}
\label{fig:OWLS}
\end{figure*}

We compare the data with simulations from the OverWhelmingly Large Simulations (OWLS) project~\citep{Schayeetal2010}. These simulations were run with a modified version of the smoothed particle hydrodynamics code Gadget 3~\citep[last described in][]{Springel2005} and include metal line cooling~\citep{Wiersmaetal2009a}, star formation~\citep{SchayeDallaVecchia2008}, chemodynamics~\citep{Wiersmaetal2009b}, supernovae feedback~\citep{DallaVecchiaSchaye2008} and supermassive black hole (SMBH) feedback~\citep{BoothSchaye2009}. This run has been shown to reproduce X-ray and optical observations of groups of galaxies~\citep{McCarthyetal2010,McCarthyetal2011} as well as halo structural properties~\citep{Duffyetal2010,Bryanetal2013} and cold gas distributions~\citep{Duffyetal2012}. The smoothing scale is 2h$^{-1}$ kpc. The OWLS simulation (AGN\textunderscore L100N512) defines the WHIM as all gas with $T>10^5$K, which accounts for 55.5 per cent of the total mass of gas and 31.3 per cent of the gas when weighted by volume. The fraction of the WHIM with pressures less than $1\times10^{-15}$ Pa is 32.0 per cent of the simulation volume and 40.1 per cent of the Universal baryon mass budget, whereas the fraction exceeding $5\times10^{-15}$ Pa (that may be probed by the X-ray studies) is 0.1 per cent by volume and 8.4 per cent by mass. Fig.~\ref{fig:OWLS}(\subfigletter{a}) shows subsets of OWLS pressure values with increasing minimum temperature thresholds. The complete GRG sample is overlaid and lies in a region with a minimum temperature between 10$^{6.5}$ to 10$^{7}$ K (see also Table~\ref{tbl:temperaturecut}). We estimate a WHIM overdensity in the vicinity of our sources of 50 to 500 (see Fig.~\ref{fig:OWLS}\subfigletter{b}; also Table~\ref{tbl:overdensitycut}). This is lower than the overdensities inferred from X-ray observations of filaments (Table~\ref{tbl:xraystudies}) and more than an order of magnitude lower than the overdensities associated with Abell clusters~\citep[$n/\bar{n}>10^3$; e.g.][]{ZakamskaNarayan2003,Akamatsuetal2012}. The suggested overdensities are substantially higher than the galaxy overdensities derived in the two previous case studies~\citep{Subrahmanyanetal2008,Safourisetal2009}. Our future redshift survey using the AAOmega will determine whether this is common and whether it is due to unusual gas densities or possible non-equilibrium conditions in these environments.

\begin{table}
 \caption{Proportion of the WHIM selected by temperature from the OWLS simulation (AGN\textunderscore L100N512) within the range of average GRG radio lobe pressures ($1.3\times10^{-15}$ to $2.0\times10^{-14}$ Pa).}\label{tbl:OWLstatisticsTemperature}
 \begin{tabular}{ccc}
 \hline
  Minimum temperature                   & \multicolumn{2}{c}{Overlap of the OWLS data sets:}         \\
  threshold                                          & Volume-weighted & Mass-weighted \\
  ($\log_{10}$ K)                                & (per cent)               & (per cent)            \\
 \hline
  5.0                                                     & 0.7                        &14.2                     \\
  5.5                                                     & 1.4                        &21.0                     \\
  6.0                                                     & 3.4                        &35.1                     \\
  6.5                                                     & 7.1                        &35.7                     \\
  7.0                                                     & 8.2                        &10.1                     \\
  \hline
 \end{tabular}
 \label{tbl:temperaturecut}
\end{table}

\begin{table}
 \caption{Proportion of the WHIM selected by overdensity from the OWLS simulation (AGN\textunderscore L100N512) within the range of average GRG radio lobe pressures ($1.3\times10^{-15}$ to $2.0\times10^{-14}$ Pa).}\label{tbl:OWLstatisticsOverdensity}
 \begin{tabular}{ccc}
  \hline
   Particle overdensity            & \multicolumn{2}{c}{Overlap of the OWLS data sets:}         \\
   interval                                  & Volume-weighted & Mass-weighted  \\
  ($\log_{10}$ $n/\bar{n}$)    & (per cent)               & (per cent)             \\
  \hline
   -1 to 0                                    & 0.0                          & 0.0                     \\
   0 to 1                                     & 0.1                          & 0.1                     \\
   1 to 2                                     & 6.5                          & 13.9                   \\
   2 to 3                                     & 62.0                        & 51.5                    \\
   3 to 4                                     & 2.4                          & 1.8                     \\
  \hline
 \end{tabular}
 \label{tbl:overdensitycut}
\end{table}

\begin{figure*}
  \centering
  \includegraphics[width=\hsize]{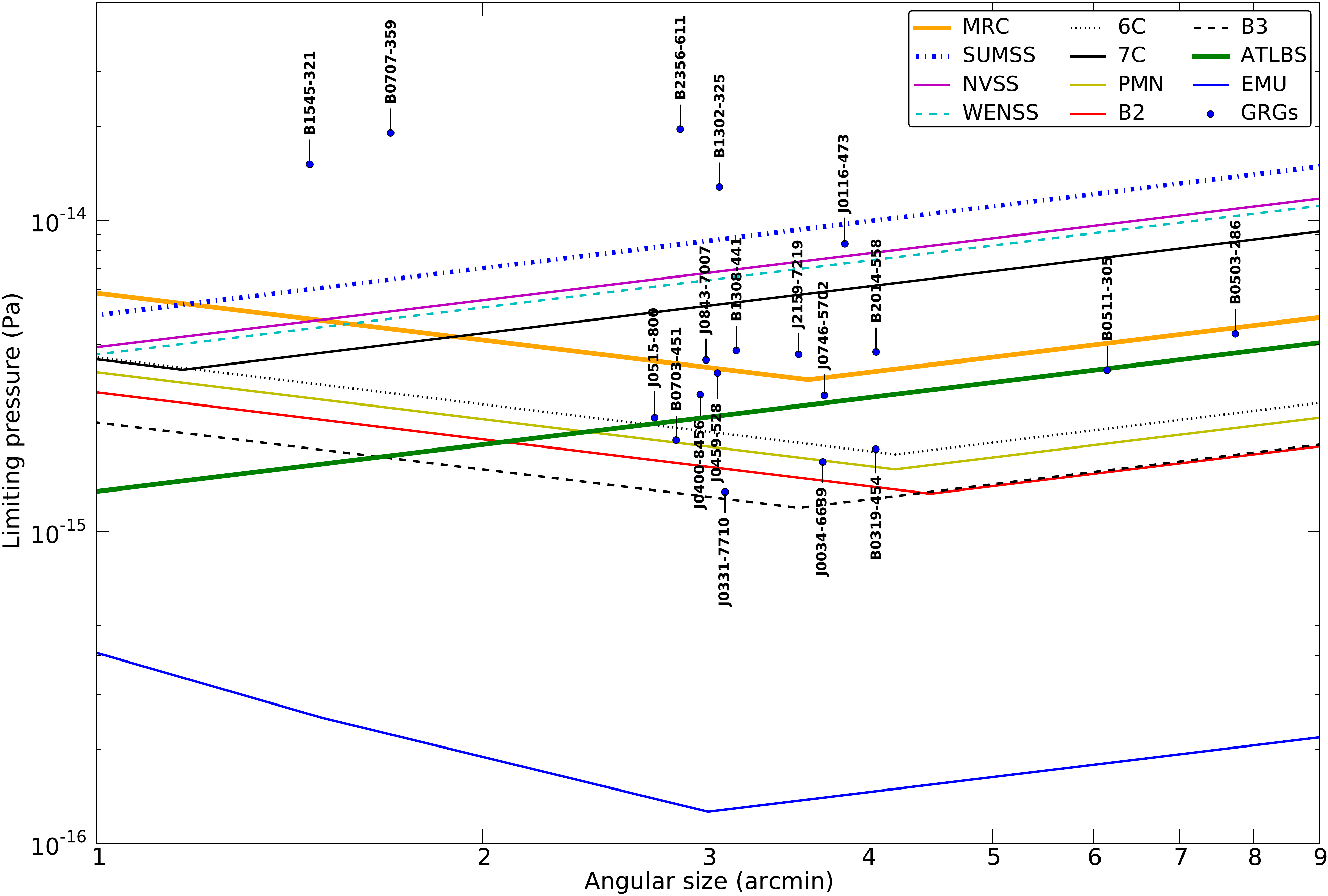}
  \caption{Lines show limiting pressure against source angular size for several surveys computed for a redshift of z = 0.105 (the median redshift of our GRG sample). Each limit is based on a corresponding 5-$\sigma$ flux sensitivity~\citep[equivalent at 1.4 GHz; see][]{Subrahmanyanetal2007}. We assume a zenith angle of \mbox{ZA = 0} and/or declination of $\pm90$ degrees in determining beam size. The projects from which our GRG sample has been compiled (MRC, SUMSS and ATLBS) are displayed in the broadest line width. Averaged lobe pressures and angular sizes for the complete sample of 19 GRGs are plotted as filled circles.}
  \label{fig:surveylimitingpressure}
\end{figure*}

Limiting pressures corresponding to the 5-$\sigma$ flux sensitivities of several surveys are presented in Fig.~\ref{fig:surveylimitingpressure}. We computed the pressures using the median GRG host redshift of 0.105 and 5-$\sigma$ flux sensitivity thresholds (equivalent at 1.4 GHz) versus angular source size provided in~\citet[][]{Subrahmanyanetal2007}. It is assumed that the flux limit is constant for unresolved sources, and the peak flux density is reduced by the ratio of source area to beam area for larger sources. On the plot we have also overlaid the Australian Square Kilometre Array Pathfinder: Evolutionary Map of the Universe (ASKAP-EMU) survey for comparison based on estimated 5-$\sigma$ flux sensitivities~\citep{Norrisetal2011}. Since both the GRG lobe pressures and the survey sensitivity plot have been computed using the same assumptions, the systematic errors will be the same in both. This plot suggests that we may discover more GRGs with lower pressures once that parameter space window is opened by forthcoming surveys like the ASKAP-EMU, which are expected to have substantially lower surface brightness.

ASKAP is expected to discover radio galaxies with lobe pressures more than an order of magnitude below what we have measured for our sample, if such sources do exist. If such radio galaxies are confined by the WHIM gas then from Fig.~\ref{fig:OWLS}, the overdensities inferred for the WHIM will be of order 50 or smaller. If gas scales with galaxies, such overdensities represent sparse environments like poor groups or even the field~(\citealp[e.g.][]{Hoggetal2003}\citealp[; also][who list overdensities for Hickson's compact groups]{Palumboetal1995}). Our comparison of GRG lobe pressures with WHIM gas simulations therefore suggest that ASKAP is likely to find radio galaxies with significantly lower pressures and hence low surface brightness. The study presented herein is, therefore, suggesting a potential science case for ASKAP-EMU, in that it has the capability to discriminate between models for the cosmological evolution of the IGM and examine the validity of simulations such as the OWLS.

\section{Discussion and Summary}
\label{sec:Discussion}
We have presented ATCA observations of a sample of 12 southern giant radio galaxies. These sources form part of a larger sample of giant radio galaxies that we will use to study the lobe-ambient medium interaction uninfluenced by either the ISM of the hosts or their coronae. Separately we have observed the large-scale galaxy distribution around these sources using the AAOmega instrument on the AAT. These observations will be presented in a separate publication.

Our wideband 2.1 GHz radio observations of the 12 giant radio galaxies have revealed several interesting aspects of these sources and of this population of radio galaxies. An interesting aspect is the faint plume of aligned emission regions seen extending beyond the previously known source extent in as many as a quarter of our sources. These emission regions have a drastically reduced surface brightness compared to the lobe that they accompany by as much as an order of magnitude. They are also mostly seen only on one side and in two cases on the distinctly closer lobe side. It is most dramatic in the case of the FRII giant, B0511--305 where the whole southern lobe is seen embedded within much fainter emission region that extends to more than its entire lobe length. Such features have previously been seen in only few sources, 4C23.56 (seen in both lobes) reported by~\citet{BlundellFabian2011} and J0031.1--6642 and J0031.8--6727 (seen only on one side) by~\citet{Saripallietal2012}. The faint emission revealed beyond the previously known extent of the bright lobes in these sources may represent relic emission from a previous active phase in these sources. The asymmetry in the faint extensions may arise either due to enhanced ambient density on that side (compared to the opposite side), which better preserves past emission regions, or projection related effects where the relic on the receding side is seen at an earlier stage (before it expands enough to drop below the brightness sensitivity limit). The optical spectra of the three GRGs that show the asymmetric faint extensions, however,  do not show evidence of projection in the form of broad emission lines. Our AAOmega observations of the fields surrounding these sources are expected to throw some light on whether environment may be responsible for the asymmetry in the relic emission.

A third of the 12 giants observed show morphologies suggesting restarted nuclear activity (J0034--6639, B0511--305, B0703--451, J2159--7219). Four others also have peculiar morphological features that could suggest restarted activity (J0331--7710, J0400--8456, J0459--528, J0746--5702). This is a large fraction and considering that the sources were selected only on the basis of their large linear size appears to be a particular characteristic of giant radio galaxies. If the giants are also the oldest of radio galaxies it is likely that there is more probability of changes in the AGN output to be experienced and recorded in the morphology.

In the sample observed there are three asymmetric giant radio galaxies that exhibit the same correspondence between the lobe extent and the lobe brightness as reported in previous studies where the closer lobe is also the brighter~\citep{Saripallietal1986,Schoenmakersetal2000b,Laraetal2001,Machalskietal2001,Saripallietal2005,Piryaetal2012}.

Pressure estimates were derived for the most relaxed components of the GRG radio lobes ($1.1\times10^{-15}$ to $2.0\times10^{-14}$ Pa; $80$ to $1500$ cm$^{-3}$ K) assuming equipartition conditions and a lower cutoff in electron Lorentz factor of $\gamma_{min}$$\sim$$10^3$. These pressures extend below those inferred by X-ray observations of galaxy filaments, which are biased towards higher density environments. Assuming that there is a pressure balance between the most diffuse and low surface brightness lobes of the giant radio sources and the ambient medium, we can also compare our results with expected IGM densities. A comparison of the lobe pressures derived for the GRG sources presented here with the WHIM gas properties in the OWLS simulation indicates that the corresponding IGM would have a temperature in excess of $10^{6.5}$--$10^{7}$ K, or a particle overdensity in the range 50 to 500. Such temperature (density) regions account for only 6 (1) per cent of the WHIM by volume or 18 (23) per cent of the WHIM by mass. These relatively small fractions suggests that, if indeed GRGs are in hydrostatic equilibrium, they must occur in relatively unrepresentative parts of the Universe, though away from the extremely rare environments of clusters and bright filaments. This is not unexpected, and is the principle of galaxy bias in $\Lambda$CDM cosmology~\citep{Kaiser1984}. However, we also have to consider the possibility that the GRG lobes are over (or under-) pressured with respect to the WHIM gas. This would modify our conclusions. In a future paper, we therefore plan to reconstruct the density field around each GRG using AAOmega spectroscopic data of the neighbouring objects, which will allow us to model the variation of dynamical pressure around each GRG and further explore the use of GRGs as intergalactic barometers.

\section*{Acknowledgments}

The Australia Telescope Compact Array is part of the Australia Telescope National Facility which is funded by the Commonwealth of Australia for operation as a National Facility managed by CSIRO. We gratefully acknowledge the Australian Astronomical Observatory for use of the Anglo-Australian Telescope and the AAOmega spectrograph. We thank R.~W.~Hunstead for bringing the GRG J0459--528 to our attention, S.~Brough for advice regarding optical data reduction, R.~Dodson for advice regarding deconvolution and the OWLS team for access to their simulation data. SuperCOSMOS was used for identifications of optical counterparts. This research has made use of NASA's Astrophysics Data System and the NASA/IPAC Extragalactic Database (NED), which is operated by the Jet Propulsion Laboratory, California Institute of Technology, under contract with the National Aeronautics and Space Administration. The CASA software is developed by the National Radio Astronomy Observatory (NRAO). Luminosity distance values were computed with E.~L.~Wright's Javascript Cosmology Calculator~\citep[][]{Wright2006}. Parts of this research were conducted by the Australian Research Council Centre of Excellence for All-sky Astrophysics (CAASTRO), through project number CE110001020.

\bibliographystyle{mn2e}
\bibliography{reference}{}

\label{lastpage}